\documentclass[3p,12pt]{elsarticle}
\usepackage{booktabs}
\usepackage[fleqn]{amsmath}
\usepackage{cases}
\usepackage{multirow}
\usepackage{xcolor}
\usepackage[normalem]{ulem}
\usepackage{tabularx}
\usepackage{siunitx}
\usepackage{comment}
\usepackage{algorithm}
\usepackage{algpseudocode}
\usepackage{algorithmicx}
\usepackage{grffile}
\usepackage{rotating}
\usepackage{lineno}
\usepackage{hyperref}
\usepackage{graphicx,enumerate,subfigure}
\usepackage{amssymb}
\usepackage{bm,amsmath}
\usepackage{caption,color,xcolor}
\usepackage{subeqnarray}
\usepackage{multirow}
\usepackage{lscape}
\usepackage{rotating}
\usepackage{graphics}
\usepackage{float}
\usepackage{threeparttable}
\usepackage{soul}
\usepackage{enumitem}
\usepackage{tikz}
\usetikzlibrary{arrows.meta,shapes.geometric,positioning}

\usepackage[percent]{overpic}

\usepackage[capitalize]{cleveref}

\floatname{algorithm}{Algorithm}
\biboptions{numbers,sort&compress} % 压制引用，[1,2,3]--> [1-3]
\journal{XXX}

\makeatletter
\let\originalps@pprintTitle\ps@pprintTitle
\def\ps@pprintTitle{%
  \originalps@pprintTitle
  \let\original@oddfoot\@oddfoot
  \def\@oddfoot{\original@oddfoot\unskip}%
  \let\@evenfoot\@oddfoot
}
\makeatother

\newcounter{bla}

\newcommand{\Kn}{\text{Kn}}

\begin{document}

\begin{frontmatter}
% \title{GSIS: a practical solver for moderate-Mach-number multiscale gas flows in complex engineering systems}
\title{A pressure-based coupled general synthetic iterative scheme for rarefied gas flow simulation}

 \author{Yanbing Zhang} % 12132428@mail.sustech.edu.cn
 \author{Yifan Wen} % 12541066@mail.sustech.edu.cn
 \author{Lei Wu\texorpdfstring{\corref{mycorrespondingauthor}}{}}
 \cortext[mycorrespondingauthor]{Corresponding authors:}
 \ead{wul@sustech.edu.cn}

\address{Department of Mechanics and Aerospace Engineering, Southern University of Science and Technology, Shenzhen 518055, China}

\begin{abstract}
Simulation of rarefied internal gas flows poses substantial computational challenges for many engineering applications with complex geometries such as divertor system for nuclear fusion. Although the general synthetic iterative scheme (GSIS) achieves fast steady-state convergence by coupling kinetic updates with macroscopic synthetic equations, its original density-based implementation suffers from degraded convergence for low-Mach-number internal flows, where large disparities between acoustic and convective time scales dominate. To address this limitation, this work develops a pressure-based coupled (PBC) GSIS. Within each GSIS iteration, a conventional kinetic solver advances the velocity distribution function, from which non-equilibrium constitutive relations are extracted. A PBC macroscopic solver then solves the steady  macroscopic synthetic equations: continuity and momentum equations are solved as a coupled pressure-velocity system under fixed temperature, followed by a segregated temperature solve. A boundary treatment is constructed to maintain strict consistency between kinetic boundary fluxes and macroscopic boundary conditions throughout the macroscopic iterations. The proposed method is validated against direct simulation Monte Carlo (DSMC) data across multiple three-dimensional test cases, including tube-to-vacuum flows, a dynamic gas lock configuration for EUV lithography, divertors in nuclear fusion, and sphere flow at Mach numbers up to 10. Numerical results demonstrate that our GSIS-PBC preserves asymptotic-preserving properties and yields satisfactory accuracy across continuum-to-transitional flow regimes. For low-speed near-continuum internal flows, GSIS-PBC delivers wall clock time speed-ups of up to two orders of magnitude compared with the conventional iterative scheme, and substantially outperforms the original density-based GSIS for challenging low-Mach internal flow problems. For nuclear-fusion divertor flows, GSIS-PBC is faster than the DSMC by four orders of magnitude.
\end{abstract}

\begin{keyword}
 rarefied gas dynamics, general synthetic iterative scheme, pressure-based coupled solver; consistent 
 mesoscopic-macroscopic boundary condition 
\end{keyword}

\end{frontmatter}

% \linenumbers

\section{Introduction}

Rarefied internal gas flows underpin a broad spectrum of modern engineering applications, ranging from extreme ultraviolet (EUV) lithography in semiconductor manufacturing to divertor systems in nuclear fusion~\cite{teng2023pollutant,vasileiadis2016modeling}. Unlike hypersonic external flows, these complex internal flows are characterized by low Mach numbers rather than strong compressibility. Once nonequilibrium effects become significant, the Navier-Stokes-Fourier (NSF) equations are no longer adequate, which motivates the use of the mesoscopic Boltzmann kinetic equations~\cite{alsmeyer1976density,ewart2007mass}. Solving such kinetic equations, however, can be computationally prohibitive. For example, the direct simulation Monte Carlo (DSMC) method needs approximately one week of computation time on \(4{,}096\) CPUs to reach steady state for each pumping-probability configuration in three-dimensional divertor simulations~\cite{tantos2024neutral}.

% 60$\sim$120 hours per case group for dynamic gas lock flows~\cite{teng2023pollutant}, and

This high computational cost arises from three distinct sources. The first is statistical sampling: at low Mach numbers, the mean flow signal is weak relative to thermal fluctuations, so extensive sampling is required to extract it, thereby increasing the computational burden~\cite{baker2005variance}. Deterministic kinetic solvers overcome this challenge~\cite{Aristov2001Direct}. The second source is the absence of the asymptotic-preserving (AP) property~\cite{Jin2022_APschemes}: in the near-continuum regime, the collision time and mean free path are much smaller than the characteristic flow scales, which incurs substantial computational time and memory overhead. Over the past decades, a wide range of multiscale numerical methods have been developed to address this issue, including hybrid NSF-DSMC coupling approaches~\cite{sun2004hybrid}, the unified gas-kinetic scheme and its variants~\cite{Xu2010unified,guo2013discrete,LiuZhu2020JCP}, implicit-explicit Runge-Kutta methods~\cite{dimarcoPareschi2013}, stochastic Fokker-Planck methods~\cite{Gorji2014JCP}, the unified stochastic particle method~\cite{Fei2020UnifiedBGK}, and the general synthetic iterative scheme (GSIS)~\cite{su2020can}.
The third source is that conventional time-marching algorithms are ill-suited to steady-state problems. Since steady-state solutions constitute the primary objective in most applications, a numerical method should avoid redundant transient evolution and converge rapidly toward the steady state. Nevertheless, the multiscale properties alone do not guarantee fast steady-state convergence: explicit implementations remain constrained by time-step stability limits, while implicit acceleration hinges on effective treatment of the coupling between macroscopic variables and numerical fluxes~\cite{zhu2016implicit}.

In addition to being noise-free and possessing the AP property \cite{su2020fast}, the GSIS directly addresses the problem of slow convergence to steady state by accelerating kinetic iterations through macroscopic synthetic equations that retain nonequilibrium corrections. The original concept was first proposed in the diffuse synthetic acceleration method for solving the neutron transport equation, as reviewed in~\cite{ADAMS20023}. This method was subsequently extended to linear rarefied gas flows with flow velocities perpendicular to the computational domain~\cite{valougeorgis2003acceleration,naris2004discrete,lihnaropoulos2007formulation,szalmas2010fast}, including typical cases such as linear Couette and Poiseuille flows. Finally, the approach was further generalized to general nonlinear rarefied gas flow problems~\cite{su2020can,zhu2021general,zhang2026efficient}.
In GSIS, because the NSF constitutive relations are explicitly embedded within the synthetic equation, its diffuse-type operator enables efficient exchange of flow information across the entire computational domain, and fast steady-state convergence is achievable within dozens of kinetic iterations\footnote{Most existing multiscale numerical methods lack this desirable property and therefore fail to fully exploit their power. Only recently have Liu and Xu incorporated this concept into their wave-particle decomposition method, achieving speedups of several orders of magnitude \cite{liu2026waveparticleii}.}. In our previous GSIS method, the macroscopic synthetic equations are solved using density-based NSF solvers. 
For hypersonic flow around an X38-like vehicle, GSIS outperforms the unified gas-kinetic wave-particle method by a factor of approximately 146 in cell-normalized CPU-hour costs at a Knudsen number of \(0.00443\); also for internal flows with simple geometry, e.g., the cavity flow, GSIS outperforms DSMC by a factor of approximately \(3.6\times10^4\) in CPU-hour costs at a Knudsen number of \(0.005\)~\cite{zhang2024efficient}. The advantages of GSIS become even more pronounced as the Knudsen number decreases further. It should be noted that the high-order/low-order method can also accelerate convergence toward the steady-state solution \cite{chacon2017multiscale,taitano2014moment}. In this approach, the stress tensor and heat flux appearing in the macroscopic synthetic equations are directly evaluated from the numerical solution of kinetic equation, supplemented by additional consistency correction terms. However, because the NSF constitutive relations are not explicitly incorporated, it suffers from stability limitations for large time‑step sizes \cite{zeng2023general,SIAM2025}.

Nevertheless, for the dynamic gas lock~\cite{he2026rarefied} and tokamak divertor~\cite{tantos2024neutral} cases, our density-based GSIS suffers from poor convergence performance. Owing to the substantial separation of acoustic and convective time scales characteristic of low-Mach internal flows, considerable stiffness mitigation is required -- a challenge exacerbated for long-duct geometries. This identical issue manifests in continuum flows described by the NSF equations. Within such continuum flow settings, pressure-based coupled (PBC) methods tackle this limitation by enforcing mass and momentum conservation in a single implicit system, wherein pressure and velocity are solved jointly and the dependence of face mass fluxes on these variables is retained~\cite{darwish2009coupled,darwish2014fully}. Optimized for the weakly compressible limit, PBC formulations alleviate the convergence difficulty caused by timescale disparity in low-speed regions~\cite {turkel1987preconditioned,xiao2017fully}. 

Based on this observation and the crucial role of macroscopic synthetic equations in driving the rapid steady-state evolution of kinetic equations, this work develops a GSIS-PBC method for efficient steady-state simulations of complex rarefied internal flows. Three novel treatments are introduced. First, the macroscopic solver first couples pressure and velocity through the mass and momentum equations, and subsequently updates the temperature field independently, while retaining the nonequilibrium stress and heat flux corrections supplied by the kinetic solution. Second, to accommodate the PBC, the GSIS framework is redesigned for improved stability. Third, although the synthetic equation accelerates the exchange of flow information within the bulk region, a new boundary treatment is developed to fully exploit the capabilities of GSIS. This framework incorporates the response of boundary fluxes to variations in macroscopic states into the coupled-macroscopic-mesoscopic solution procedure, rigorously enforces consistency between mesoscopic and macroscopic quantities, and permits synchronous adjustment of boundary states and interior flow fields. The proposed method aims to achieve several orders of magnitude speedup in wall-clock time compared with the conventional density-based GSIS for internal rarefied flow problems. 

The remainder of this paper is organized as follows. The governing equations are presented in Section~\ref{sec:Governing_equations}, followed by a detailed exposition of the GSIS-PBC numerical framework and associated boundary treatment in Section~\ref{sec:gsis}. Section~\ref{sec:num_example} validates the performance of GSIS-PBC via a suite of three‑dimensional test cases encompassing dynamic gas lock and tokamak divertor flows. Finally, conclusions are summarized in Section~\ref{sec:conclusion}.

\section{Governing equations}\label{sec:Governing_equations}

\subsection{The gas kinetic model}\label{subsec:kinetic_model}

Let $f(t,\bm{x},\bm{\xi})$ be the velocity distribution function and $f\,\mathrm{d}\bm{x}\,\mathrm{d}\bm{\xi}$ be the gas mass contained in the corresponding phase-space element, where $t$ denotes time, $\bm{x}\in\mathbb{R}^{3}$ is the spatial coordinate, and $\bm{\xi}\in\mathbb{R}^{3}$ is the molecular velocity. For a monatomic gas, the evolution of $f$ can be described by the kinetic equation following the relaxation-time approximation:
\begin{equation}
 \frac{\partial f}{\partial t}
 +\bm{\xi}\cdot\nabla_{\bm{x}}f
 =\frac{f^{S}-f}{\tau},
 \label{eq:shakhov_model}
\end{equation}
where $\tau$ is the collision relaxation time. The target distribution $f^{S}$ in Eq.~\eqref{eq:shakhov_model} is specified by the Shakhov
model~\cite{shakhov1968} as
\begin{equation}
 f^{S}=f^{\mathrm{eq}}\left[
  1+(1-\Pr)
  \frac{\bm{c}\cdot\bm{q}}{5pRT}
  \left(\frac{|\bm{c}|^2}{RT}-5\right)
 \right].
 \label{eq:shakhov_gS}
\end{equation}
Here $R$ is the specific gas constant and $\Pr$ is the Prandtl number. For the
monatomic ideal gas considered in this work, $\Pr=2/3$. The peculiar molecular
velocity is
$\bm{c}=\bm{\xi}-\bm{u}$, where $\bm{u}$ is the bulk velocity. The local
Maxwell equilibrium distribution is
\begin{equation}
 f^{\mathrm{eq}}
 =\frac{\rho}{(2\pi RT)^{3/2}}
  \exp\!\left(-\frac{|\bm{c}|^2}{2RT}\right).
 \label{eq:equilibrium_distribution}
\end{equation}
The density $\rho$, temperature $T$,
thermodynamic pressure $p$, and heat-flux $\bm{q}$ appearing in Eqs.~\eqref{eq:shakhov_gS} and \eqref{eq:equilibrium_distribution}, together with $\bm{u}$, are obtained from the moments of $f$ defined in section~\ref{subsec:macroscopic_moments}.

% The Shakhov correction preserves the collision-invariant moments while
% adjusting the heat-flux relaxation to the prescribed Prandtl number.

\subsection{Macroscopic equations}
\label{subsec:macroscopic_moments}

Introduce the collision-invariant vector
\begin{equation}
 \bm{\psi}(\bm{\xi})=
 \left(1,\bm{\xi},\frac{|\bm{\xi}|^2}{2}\right)^{\mathsf T}.
 \label{eq:collision_invariants}
\end{equation}
The conservative variables are the corresponding velocity moments,
\begin{equation}
 \bm{U}
 =\begin{pmatrix}
   \rho\\[1mm]
   \rho\bm{u}\\[1mm]
   \rho E
  \end{pmatrix}
 =\int_{\mathbb{R}^{3}}\bm{\psi}(\bm{\xi})
 f(t,\bm{x},\bm{\xi})\,\mathrm{d}\bm{\xi},
 \label{eq:moment_mono}
\end{equation}
where $E$ is the total energy per unit mass. For monatomic gas, the energy moment gives $E=3RT/2+|\bm{u}|^2/2$, while the scalar pressure is $p=\rho RT$.

The pressure tensor $\bm{\Pi}$ and heat-flux vector $\bm{q}$ follow from the
higher-order moments of the velocity distribution function: 
\begin{equation} \label{eq:stress_q_mono}
\begin{aligned}
 \bm{\Pi}
 &=\int_{\mathbb{R}^{3}}\bm{c}\bm{c}^{\mathsf T}f\,
   \mathrm{d}\bm{\xi}
 =p\bm{I}+\bm{\sigma},
 \\
 \bm{q}
& =\int_{\mathbb{R}^{3}}\frac{|\bm{c}|^2}{2}\bm{c}f\,
   \mathrm{d}\bm{\xi},
\end{aligned}
\end{equation}
where $\bm{I}$ is the identity tensor and
$\bm{\sigma}$ is the deviatoric part of the pressure tensor.

Taking the moments of Eq.~\eqref{eq:shakhov_model} with respect to
$\bm{\psi}$ therefore yields the exact macroscopic balance equations
\begin{equation}
 \begin{aligned}
  \frac{\partial\rho}{\partial t}
  +\nabla\cdot(\rho\bm{u})&=0,\\
  \frac{\partial(\rho\bm{u})}{\partial t}
  +\nabla\cdot\left(
    \rho\bm{u}\bm{u}^{\mathsf T}+p\bm{I}+\bm{\sigma}
   \right)&=\bm{0},\\
  \frac{\partial(\rho E)}{\partial t}
  +\nabla\cdot\left[
    (\rho E+p)\bm{u}+\bm{\sigma}\cdot\bm{u}+\bm{q}
   \right]&=0,
 \end{aligned}
 \label{eq:macro_mono}
\end{equation}
which contains no constitutive approximation: the
stress and heat flux are only the velocity moments in Eq.~\eqref{eq:stress_q_mono}, rather then expressed in terms of the conservative variables and their spatial derivatives.

The degree of gas rarefaction is characterized by the Knudsen number
\begin{equation}
 \Kn
  =\frac{\lambda}{L_{\mathrm{ref}}}
 =\frac{\mu}{p L_{\mathrm{ref}}}
  \sqrt{\frac{\pi RT}{2}},
 \label{eq:knudsen_number}
\end{equation}
where $L_{\mathrm{ref}}$ is the characteristic length, $\lambda$ is the molecular mean free path, and $\mu$ is the dynamic viscosity. In the near-continuum regime, where $\Kn\ll1$, a first-order Chapman--Enskog expansion of Eq.~\eqref{eq:shakhov_model} recovers the NSF constitutive relations~\cite{chapman1990mathematical},
\begin{equation}
 \begin{aligned}
  \bm{\sigma}^{\mathrm{NSF}}
  &=-\mu\left[
   \nabla\bm{u}+(\nabla\bm{u})^{\mathsf T}
   -\frac{2}{3}(\nabla\cdot\bm{u})\bm{I}
  \right],\\
  \bm{q}^{\mathrm{NSF}}
  &=-\kappa\nabla T.
 \end{aligned}
 \label{eq:nsf_constitutive_relations}
\end{equation}
The corresponding transport coefficients satisfy
\begin{equation}
 \mu=p\tau,
 \qquad
 \kappa=\frac{5}{2}\frac{\mu R}{\Pr}.
 \label{eq:ce_transport_coefficients}
\end{equation}
The first relation connects the relaxation time to the viscosity, whereas the
Shakhov correction enforces the prescribed Prandtl number and thereby fixes
$\kappa$ relative to $\mu$. The temperature dependence of the viscosity is
modeled by the power law
\begin{equation}
 \mu(T)=\mu_{\mathrm{ref}}
 \left(\frac{T}{T_{\mathrm{ref}}}\right)^{\omega},
 \label{eq:viscosity_powerlaw}
\end{equation}
where $T_{\mathrm{ref}}$ is the reference temperature,
$\mu_{\mathrm{ref}}=\mu(T_{\mathrm{ref}})$ is the reference viscosity, and
$\omega$ is the viscosity index.

\section{Pressure-based coupled GSIS}\label{sec:gsis}

The GSIS accelerates steady kinetic calculations by coupling the kinetic iteration with macroscopic synthetic equations \cite{su2020can,zhang2024efficient}. In the present method, a PBC solver advances the macroscopic synthetic equations by treating the continuity and momentum equations as a coupled pressure--velocity system and then solving the temperature equation. To accommodate the PBC solver, the GSIS framework is modified for numerical stability, followed by the detailed PBC formulation and the kinetic-macroscopic boundary treatment.

\subsection{The GSIS framework}\label{subsec:GSIS}

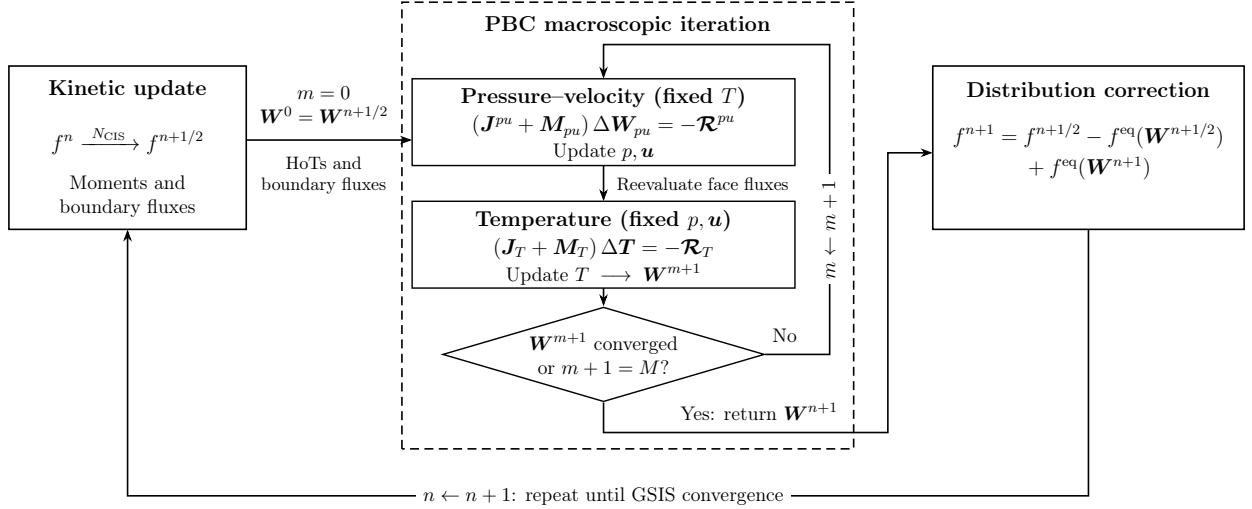
\begin{figure}[!t]
\centering
\resizebox{\linewidth}{!}{%
\begin{tikzpicture}[font=\normalsize,
 flow/.style={-{Stealth[length=2.4mm,width=1.7mm]},line width=.9pt},
 heading/.style={font=\normalsize\bfseries,align=center},
 note/.style={font=\small,align=center},
 fine/.style={font=\footnotesize,align=center},
 decision/.style={diamond,draw,line width=.8pt,aspect=3.4,
   align=center,font=\small,inner sep=1.8pt,
   minimum width=5.3cm,minimum height=1.15cm}]

% Kinetic stage: the step count belongs on the transition, not in a paragraph.
\draw[line width=.9pt] (0,5.10) rectangle (4.65,8.30);
\node[heading] at (2.325,7.83) {Kinetic update};
\node at (2.325,6.85) {$\displaystyle f^n\xrightarrow{\ N_{\mathrm{CIS}}\ }f^{n+1/2}$};
\node[note] at (2.325,5.75) {Moments and\\boundary fluxes};

% The PBC sequence subsumes the former Figure 2.
\draw[line width=.9pt,dash pattern=on 5pt off 3pt]
 (7.70,.80) rectangle (16.55,9.55);
\node[heading] at (12.125,9.10) {PBC macroscopic iteration};
% \node[note,inner xsep=9pt,inner ysep=5pt] at (11.65,9.15)
%  {Kinetic-consistent boundary fluxes};

\draw[line width=.8pt] (7.90,6.35) rectangle (15.40,8.05);
\node[heading] at (11.65,7.68) {Pressure--velocity (fixed $T$)};
\node at (11.65,7.12)
 {$(\bm J^{pu}+\bm M_{pu})\,\Delta\bm W_{pu}=-\boldsymbol{\mathcal R}^{pu}$};
\node[note] at (11.65,6.62) {Update $p,\bm u$};

\draw[line width=.8pt] (7.90,3.95) rectangle (15.40,5.65);
\node[heading] at (11.65,5.28) {Temperature (fixed $p,\bm u$)};
\node at (11.65,4.73)
 {$(\bm J_T+\bm M_T)\,\Delta\bm T=-\boldsymbol{\mathcal R}_T$};
\node[note] at (11.65,4.22) {Update $T\;\longrightarrow\;\bm W^{m+1}$};
\draw[flow] (11.65,6.35) -- (11.65,5.65);
\node[fine,anchor=west,inner sep=2pt] at (11.83,6.0) {Reevaluate face fluxes};

\node[decision] (stop) at (11.65,2.65)
 {$\bm W^{m+1}$ converged\\or $m+1=M$?};
\draw[flow] (11.65,3.95) -- (stop.north);
\draw[flow] (stop.east) -- (16.08,2.65) -- (16.08,8.72) -- (11.65,8.72) -- (11.65,8.05);
\node[note,anchor=south] at (15.2,2.74) {No};
\node[note,rotate=90,fill=white,inner sep=3pt] at (16.08,5.1) {$m\leftarrow m+1$};

% Initial state and closure are supplied once at the start of each PBC stage.
\draw[flow] (4.65,6.85) -- (7.90,6.85);
\node[note,anchor=south] at (6.17,7.04) {$m=0$\\$\bm W^0=\bm W^{n+1/2}$};
\node[fine,anchor=north] at (6.17,6.67) {HoTs and\\boundary fluxes};

\draw[line width=.9pt] (18.10,5.10) rectangle (24.20,8.30);
\node[heading] at (21.15,7.83) {Distribution correction};
\node at (21.15,6.60) {\setlength{\jot}{3pt}$\displaystyle
 \begin{aligned}
 f^{n+1} ={}& f^{n+1/2}-f^{\mathrm{eq}}(\bm W^{n+1/2})\\
 & {}+f^{\mathrm{eq}}(\bm W^{n+1})
 \end{aligned}$};
% \node[note] at (21.15,4.75) {Cellwise distribution check};
\draw[flow] (stop.south) -- (11.65,1.12) -- (17.22,1.12) -- (17.22,6.60) -- (18.10,6.60);
\node[note,anchor=south] at (14.70,1.23) {Yes: return $\bm W^{n+1}$};

\draw[flow] (21.15,5.10) -- (21.15,-.12) -- (2.325,-.12) -- (2.325,5.10);
\node[note,fill=white,inner sep=4pt] at (11.65,-.12)
 {$n\leftarrow n+1$: repeat until GSIS convergence};

\end{tikzpicture}
}
\caption{GSIS-PBC solution procedure.
Here \(n\) and \(m\) index GSIS and PBC iterations, respectively;
\(N_{\mathrm{CIS}}\) is the number of CIS steps per GSIS iteration, and
\(M\) is the maximum number of PBC iterations.
The primitive state is \(\boldsymbol W=(p,\boldsymbol u,T)^{\mathsf T}\);
\(\Delta\boldsymbol W_{pu}\) and \(\Delta\boldsymbol T\) are the
pressure--velocity and temperature corrections.
The symbols \(\boldsymbol{\mathcal R}\), \(\boldsymbol J\), and
\(\boldsymbol M\) denote residual vectors, approximate Jacobians, and
pseudo-time matrices, respectively.
Here \(f^{\mathrm{eq}}\) denotes the Maxwellian equilibrium distribution.}
\label{fig:gsis_framework}
\end{figure}

As illustrated in Fig.~\ref{fig:gsis_framework}, GSIS-PBC alternates between kinetic and macroscopic updates. Multiple kinetic updates are performed between successive macroscopic corrections to reduce numerical oscillations in the velocity distribution function and improve the stability of the coupled iteration. At the \(n\)th GSIS iteration, the conventional iterative scheme (CIS) performs \(N_{\mathrm{CIS}}\) kinetic iterations to advance the velocity distribution function from \(f^n\) to \(f^{n+1/2}\), from which the macroscopic state, high-order terms (HoTs), and kinetic boundary fluxes are evaluated. These quantities supply the initial state, closure, and boundary information for the PBC macroscopic solver. After the PBC iterations, the updated macroscopic state \(\boldsymbol W^{n+1}\) is used to correct the equilibrium part of \(f^{n+1/2}\), yielding \(f^{n+1}\). The process is repeated until convergence.

For cell \(i\), the macroscopic quantities supplied to the PBC solver are
obtained by taking velocity moments of \(f^{n+1/2}\) according to
Eqs.~\eqref{eq:moment_mono} and \eqref{eq:stress_q_mono}, giving the
conservative state \(\boldsymbol U_i^{n+1/2}\), stress
\(\boldsymbol\sigma_i^{n+1/2}\), and heat flux \(\boldsymbol q_i^{n+1/2}\).
The associated primitive state is denoted by
\(\boldsymbol W_i^{n+1/2}=(p_i^{n+1/2},\boldsymbol u_i^{n+1/2},
T_i^{n+1/2})^{\mathsf T}\).

The NSF stress and heat flux are evaluated from the resulting spatial field
\(\boldsymbol W^{n+1/2}\) using Eq.~\eqref{eq:nsf_constitutive_relations}.
The HoTs, which characterize the non-equilibrium constitutive relations beyond the NSF level, are obtained by subtracting these NSF contributions from the corresponding kinetic moments:
\begin{equation}
 \begin{aligned}
  \mathrm{HoT}_{\sigma,i}^{n+1/2}
  &=
  \boldsymbol\sigma_i^{n+1/2}
  -\left[\boldsymbol\sigma^{\mathrm{NSF}}
  (\boldsymbol W^{n+1/2})\right]_i,\\
  \mathrm{HoT}_{q,i}^{n+1/2}
  &=
  \boldsymbol q_i^{n+1/2}
  -\left[\boldsymbol q^{\mathrm{NSF}}
  (\boldsymbol W^{n+1/2})\right]_i,
 \end{aligned}
 \label{eq:gsis_hot}
\end{equation}
where \([\,\cdot\,]_i\) denotes the cell-wise NSF value evaluated using
gradients reconstructed from neighboring cells. Combining the NSF contributions with the HoTs and omitting the time derivative yields the steady-state macroscopic synthetic equations:
\begin{equation}
 \nabla\cdot
 \begin{bmatrix}
  \rho\boldsymbol u\\[1mm]
  \rho\boldsymbol u\boldsymbol u^{\mathsf T}
  +p\boldsymbol I
  +\boldsymbol\sigma^{\mathrm{NSF}}(\boldsymbol W)
  +\mathrm{HoT}_{\sigma}^{n+1/2}\\[1mm]
  (\rho E+p)\boldsymbol u
  +\left[
   \boldsymbol\sigma^{\mathrm{NSF}}(\boldsymbol W)
   +\mathrm{HoT}_{\sigma}^{n+1/2}
  \right]\cdot\boldsymbol u
  +\boldsymbol q^{\mathrm{NSF}}(\boldsymbol W)
  +\mathrm{HoT}_{q}^{n+1/2}
 \end{bmatrix}
 =\boldsymbol0.
 \label{eq:steady_macroscopic_asymptotic_equations}
\end{equation}
The PBC solver advances these equations with the HoTs held fixed, while updating the NSF stress and heat flux with the current macroscopic state. 
The resulting state \(\boldsymbol W^{n+1}\) is then used to update the kinetic
distribution by replacing its equilibrium contribution. At velocity point
\(k\), this gives
\begin{equation}
 f_{i,k}^{n+1}
 =
 f_{i,k}^{n+1/2}
 -f^{\mathrm{eq}}(\boldsymbol W_i^{n+1/2};\boldsymbol\xi_k)
 +f^{\mathrm{eq}}(\boldsymbol W_i^{n+1};\boldsymbol\xi_k),
 \label{eq:gsis_equilibrium_correction}
\end{equation}
where \(f^{\mathrm{eq}}(\boldsymbol W;\boldsymbol\xi_k)\) denotes the Maxwell
equilibrium distribution at velocity point \(k\). Since this correction does
not guarantee non-negativity, the corrected distribution is checked cellwise.
In practice, if
\(\min_k f_{i,k}^{n+1}<-10^{-8}\rho_i^{n+1}/(RT_i^{n+1})^{3/2}\), all
velocity-space values in cell \(i\) are replaced by
\(f^{\mathrm{eq}}(\boldsymbol W_i^{n+1};\boldsymbol\xi_k)\); otherwise, the
correction in Eq.~\eqref{eq:gsis_equilibrium_correction} is retained. Across the numerical cases considered in this work, negative values arose primarily following the first GSIS correction, were seldom observed after the second correction, and were absent in the converged distribution.

Once GSIS has converged,
the macroscopic state remains unchanged between successive GSIS iterations,
and hence \(\boldsymbol\sigma^{\mathrm{NSF}}\) and
\(\boldsymbol q^{\mathrm{NSF}}\) no longer change. Consequently, the converged
solution of the synthetic equation~\eqref{eq:steady_macroscopic_asymptotic_equations} is exactly the moment equation \eqref{eq:macro_mono} loyal to the kinetic equation \eqref{eq:shakhov_model} in the steady state. 

\subsection{The CIS iteration}\label{subsec:CIS}

At the \(n\)th GSIS iteration, the CIS advances \(f^n\) to the
intermediate distribution \(f^{n+1/2}\). The distribution after \(\ell\)
kinetic updates is denoted by \(f^\ell\), with \(f^0=f^n\) and
\(f^{N_{\mathrm{CIS}}}=f^{n+1/2}\). Let \(\Omega_i\) be a control volume
of volume \(V_i\) and centroid \(\boldsymbol x_i\), and let \(N_i\) denote
the set of face-neighbouring cells of cell \(i\), including the ghost cells
associated with physical boundary faces. The common face of cells \(i\) and \(j\)
is denoted by \(ij\), with centroid \(\boldsymbol x_{ij}\) and area vector
\(\boldsymbol S_{ij}=A_{ij}\boldsymbol n_{ij}\) directed from \(i\) to \(j\).
At velocity point \(k\), the finite-volume form of
Eq.~\eqref{eq:shakhov_model} is
\begin{equation}
 \begin{aligned}
  &\frac{V_i}{\Delta t}
  \left(f_{i,k}^{\ell+1}-f_{i,k}^{\ell}\right)
  +\sum_{j\in N_i}
  \left(\boldsymbol\xi_k\cdot\boldsymbol S_{ij}\right)
  f_{ij,k}^{\ell+1}
  =
  \frac{V_i}{\tau_i^{\ell}}
  \left(f_{i,k}^{S,\ell}-f_{i,k}^{\ell+1}\right),\\
  &f_{ij,k}
  =
  \begin{cases}
   f_{i,k}
   +\varphi_{i,k}\nabla f_{i,k}\cdot
   (\boldsymbol x_{ij}-\boldsymbol x_i),&
   \boldsymbol\xi_k\cdot\boldsymbol S_{ij}\geq0,\\
   f_{j,k}
   +\varphi_{j,k}\nabla f_{j,k}\cdot
   (\boldsymbol x_{ij}-\boldsymbol x_j),&
   \boldsymbol\xi_k\cdot\boldsymbol S_{ij}<0,
  \end{cases}
 \end{aligned}
 \label{eq:cis_fv}
\end{equation}
Here \(\Delta t\) is a pseudo-time step introduced to enhance numerical
stability, and \(\varphi_{i,k}\in[0,1]\) is the Venkatakrishnan
limiter~\cite{venkatakrishnan1995} applied to the cell-centred gradient. The
upwind reconstructed value is selected according to the sign of
\(\boldsymbol\xi_k\cdot\boldsymbol S_{ij}\).

For a physical boundary face \(b\) of area \(A_b\) adjacent to cell \(i\), let
\(\boldsymbol n_b\) be the unit normal pointing out of the gas domain and set
\begin{equation}
 \Xi_b^\pm
 =\{\boldsymbol\xi:\,
 \pm\boldsymbol\xi\cdot\boldsymbol n_b>0\}.
 \label{eq:cis_boundary_half_spaces}
\end{equation}
The boundary distribution used by the CIS is
\begin{equation}
 f_{b,k}
 =
 \begin{cases}
  f_{i,k}^{\mathrm{rec}},
  &\boldsymbol\xi_k\in\Xi_b^+,\\
  f_{b,k}^{\mathrm{bc}},
  &\boldsymbol\xi_k\in\Xi_b^-,
 \end{cases}
 \label{eq:cis_kinetic_boundary_distribution}
\end{equation}
where \(f_{i,k}^{\mathrm{rec}}\) is reconstructed from the adjacent cell
with the same limited reconstruction as in Eq.~\eqref{eq:cis_fv}, while
\(f_{b,k}^{\mathrm{bc}}\) is specified by the prescribed mass-flow, stationary
fully diffuse wall, pressure inlet or outlet, or far-field boundary condition.
Their discrete constructions are given in \ref{app:kinetic_boundary}.

\subsection{PBC solution of the macroscopic synthetic equations}\label{subsec:pbc_solver}

The PBC formulation is used to efficiently solve the steady macroscopic synthetic equations~\eqref{eq:steady_macroscopic_asymptotic_equations} when large disparities between acoustic and convective time scales dominate. Let \(m=0,\ldots,M-1\) denote the PBC
iteration index and \(\boldsymbol W^m=(p,\boldsymbol u,T)^{\mathsf T}\) the current primitive state, where \(M\) is the prescribed maximum number of PBC iterations. The continuity and momentum equations are first solved as a coupled pressure--velocity system~\cite{darwish2009coupled,darwish2014fully}, with temperature held fixed. After pressure and velocity are updated, the face quantities are reevaluated and the scalar temperature equation is solved to obtain \(\boldsymbol W^{m+1}\). Figure~\ref{fig:gsis_framework} summarizes this sequence.

The following subsections construct the internal and boundary face fluxes,
derive the coupled pressure--velocity and temperature systems, and present
their pseudo-time-stabilized solution.

\subsubsection{Finite-volume discretization of internal-face fluxes}
\label{subsubsec:pbc_face_flux}

For an internal face \(ij\), let \(\boldsymbol n_{ij}\) be the unit normal
directed from cell \(i\) to cell \(j\), and let
\(u_n=\boldsymbol u\cdot\boldsymbol n_{ij}\). The face flux corresponding to
Eq.~\eqref{eq:steady_macroscopic_asymptotic_equations} is
\begin{equation}
 \boldsymbol F_{ij}
 =\boldsymbol F_{ij}^{\mathrm{adv}}
 +\boldsymbol F_{ij}^{\mathrm{vis}},
 \label{eq:pbc_conventional_face_flux}
\end{equation}
where
\begin{equation}
 \boldsymbol F_{ij}^{\mathrm{adv}}
 =A_{ij}
 \begin{bmatrix}
  \rho u_n\\[1mm]
  \rho u_n\boldsymbol u+p\boldsymbol n_{ij}\\[1mm]
  (\rho E+p)u_n
 \end{bmatrix},
 \label{eq:pbc_conventional_advective_flux}
\end{equation}
and
\begin{equation}
 \boldsymbol F_{ij}^{\mathrm{vis}}
 =A_{ij}
 \begin{bmatrix}
  0\\[1mm]
  \left[
   \boldsymbol\sigma^{\mathrm{NSF}}
   +\mathrm{HoT}_{\sigma}^{n+1/2}
  \right]\cdot\boldsymbol n_{ij}\\[1mm]
  \left\{
   \left[
    \boldsymbol\sigma^{\mathrm{NSF}}
    +\mathrm{HoT}_{\sigma}^{n+1/2}
   \right]\cdot\boldsymbol n_{ij}
  \right\}\cdot\boldsymbol u
  +\left[
   \boldsymbol q^{\mathrm{NSF}}
   +\mathrm{HoT}_{q}^{n+1/2}
  \right]\cdot\boldsymbol n_{ij}
 \end{bmatrix}.
 \label{eq:pbc_conventional_viscous_flux}
\end{equation}
In Eqs.~\eqref{eq:pbc_conventional_advective_flux} and
\eqref{eq:pbc_conventional_viscous_flux}, the face index is retained only on
the fluxes and geometric quantities; the unindexed flow variables denote
their face values. The
numerical construction below prevents pressure--velocity decoupling associated
with the collocated variable arrangement, applies upwinding to transported
quantities, and accounts for mesh non-orthogonality in the NSF stress and heat
flux.

For the advective flux \(\boldsymbol F_{ij}^{\mathrm{adv}}\),
evaluating \(A_{ij}u_n\) requires the normal velocity at the face. With both
pressure and velocity stored at cell centres, direct velocity interpolation
can leave a checkerboard pressure mode undetected: the discrete pressure
gradients cancel in the cell-centred momentum equations.
The velocity field then does not respond to the pressure jump \(p_j-p_i\)
between adjacent cells, and the interpolated face volume flow, and hence the
mass flux in the continuity equation, contains no direct coupling between
their pressures. Following Rhie and Chow~\cite{rhie1983numerical}, this missing
coupling is restored by adding a compact face-flux correction involving
\(p_j-p_i\). Accordingly, \(A_{ij}u_n\) in
Eq.~\eqref{eq:pbc_conventional_advective_flux} is replaced by
\begin{equation}
 \dot{\mathcal V}_{ij}^{RC}
 =
 \boldsymbol u_{ij}\cdot\boldsymbol S_{ij}
 -\beta_{ij}^{RC}\Delta_{ij}p,
 \label{eq:pbc_rc_volume_flow}
\end{equation}
where \(\boldsymbol u_{ij}
=w_i\boldsymbol u_i+w_j\boldsymbol u_j\).
The cell-centre displacement and its normal projection are
\begin{equation}
 \boldsymbol d_{ij}=\boldsymbol x_j-\boldsymbol x_i,
 \qquad
 d_{n,ij}=\boldsymbol n_{ij}\cdot\boldsymbol d_{ij}>0,
 \label{eq:pbc_face_geometry}
\end{equation}
and the interpolation weights are
\begin{equation}
 w_j
 =\min\!\left\{1,\max\!\left[0,
 \frac{\boldsymbol n_{ij}\cdot(\boldsymbol x_{ij}-\boldsymbol x_i)}
      {d_{n,ij}}
 \right]\right\},
 \qquad
 w_i=1-w_j.
 \label{eq:pbc_interpolation_weights}
\end{equation}
In Eq.~\eqref{eq:pbc_rc_volume_flow}, the first term is the volume flow
obtained from the interpolated velocity, and the second is the Rhie--Chow
pressure correction. The part of the two-cell pressure difference that is not
represented by the reconstructed pressure gradient is
\begin{equation}
 \Delta_{ij}p
 =
 (p_j-p_i)-\overline{\nabla p}_{ij}\cdot\boldsymbol d_{ij},
 \label{eq:pbc_rc_pressure_difference}
\end{equation}
where \(\overline{\nabla p}_{ij}
=w_i(\nabla p)_i+w_j(\nabla p)_j\).
For a linear pressure field, the two terms in  
Eq.~\eqref{eq:pbc_rc_pressure_difference} are equal and the correction 
vanishes. For an odd--even pressure mode, their difference provides the 
missing direct coupling between \(p_i\) and \(p_j\). The correction 
coefficient in Eq.~\eqref{eq:pbc_rc_volume_flow} is
\begin{equation}
 \beta_{ij}^{RC}
 =
 \frac{A_{ij}}{d_{n,ij}}
 \left(
  w_iD_i^{RC}
  +w_jD_j^{RC}
 \right),
 \label{eq:pbc_rc_coefficient}
\end{equation}
where \(D_i^{RC}\) is the momentum mobility of cell \(i\). To motivate its
construction, consider a cell-local approximation of the steady momentum
balance~\cite{rhie1983numerical},
\(\boldsymbol A_{u,i}\delta\boldsymbol u_i\simeq
-V_i\delta(\nabla p)_i\), where \(\boldsymbol A_{u,i}\) represents the local
momentum response to a velocity perturbation.
Approximating this local response by an isotropic scalar scale \(a_{u,i}\) gives \(\delta\boldsymbol u_i\simeq-(V_i/a_{u,i})\delta(\nabla p)_i\), which motivates the mobility \(D_i^{RC}=V_i/a_{u,i}\).
Let \(\mathcal N_i^{\mathrm{int}}\) denote the set
of cells sharing an internal face with cell \(i\), and let
\(\mathcal F_i^{\mathrm b}\) denote its physical boundary faces. The scalar
response is constructed as
\begin{equation}
 \begin{aligned}
 a_{u,i}
 ={}&
 \sum_{j\in\mathcal N_i^{\mathrm{int}}}
 \left[
  \max\!\left(\widehat{\dot m}_{ij},0\right)
  +\frac{4\mu_{ij}A_{ij}}{3d_{n,ij}}
 \right]
 +\sum_{b\in\mathcal F_i^{\mathrm b}}
 \left\|\boldsymbol B_{u,ib}\right\|_2,
%  \\
%  D_i^{RC}={}&\frac{V_i}{a_{u,i}},
 \end{aligned}
 \label{eq:pbc_steady_rc_mobility}
\end{equation}
where \(\mu_{ij}=(\mu_i+\mu_j)/2\), and \(\widehat{\dot m}_{ij}\) is the
most recently evaluated shared face mass flow rate~\cite{denner2020conservative}, including the RC pressure
correction and the selected spatial reconstruction. When no preceding value is available, it is initialized as
\(\widehat{\dot m}_{ij}^{\,0}=(\rho_i+\rho_j)
[(\boldsymbol u_i+\boldsymbol u_j)\cdot\boldsymbol n_{ij}]A_{ij}/4\).
The first internal-face term estimates the convective self-response with the
lagged flow held fixed and its sign determining the donor. The second is the
largest eigenvalue of the two-point viscous velocity block in
Eq.~\eqref{eq:pbc_viscous_block}, representing its maximum directional response.
For a physical boundary face,
\(\boldsymbol B_{u,ib}=[\boldsymbol J_{b,i}^{pu}]_{\mathrm{mom},\boldsymbol u}\)
is the \(3\times3\) velocity subblock of the boundary Jacobian. It is evaluated
at the current cell-centred state with pressure, temperature, and the boundary
consistency corrections held fixed, while retaining the derivatives of the
boundary-condition-dependent incoming-flux mapping. Its matrix \(2\)-norm is
the largest singular value and gives the maximum magnitude of the linearized
momentum-flux response over all unit velocity perturbations. The corresponding
face-integrated boundary Jacobians are given in
\ref{app:boundary_jacobians}.

The advected quantities are evaluated by donor-cell upwinding. The donor
cell and mass flow rate are
\begin{equation}
 \begin{aligned}
 i_{\mathrm{up}}
 &=
 \begin{cases}
  i,&\dot{\mathcal V}_{ij}^{RC}\geq0,\\
  j,&\dot{\mathcal V}_{ij}^{RC}<0,
 \end{cases}
 &\qquad
 \dot m_{ij}
 &=\rho_{ij}\dot{\mathcal V}_{ij}^{RC}.
 \end{aligned}
 \label{eq:pbc_face_mass_flow}
\end{equation}
Here, \(\rho_{ij}=p_{i_{\mathrm{up}}}/(RT_{i_{\mathrm{up}}})\), and the
subscript \(\mathrm{up}\) denotes a quantity evaluated in the donor cell
\(i_{\mathrm{up}}\). The pressure in the momentum flux is interpolated as
\(p_{ij}=w_ip_i+w_jp_j\), and the total enthalpy per unit mass is
\(H=\frac{5}{2}RT+\frac{|\boldsymbol u|^2}{2}\). With the Rhie--Chow volume
flow and donor-cell upwinding, the advective flux in
Eq.~\eqref{eq:pbc_conventional_advective_flux} is evaluated numerically as
\begin{equation}
 \boldsymbol F_{ij}^{\mathrm{adv}}
 =
 \begin{bmatrix}
  \dot m_{ij}\\[1mm]
  \dot m_{ij}\boldsymbol u_{\mathrm{up}}
  +p_{ij}\boldsymbol S_{ij}\\[1mm]
  \dot m_{ij}H_{\mathrm{up}}
 \end{bmatrix},
 \label{eq:pbc_advective_face_flux}
\end{equation}

For the viscous flux \(\boldsymbol F_{ij}^{\mathrm{vis}}\), the cell-centred
velocity and temperature gradients are reconstructed by the least-squares
method. On a non-orthogonal mesh, \(\boldsymbol d_{ij}\) is not generally
aligned with \(\boldsymbol n_{ij}\), so direct interpolation of the
cell-centred gradients does not necessarily reproduce the difference between
the adjacent cell values. For a velocity component or temperature, denoted by
\(\phi\), the resulting face gradient is evaluated using the standard
non-orthogonal treatment~\cite{syrakos2017critical,mathur1997pressure},
\begin{equation}
 \begin{aligned}
 \nabla\phi_{ij}
 &=\overline{\nabla\phi}_{ij}
 +\frac{(\phi_j-\phi_i)
 -\overline{\nabla\phi}_{ij}\cdot\boldsymbol d_{ij}}
 {d_{n,ij}}\boldsymbol n_{ij},\\
 \overline{\nabla\phi}_{ij}
 &=w_i\nabla\phi_i
  +w_j\nabla\phi_j.
 \end{aligned}
 \label{eq:pbc_nonorthogonal_face_gradient}
\end{equation}
The added normal component enforces
\(\nabla\phi_{ij}\cdot\boldsymbol d_{ij}=\phi_j-\phi_i\) without altering the
reconstructed tangential gradient.
The component gradients form the face velocity-gradient tensor used in the
NSF stress, while the temperature gradient gives the NSF heat flux through
Eq.~\eqref{eq:nsf_constitutive_relations}. The face thermal conductivity is
\(\kappa_{ij}=(\kappa_i+\kappa_j)/2\).

The HoTs are interpolated with the same geometric weights,
\begin{equation}
\begin{aligned}
 \mathrm{HoT}_{\sigma,ij}^{n+1/2}
 =&
 w_i\mathrm{HoT}_{\sigma,i}^{n+1/2}
 +w_j\mathrm{HoT}_{\sigma,j}^{n+1/2},
 \\
 \mathrm{HoT}_{q,ij}^{n+1/2}
 =&
 w_i\mathrm{HoT}_{q,i}^{n+1/2}
 +w_j\mathrm{HoT}_{q,j}^{n+1/2}.
 \label{eq:pbc_face_high_order_terms} 
\end{aligned}
\end{equation}
The face-integrated traction and heat flow rate are defined by
\begin{equation}
 \begin{aligned}
 \boldsymbol t_{ij}
 &=
 \left[
  \boldsymbol\sigma_{ij}^{\mathrm{NSF}}
  +\mathrm{HoT}_{\sigma,ij}^{n+1/2}
 \right]\cdot\boldsymbol S_{ij},\\
 Q_{ij}
 &=
 \left[
  \boldsymbol q_{ij}^{\mathrm{NSF}}
  +\mathrm{HoT}_{q,ij}^{n+1/2}
 \right]\cdot\boldsymbol S_{ij}.
 \end{aligned}
 \label{eq:pbc_face_traction_heat}
\end{equation}
Using the centred face velocity
\(\overline{\boldsymbol u}_{ij}=(\boldsymbol u_i+\boldsymbol u_j)/2\), the numerical
viscous flux is
\begin{equation}
 \boldsymbol F_{ij}^{\mathrm{vis}}
 =
 \begin{bmatrix}
  0\\[1mm]
  \boldsymbol t_{ij}\\[1mm]
  \boldsymbol t_{ij}\cdot\overline{\boldsymbol u}_{ij}+Q_{ij}
 \end{bmatrix}.
 \label{eq:pbc_viscous_face_flux}
\end{equation}

For clarity, \(\rho_{ij}\) in Eq.~\eqref{eq:pbc_face_mass_flow} and
\(\boldsymbol u_{\mathrm{up}}\) and \(H_{\mathrm{up}}\) in
Eq.~\eqref{eq:pbc_advective_face_flux} are evaluated from the donor-cell
state. For a higher-order evaluation of the advective flux, the
primitive variables \((p,\boldsymbol u,T)\) may instead be reconstructed from
cells \(i\) and \(j\) to face \(ij\) using a slope-limited reconstruction.
The sign of \(\dot{\mathcal V}_{ij}^{RC}\) selects the reconstructed upwind
state, from which \(\rho_{ij}\), \(\boldsymbol u_{\mathrm{up}}\), and
\(H_{\mathrm{up}}\) are evaluated. The Rhie--Chow volume flow, the centred
pressure \(p_{ij}\), and the viscous flux in
Eq.~\eqref{eq:pbc_viscous_face_flux} retain the forms given above.
Together, Eqs.~\eqref{eq:pbc_advective_face_flux} and
\eqref{eq:pbc_viscous_face_flux} define the numerical flux on an internal
face. 
% The boundary fluxes are specified in Section~\ref{subsubsec:pbc_boundary_flux}.

\subsubsection{Consistent mesoscopic--macroscopic boundary flux}
\label{subsubsec:pbc_boundary_flux}

While the macroscopic synthetic equations boost flow information exchange within the bulk region, a consistent mesoscopic--macroscopic boundary would fully leverage GSIS to improve convergence toward steady state.
Following the work of Zhang et
al.~\cite{zhang2026efficient}, the macroscopic boundary flux is updated during
the PBC iterations while remaining consistent with the kinetic boundary
condition imposed in the CIS. In the half-range integrals below, the
superscripts \(+\) and \(-\)
indicate integration over the outgoing and incoming velocity half-spaces
\(\Xi_b^+\) and \(\Xi_b^-\), respectively, as defined in
Eq.~\eqref{eq:cis_boundary_half_spaces}. The same signs label the
corresponding face-integrated boundary-flux components. The continuous
equilibrium half-range flux per unit area is
\begin{equation}
 \boldsymbol F_{\mathrm{eq}}^\pm(\boldsymbol W;\boldsymbol n_b)
 =
 \int_{\Xi_b^\pm}
 (\boldsymbol\xi\cdot\boldsymbol n_b)
 \boldsymbol\psi(\boldsymbol\xi)
 f^{\mathrm{eq}}(\boldsymbol W;\boldsymbol\xi)\,\mathrm d\boldsymbol\xi.
 \label{eq:pbc_boundary_half_moment}
\end{equation}
The corresponding face-integrated continuous outgoing flux is
\begin{equation}
 \widetilde{\boldsymbol F}_b^+(\boldsymbol W)
 =A_b\boldsymbol F_{\mathrm{eq}}^+
 (\boldsymbol W;\boldsymbol n_b).
 \label{eq:pbc_boundary_continuous_outgoing_flux}
\end{equation}
At the start of the PBC stage, set \(\boldsymbol W_i^0=
\boldsymbol W_i^{n+1/2}\). The face-integrated outgoing and incoming kinetic
boundary fluxes are stored as
\begin{equation}
 \begin{aligned}
  \boldsymbol F_b^{+,m=0}
  &=A_b\sum_{\boldsymbol\xi_k\in\Xi_b^+}
  \varpi_k(\boldsymbol\xi_k\cdot\boldsymbol n_b)
  \boldsymbol\psi(\boldsymbol\xi_k)f_{b,k}^{n+1/2},\\
  \boldsymbol F_b^{-,m=0}
  &=A_b\sum_{\boldsymbol\xi_k\in\Xi_b^-}
  \varpi_k(\boldsymbol\xi_k\cdot\boldsymbol n_b)
  \boldsymbol\psi(\boldsymbol\xi_k)f_{b,k}^{n+1/2},
 \end{aligned}
 \label{eq:pbc_boundary_initial_half_fluxes}
\end{equation}
where \(\varpi_k\) is the velocity-space quadrature weight. A tilde denotes a
face-integrated continuous half-space flux, whereas
\(\boldsymbol F_b^{\pm,m}\) denotes the corresponding PBC boundary flux. The
difference between each stored discrete flux and its continuous counterpart
at \(m=0\) is retained as the boundary consistency correction during the PBC
iterations.

During PBC iteration \(m\), the outgoing flux is written as the current
continuous equilibrium half-range flux plus the fixed boundary consistency
correction:
\begin{equation}
 \boldsymbol F_b^{+,m}
 =\widetilde{\boldsymbol F}_b^+(\boldsymbol W_{i,b}^m)
 +\left[
  \boldsymbol F_b^{+,m=0}
  -\widetilde{\boldsymbol F}_b^+(\boldsymbol W_{i,b}^0)
 \right].
 \label{eq:pbc_boundary_outgoing_update}
\end{equation}
Here \(\boldsymbol W_{i,b}^m\) is interpolated to boundary face \(b\) from the
adjacent cell \(i\), using the same limited reconstruction as in
Eq.~\eqref{eq:cis_fv}. Let
\(\widetilde{\boldsymbol F}_b^-
(\boldsymbol W_{i,b},\boldsymbol F_b^+)\) denote the
boundary-condition-dependent mapping for the continuous incoming half-space
flux. The incoming flux is constructed analogously as
\begin{equation}
 \boldsymbol F_b^{-,m}
 =\widetilde{\boldsymbol F}_b^-\!\left(
  \boldsymbol W_{i,b}^m,
  \boldsymbol F_b^{+,m}
 \right)
 +\left[
  \boldsymbol F_b^{-,m=0}
  -\widetilde{\boldsymbol F}_b^-\!\left(
   \boldsymbol W_{i,b}^0,
   \boldsymbol F_b^{+,m=0}
  \right)
 \right].
 \label{eq:pbc_boundary_incoming_update}
\end{equation}
The resulting boundary flux used in the PBC residual is
\begin{equation}
 \boldsymbol F_b^m
 =\boldsymbol F_b^{+,m}
 +\boldsymbol F_b^{-,m}.
 \label{eq:pbc_consistent_boundary_flux}
\end{equation}
At a common fixed point of the kinetic and macroscopic iterations,
the state returned by the PBC solver satisfies
\(\boldsymbol W_i^{n+1}=\boldsymbol W_i^0\); consequently, the reconstructed
boundary state used in the final PBC flux coincides with
\(\boldsymbol W_{i,b}^{0}\).
Equations~\eqref{eq:pbc_boundary_outgoing_update} and
\eqref{eq:pbc_boundary_incoming_update} then recover the stored discrete
reference values of both half-space fluxes. The total macroscopic boundary
flux therefore coincides with the discrete kinetic boundary flux.

The closed-form expressions for \(\boldsymbol F_{\mathrm{eq}}^\pm\) and the
boundary-specific forms of \(\widetilde{\boldsymbol F}_b^-\) for prescribed
mass-flow boundaries, a stationary fully diffuse wall, pressure inlets and
outlets, and far-field boundaries are given in
\ref{app:macroscopic_boundary}.

With the internal and boundary face fluxes specified, the
conservative-variable residual of cell \(i\) is
\begin{equation}
 \boldsymbol{\mathcal R}_i
 =
 \begin{bmatrix}
  \mathcal R_{\rho,i}\\
  \boldsymbol{\mathcal R}_{\mathrm{mom},i}\\
  \mathcal R_{E,i}
 \end{bmatrix}
 =\sum_{j\in\mathcal N_i^{\mathrm{int}}}
  \left(\boldsymbol F_{ij}^{\mathrm{adv}}
  +\boldsymbol F_{ij}^{\mathrm{vis}}\right)
  +\sum_{b\in\mathcal F_i^{\mathrm b}}\boldsymbol F_b.
 \label{eq:pbc_five_moment_residual}
\end{equation}

\subsubsection{Coupled pressure--velocity solution}
\label{subsubsec:pbc_pressure_velocity}

The pressure--velocity residual is formed from the mass and momentum
components of Eq.~\eqref{eq:pbc_five_moment_residual}:
\begin{equation}
 \boldsymbol{\mathcal R}_i^{pu}
 \equiv
 \begin{bmatrix}
  \mathcal R_{\rho,i}\\
  \boldsymbol{\mathcal R}_{\mathrm{mom},i}
 \end{bmatrix},
 \label{eq:pbc_pu_residual_vector}
\end{equation}
where the subscripts \(\rho\) and \(\mathrm{mom}\) select the mass and momentum
components, respectively. The global vector \(\boldsymbol{\mathcal R}^{pu}\)
is assembled from the cell residuals \(\boldsymbol{\mathcal R}_i^{pu}\). Let
\(\boldsymbol W_{pu}\)
collect the cell states
\(\boldsymbol W_{pu,i}=(p_i,\boldsymbol u_i)^{\mathsf T}\), and let
\(\Delta\boldsymbol W_{pu}\) collect the corresponding corrections
\(\Delta\boldsymbol W_{pu,i}=(\Delta p_i,\Delta\boldsymbol u_i)^{\mathsf T}\).
At fixed temperature, the approximate linearization is
\begin{equation}
 \begin{aligned}
  \boldsymbol J^{pu}\Delta\boldsymbol W_{pu}
  &=-\boldsymbol{\mathcal R}^{pu},\qquad
  \boldsymbol J^{pu}=\boldsymbol J_{\mathrm{int}}^{pu}
  +\boldsymbol J_{\mathrm{bnd}}^{pu}
  \approx\frac{\partial\boldsymbol{\mathcal R}^{pu}}
  {\partial\boldsymbol W_{pu}}.
 \end{aligned}
 \label{eq:pbc_spatial_pu_system}
\end{equation}
The nonlinear residual uses the selected spatial reconstruction, whereas the
approximate Jacobian is evaluated from the first-order cell-centred flux
representation.
Here \(\boldsymbol J_{\mathrm{int}}^{pu}\) and
\(\boldsymbol J_{\mathrm{bnd}}^{pu}\) denote the contributions to the approximate
Jacobian from internal and boundary faces, respectively; these two
contributions are constructed below.

For the internal-face part \(\boldsymbol J_{\mathrm{int}}^{pu}\), the mass and
momentum components of Eqs.~\eqref{eq:pbc_advective_face_flux} and
\eqref{eq:pbc_viscous_face_flux} give the cellwise Jacobian action
\begin{equation}
 \left(\boldsymbol J_{\mathrm{int}}^{pu}
 \Delta\boldsymbol W_{pu}\right)_i
 =
 \sum_{j\in\mathcal N_i^{\mathrm{int}}}\sum_{s\in\{i,j\}}
 \begin{bmatrix}
  \boldsymbol J_{\rho,ij,s}\\
  \boldsymbol J_{\mathrm{mom},ij,s}
 \end{bmatrix}
 \Delta\boldsymbol W_{pu,s},
 \label{eq:pbc_pu_internal_jacobian_sum}
\end{equation}
where the \(1\times4\) continuity row and \(3\times4\) momentum block are
\begin{equation}
 \boldsymbol J_{\rho,ij,s}
 =
 \frac{\partial\dot m_{ij}}{\partial\boldsymbol W_{pu,s}},
 \qquad
 \boldsymbol J_{\mathrm{mom},ij,s}
 =
 \frac{\partial}{\partial\boldsymbol W_{pu,s}}
 \left(
  \dot m_{ij}\boldsymbol u_{\mathrm{up}}
  +p_{ij}\boldsymbol S_{ij}
  +\boldsymbol t_{ij}
 \right).
 \label{eq:pbc_pu_face_jacobians}
\end{equation}

The continuity row is evaluated with the face weights,
\(\beta_{ij}^{RC}\), and the reconstructed pressure gradient fixed at the
current nonlinear state. Keeping \(i_{\mathrm{up}}\) fixed during the
linearization gives
\begin{equation}
 \begin{aligned}
  \frac{\partial\dot m_{ij}}{\partial p_i}
  &=\rho_{ij}\beta_{ij}^{RC}
  +\delta_{i,i_{\mathrm{up}}}
  \frac{\dot{\mathcal V}_{ij}^{RC}}{RT_i},
  &\qquad
  \frac{\partial\dot m_{ij}}{\partial p_j}
  &=-\rho_{ij}\beta_{ij}^{RC}
  +\delta_{j,i_{\mathrm{up}}}
  \frac{\dot{\mathcal V}_{ij}^{RC}}{RT_j},\\
  \frac{\partial\dot m_{ij}}{\partial\boldsymbol u_i}
  &=\rho_{ij}w_i\boldsymbol S_{ij}^{\mathsf T},
  &
  \frac{\partial\dot m_{ij}}{\partial\boldsymbol u_j}
  &=\rho_{ij}w_j\boldsymbol S_{ij}^{\mathsf T},
 \end{aligned}
 \label{eq:pbc_mass_flow_jacobian}
\end{equation}
where \(\delta_{s,i_{\mathrm{up}}}\), with \(s\in\{i,j\}\), is the Kronecker
delta.

The NSF part of the face-integrated stress \(\boldsymbol t_{ij}\) in
Eq.~\eqref{eq:pbc_face_traction_heat} is evaluated using the non-orthogonally
corrected gradient in
Eq.~\eqref{eq:pbc_nonorthogonal_face_gradient}. Its approximate linearization
retains the two-point normal response
\begin{equation}
 \Delta\boldsymbol t_{ij}^{\mathrm{NSF}}
 \simeq
 \boldsymbol C_{\mu,ij}
 (\Delta\boldsymbol u_i-\Delta\boldsymbol u_j),\qquad
 \boldsymbol C_{\mu,ij}
 =
 \frac{\mu_{ij}A_{ij}}{d_{n,ij}}
 \left(
  \boldsymbol I_3+\frac{1}{3}
  \boldsymbol n_{ij}\boldsymbol n_{ij}^{\mathsf T}
 \right).
 \label{eq:pbc_viscous_block}
\end{equation}
Here \(\boldsymbol I_3\) is the \(3\times3\) identity matrix. The
non-orthogonal part, transport-coefficient derivatives, and
\(\mathrm{HoT}_{\sigma,ij}^{n+1/2}\) remain in the nonlinear residual but are
not differentiated in this approximate Jacobian. To expand the momentum block,
define
\begin{equation}
 \boldsymbol E_u=
 \begin{bmatrix}\boldsymbol0_{3\times1}&\boldsymbol I_3\end{bmatrix},
 \qquad
 \boldsymbol e_p^{\mathsf T}
 =
 \begin{bmatrix}1&0&0&0\end{bmatrix}.
 \label{eq:pbc_momentum_jacobian_notation}
\end{equation}
Here \(\boldsymbol E_u\) selects the three velocity components of a
pressure--velocity increment, while \(\boldsymbol e_p^{\mathsf T}\) selects
its pressure component.
With \(\zeta_i=1\), \(\zeta_j=-1\), the \(3\times4\) derivative of the oriented
momentum flux with respect to cell \(s\in\{i,j\}\) is
\begin{equation}
 \boldsymbol J_{\mathrm{mom},ij,s}
 =
 \boldsymbol u_{\mathrm{up}}\boldsymbol J_{\rho,ij,s}
 +\dot m_{ij}\delta_{s,i_{\mathrm{up}}}\boldsymbol E_u
 +w_s\boldsymbol S_{ij}\boldsymbol e_p^{\mathsf T}
 +\zeta_s\boldsymbol C_{\mu,ij}\boldsymbol E_u.
 \label{eq:pbc_momentum_face_jacobian}
\end{equation}
The four terms represent, respectively, the response of the mass flow, the
donor velocity, the face pressure, and the two-point NSF stress.

For the boundary part \(\boldsymbol J_{\mathrm{bnd}}^{pu}\), the consistency
corrections in Eqs.~\eqref{eq:pbc_boundary_outgoing_update} and
\eqref{eq:pbc_boundary_incoming_update} are fixed during the PBC
iterations. The boundary residual retains the reconstructed face state
\(\boldsymbol W_{i,b}\), whereas the approximate Jacobian is evaluated with
\(\boldsymbol W_{i,b}=\boldsymbol W_i\). Omitting the iteration superscript,
the cellwise boundary contribution is
\begin{equation}
 \left(\boldsymbol J_{\mathrm{bnd}}^{pu}
 \Delta\boldsymbol W_{pu}\right)_i
 =\sum_{b\in\mathcal F_i^{\mathrm b}}
 \boldsymbol J_{b,i}^{pu}\Delta\boldsymbol W_{pu,i},
 \label{eq:pbc_boundary_pu_sum}
\end{equation}
where the \(4\times4\) boundary-face block is
\begin{equation}
 \boldsymbol J_{b,i}^{pu}
 =\left[
  \left(
   \boldsymbol I_5+
   \frac{\partial\widetilde{\boldsymbol F}_b^-
   (\boldsymbol W_i,\boldsymbol F_b^+)}
        {\partial\boldsymbol F_b^+}
  \right)
  \frac{\partial\widetilde{\boldsymbol F}_b^+(\boldsymbol W_i)}
       {\partial\boldsymbol W_{pu,i}}
  +\frac{\partial\widetilde{\boldsymbol F}_b^-
   (\boldsymbol W_i,\boldsymbol F_b^+)}
        {\partial\boldsymbol W_{pu,i}}
 \right]_{\rho,\mathrm{mom}}.
 \label{eq:pbc_boundary_pu_jacobian}
\end{equation}
Here \(\boldsymbol I_5\) is the \(5\times5\) identity matrix, and the subscript
\((\rho,\mathrm{mom})\) retains the mass and momentum rows. Detailed
expressions for the boundary-specific derivatives are given in
~\ref{app:boundary_jacobians}. Together, the internal and boundary
face sums form \(\boldsymbol J^{pu}\) in
Eq.~\eqref{eq:pbc_spatial_pu_system}.

\subsubsection{Temperature solution}\label{subsubsec:pbc_temperature}

Following the coupled pressure--velocity solve, \(p\) and \(\boldsymbol u\)
are held fixed during a separate temperature update, consistent with the
segregated energy treatment in pressure-based coupled
methods~\cite{mangani2014development}. At fixed pressure, however, the
equation of state couples \(T\) and \(\rho\), so a temperature correction also
changes the density and the bulk kinetic-energy density. To isolate
the thermal balance, the total-energy residual \(\mathcal R_E\) is
combined with the mass and momentum residuals to form the temperature
residual \(\mathcal R_T\), in line with related conservative-residual
transformations used for enthalpy updates~\cite{miettinen2015application}.
The relation \(\rho E=3\rho RT/2+\rho|\boldsymbol u|^2/2\) gives
\begin{equation}
 \frac{3}{2}\rho R\,\mathrm dT
 =
 \mathrm d(\rho E)
 -\boldsymbol u\cdot\mathrm d(\rho\boldsymbol u)
 +\left(\frac{|\boldsymbol u|^2}{2}-\frac{3}{2}RT\right)\mathrm d\rho.
 \label{eq:pbc_temperature_differential}
\end{equation}
Using the conservative-variable transformation in
Eq.~\eqref{eq:pbc_temperature_differential}, with an additional correction for
the discrete kinetic-energy defect, the cell temperature residual is defined as
\begin{equation}
 \begin{aligned}
  \mathcal R_{T,i}
  ={}&\mathcal R_{E,i}
  -\boldsymbol u_i\cdot\boldsymbol{\mathcal R}_{\mathrm{mom},i}
  +\left(K_i-\frac{3}{2}RT_i\right)\mathcal R_{\rho,i}
  -\sum_{j\in\mathcal N_i^{\mathrm{int}}}
   \mathcal K_{i,ij}.
 \end{aligned}
 \label{eq:pbc_temperature_residual}
\end{equation}
Here \(K_i=|\boldsymbol u_i|^2/2\). The conservative-to-temperature residual
transformation introduces the discrete kinetic-energy defect
\(\mathcal K_{i,ij}=\dot m_{ij}|\boldsymbol u_{\mathrm{up}}-\boldsymbol u_i|^2/2\),
where \(\boldsymbol u_{\mathrm{up}}\) is the velocity used in the numerical
advective flux. This defect is subtracted in
Eq.~\eqref{eq:pbc_temperature_residual} to exclude this discretization-induced
contribution from the thermal balance.

Applying the corrected residual definition in
Eq.~\eqref{eq:pbc_temperature_residual} to the internal-face fluxes in
Eqs.~\eqref{eq:pbc_advective_face_flux} and
\eqref{eq:pbc_viscous_face_flux}, and to the complete mesoscopic–macroscopic
boundary flux in Eq.~\eqref{eq:pbc_consistent_boundary_flux}, gives,
respectively,
\begin{equation}
 \begin{aligned}
  F_{T,ij}={}&
  \dot m_{ij}\left(\frac{5}{2}RT_{\mathrm{up}}
    -\frac{3}{2}RT_i\right)
  -p_{ij}\boldsymbol u_i\cdot\boldsymbol S_{ij}
  +\boldsymbol t_{ij}\cdot(\overline{\boldsymbol u}_{ij}-\boldsymbol u_i)
  +Q_{ij},\\
  F_{T,i,b}={}&
  F_{E,b}
  -\boldsymbol u_i\cdot\boldsymbol F_{\mathrm{mom},b}
  +\left(K_i-\frac{3}{2}RT_i\right)F_{\rho,b}.
 \end{aligned}
 \label{eq:pbc_temperature_face_fluxes}
\end{equation}

Let \(\boldsymbol{\mathcal R}_T\), \(\boldsymbol T\), and
\(\Delta\boldsymbol T\) collect the cell residuals, temperatures, and
temperature corrections, respectively. At fixed pressure and velocity, the
approximate Jacobian \(\boldsymbol J_T\approx
\partial\boldsymbol{\mathcal R}_T/\partial\boldsymbol T\) is assembled
cellwise as
\begin{equation}
 \left(\boldsymbol J_T\Delta\boldsymbol T\right)_i
 =\sum_{j\in\mathcal N_i^{\mathrm{int}}}
 \boldsymbol J_{T,ij}^{\mathrm{int}}
 (\Delta T_i,\Delta T_j)^{\mathsf T}
 +\sum_{b\in\mathcal F_i^{\mathrm b}}J_{T,i,b}\Delta T_i
 =-\mathcal R_{T,i}.
 \label{eq:pbc_spatial_temperature_system}
\end{equation}

To construct \(\boldsymbol J_{T,ij}^{\mathrm{int}}\) in
Eq.~\eqref{eq:pbc_spatial_temperature_system}, all quantities in
\(F_{T,ij}\) other than \(T_i\) and \(T_j\) are frozen at the start of the
temperature update, including the mass flow, donor-cell selection, face
traction, transport coefficients, HoTs, and non-orthogonal heat-flux
correction. With the same first-order approximation as in the pressure--velocity
system, the resulting \(1\times2\) row
\(\boldsymbol J_{T,ij}^{\mathrm{int}}
=\partial F_{T,ij}/\partial(T_i,T_j)\) is
\begin{equation}
 \begin{aligned}
  \boldsymbol J_{T,ij}^{\mathrm{int}}
  ={}&R\dot m_{ij}
  \begin{cases}
   \begin{bmatrix}1&0\end{bmatrix},&\dot m_{ij}\geq0,\\[3mm]
   \begin{bmatrix}-\frac{3}{2}&\frac{5}{2}\end{bmatrix},
   &\dot m_{ij}<0
  \end{cases}
  +C_{\kappa,ij}
  \begin{bmatrix}1&-1\end{bmatrix},
 \end{aligned}
 \label{eq:pbc_temperature_internal_jacobian}
\end{equation}
where \(C_{\kappa,ij}=\kappa_{ij}A_{ij}/d_{n,ij}\).
For the boundary face \(b\), the corresponding Jacobian is
\begin{equation}
 \begin{aligned}
  J_{T,i,b}
  ={}&\frac{\partial F_{E,b}}{\partial T_i}
  -\boldsymbol u_i\cdot
   \frac{\partial\boldsymbol F_{\mathrm{mom},b}}{\partial T_i}
  +\left(K_i-\frac{3}{2}RT_i\right)
   \frac{\partial F_{\rho,b}}{\partial T_i}
  -\frac{3}{2}RF_{\rho,b}.
 \end{aligned}
 \label{eq:pbc_boundary_temperature_jacobian}
\end{equation}
Here \(\partial\boldsymbol F_b/\partial T_i\) is the final column of the full
boundary-flux Jacobian in
Eq.~\eqref{eq:app_full_boundary_flux_jacobian}. Its boundary-specific forms
are given by Eqs.~\eqref{eq:app_massflow_boundary_jacobian},
\eqref{eq:app_pressure_boundary_jacobian}, and
\eqref{eq:app_farfield_boundary_jacobian} in
\ref{app:boundary_jacobians}.

\subsubsection{Pseudo-time stabilization and linear solution}
\label{subsubsec:pbc_linear_solution}

The spatial systems derived above are stabilized by block-diagonal local
pseudo-time terms, following the general idea of pseudo-transient continuation~\cite{kelley1998convergence}. For cell \(i\), the pseudo-time step is
\begin{equation}
 \begin{aligned}
  \Delta t_i^{\mathrm{pse}}
  &=
  \frac{\mathrm{CFL}}
  {\Lambda_i^{\mathrm{spec}}}V_i,\\
  \Lambda_i^{\mathrm{spec}}
  &=
  \sum_{j\in N_i}A_{ij}
  \left[
   \left|\boldsymbol u_{ij}\cdot\boldsymbol n_{ij}\right|
   +\chi_{a,ij}c_{s,ij}
   +\frac{8\mu_{ij}}{3\rho_i d_{n,ij}}
  \right],\\
  \chi_{a,ij}
  &=
  \min\!\left[
   1,
   \max\!\left(0.01,\frac{U_{ij}}{c_{s,ij}}\right)
  \right],
  \qquad
  U_{ij}=\sqrt{\frac{|\boldsymbol u_i|^2+|\boldsymbol u_j|^2}{2}},
  \qquad
  c_{s,ij}=\frac{c_{s,i}+c_{s,j}}{2}.
 \end{aligned}
 \label{eq:pbc_local_pseudotime}
\end{equation}
Here \(\Delta t_i^{\mathrm{pse}}\) is the local pseudo-time step,
\(\mathrm{CFL}\) is the prescribed Courant number, and
\(\Lambda_i^{\mathrm{spec}}\) is the corresponding local spectral-radius
estimate. The sound speed is \(c_{s,i}=\sqrt{\gamma RT_i}\), with
\(\gamma=5/3\). The density in the viscous term remains the
cell density \(\rho_i\), consistent with the momentum inertia of cell \(i\).
At a physical boundary face, the velocity, sound speed, and viscosity are
evaluated from the adjacent interior cell. For a boundary face,
\(d_{n,ij}=2|\boldsymbol x_{ij}-\boldsymbol x_i|\).
The prescribed lower bound of
\(0.01\) retains a nonzero acoustic contribution as the local velocity vanishes.
The corresponding cell-local pseudo-time blocks for the pressure--velocity
and temperature systems are
\begin{equation}
 \begin{aligned}
  \boldsymbol M_{pu,i}
  &=
  \frac{V_i}{\Delta t_i^{\mathrm{pse}}}
  \begin{bmatrix}
   0&\boldsymbol0^{\mathsf T}\\
   \boldsymbol0&\rho_i\boldsymbol I_3
  \end{bmatrix},\qquad
  M_{T,i}=
  \frac{3\rho_i R}{2\Delta t_i^{\mathrm{pse}}}V_i.
 \end{aligned}
 \label{eq:pbc_pseudotime_blocks}
\end{equation}
Assembling the cell-local blocks in Eq.~\eqref{eq:pbc_pseudotime_blocks} into
the global block-diagonal matrices \(\boldsymbol M_{pu}\) and
\(\boldsymbol M_T\) gives the stabilized linear systems
\begin{equation}
 \left(\boldsymbol J^{pu}+\boldsymbol M_{pu}\right)
 \Delta\boldsymbol W_{pu}
 =-\boldsymbol{\mathcal R}^{pu},
 \qquad
 \left(\boldsymbol J_T+\boldsymbol M_T\right)
 \Delta\boldsymbol T
 =-\boldsymbol{\mathcal R}_T,
 \label{eq:pbc_global_systems}
\end{equation}
% In practice, these systems can be solved using matrix-free iterative methods available in
% established numerical libraries such as PETSc~\cite{balay1997petsc}.
The resulting linear systems can be solved using standard iterative methods, either implemented directly or provided by existing numerical libraries.

\section{Numerical examples}\label{sec:num_example}

This section assesses the accuracy, robustness, and computational efficiency of GSIS-PBC through four three-dimensional test cases. The tube flow case examines the performance of GSIS-PBC over a wide range of gas rarefaction. The dynamic gas lock tests convergence under prescribed mass-flow boundary conditions, while the divertor case examines large-scale simulation in a complex internal geometry. The sphere flow case evaluates the capability of GSIS-PBC for high-speed flows.

Note that the axisymmetric geometries used here are chosen solely to facilitate validation against reference solutions. All GSIS-PBC simulations use full three-dimensional meshes, and the method is applicable to both axisymmetric and non-axisymmetric configurations.
The DSMC simulations performed in this work use the two-dimensional axisymmetric formulation in SPARTA~\cite{plimpton2019direct} with the variable-hard-sphere collision model.

Convergence is reached when the relative change between two consecutive iterations satisfies
\begin{equation}
 \epsilon
 =\max_{Q\in\{\rho,\boldsymbol u,T\}}
 \left[
 \frac{\sum_i\left|Q_i^n-Q_i^{n-1}\right|^2V_i}
 {\sum_i\left|Q_i^{n-1}\right|^2V_i}
 \right]^{1/2}
 <10^{-5}.
 \label{eq:numerical_convergence}
\end{equation}
A CFL number of \(500\) is used for the PBC pseudo-time step in all cases.
The sphere and tube simulations are performed on an Intel Xeon Silver 4208 CPU@2.10\,GHz platform, while the dynamic gas lock and divertor simulations are performed on an AMD EPYC 7452 CPU@2.35\,GHz platform.
The reported GSIS-PBC wall-clock times include initialization and iterations.
The cost per GSIS iteration varies with the number of linear-solver iterations required for the macroscopic systems in Eq.~\eqref{eq:pbc_global_systems}.

% \subsection{Thermal creep in long curved channel}

% Commun. Comput. Phys. Vol. 6, No. 5, pp. 919-954
% (2009)
% Numerical Simulations of Rarefied Gases in Curved
% Channels: Thermal Creep, Circulating Flow,
% and Pumping Effect

% \subsection{Long complex channel?}

% Using Murray’s law to design artificial
% vascular microfluidic networks
% R. W. Barber, K. Cieślicki  D. R. Emerson

\subsection{Flow through a circular tube into vacuum}\label{subsec:tube}

\begin{figure}[!t]
\centering
\subfigure[Computational domain.]{\includegraphics[width=0.44\linewidth]{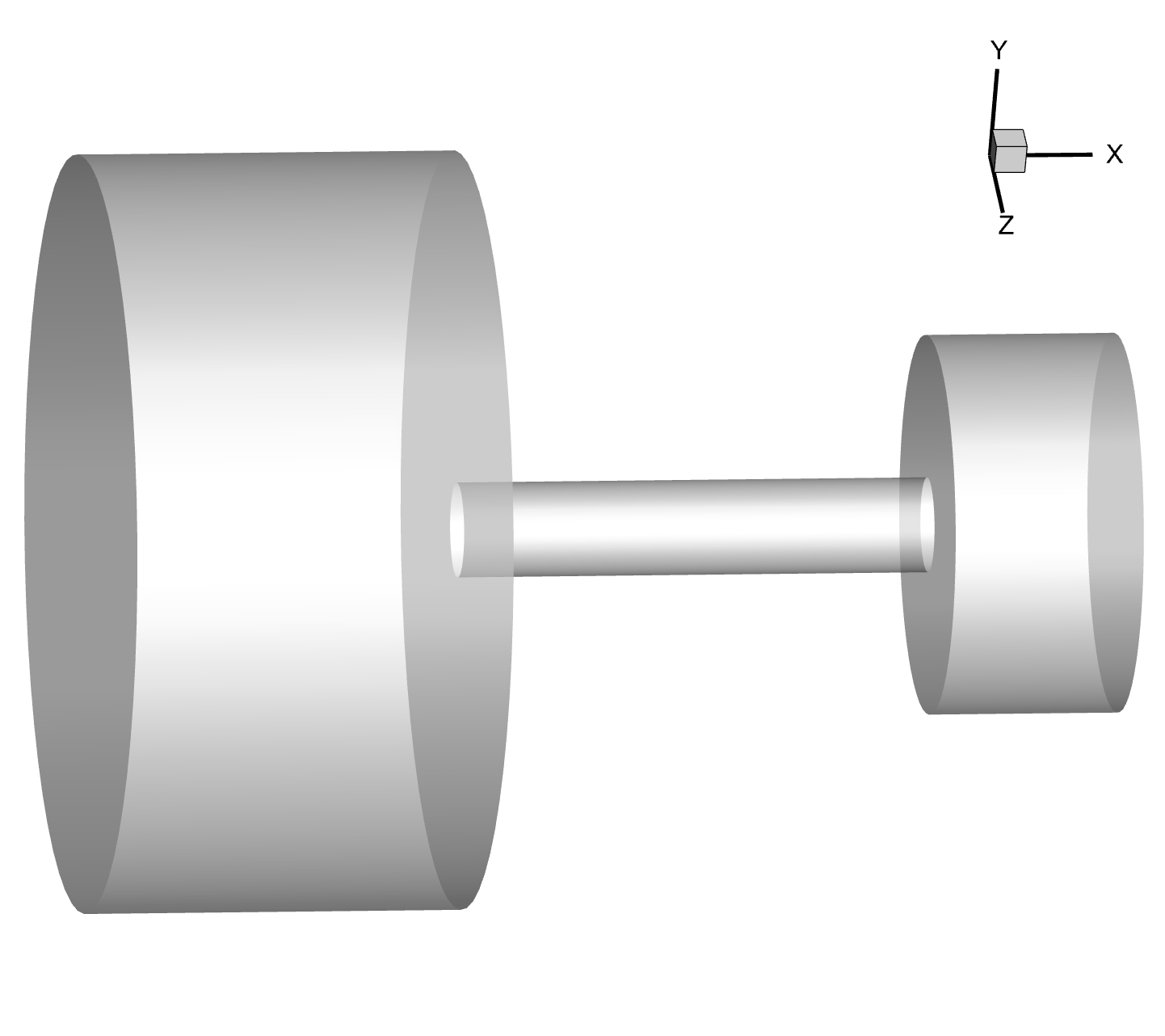}}
\hfill
\subfigure[Dimensionless geometry.]{\includegraphics[width=0.54\linewidth]{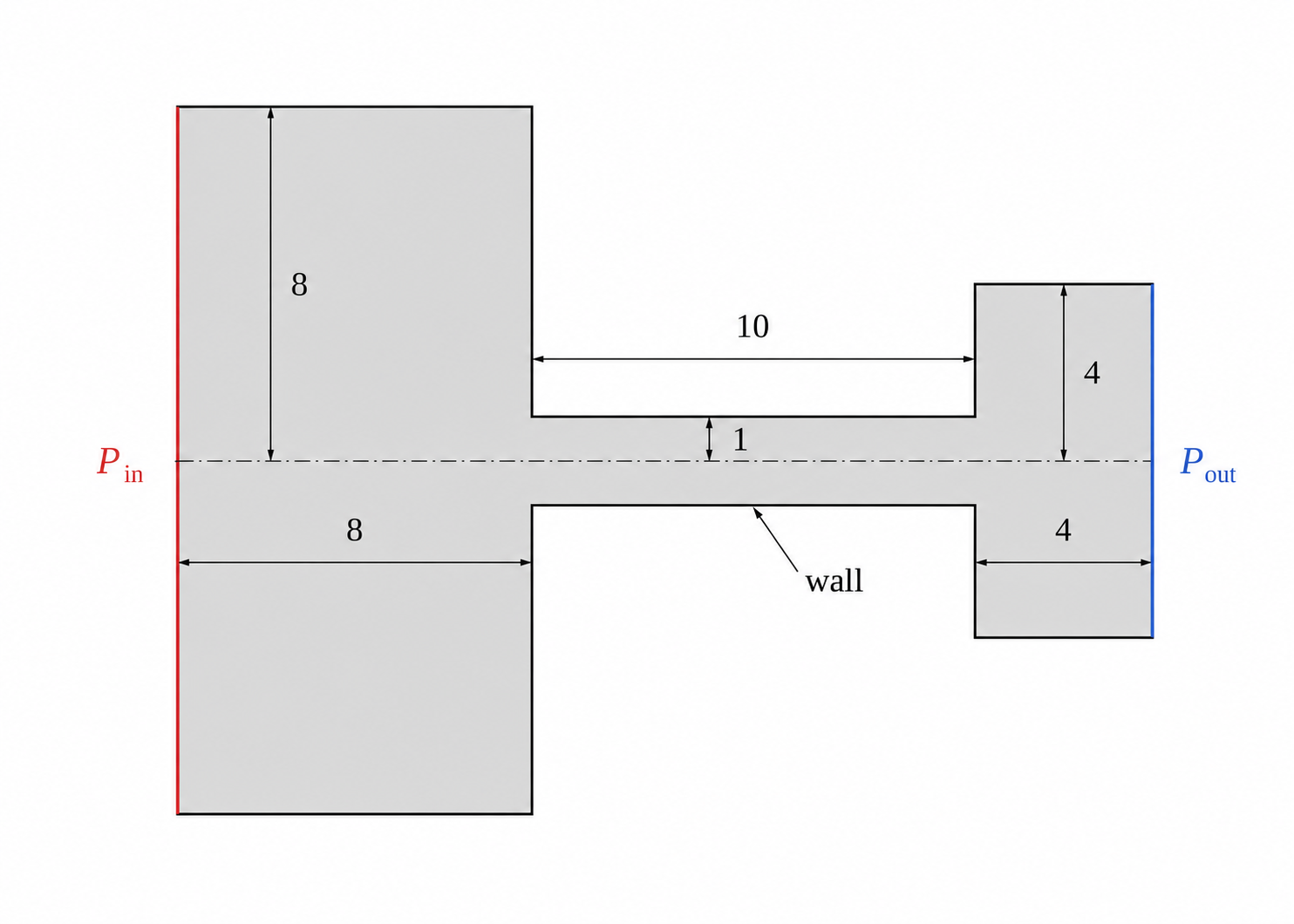}}
\caption{Geometry and dimensions of the tube-flow configuration.
All dimensions are normalized by the tube radius.}
\label{fig:tube_geometry}
\end{figure}

To assess the applicability of GSIS-PBC over a wide range of gas rarefaction, we consider helium flow through a circular tube into vacuum~\cite{varoutis2008rarefied}, see Fig.~\ref{fig:tube_geometry}. The tube radius is taken the \(L_{\mathrm{ref}}\), giving a tube length of 10. The upstream reservoir has a radius and length of 8, while the corresponding dimensions of the downstream reservoir are both 4. The molecular mass of helium is
\(m=6.65\times10^{-27}\,\mathrm{kg}\).
Its viscosity at \(T_{\mathrm{ref}}=T_0=273.15\,\mathrm{K}\) is
\(\mu_0=\mu(T_0)=1.916\times10^{-5}\,\mathrm{Pa\,s}\),
and the viscosity index is \(\omega=0.66\).
At the upstream end, the temperature is fixed at \(T_0\), and the
inlet pressure \(p_{\mathrm{in}}\) is adjusted to obtain the prescribed
rarefaction parameter \(\delta_{\mathrm{rp}}\). A vacuum boundary is applied at
the downstream end. All remaining boundaries are stationary, fully
diffuse walls maintained at \(T_0\).
The parameter \(\delta_{\mathrm{rp}}\) is related to the Knudsen number in
Eq.~\eqref{eq:knudsen_number},
evaluated using the inlet reservoir state and the tube radius, by
\begin{equation}
 \delta_{\mathrm{rp}}=\frac{\sqrt{\pi}}{2\Kn}.
 \label{eq:tube_rarefaction}
\end{equation}
Four cases with \(\delta_{\mathrm{rp}}=1\), \(10\), \(100\), and
\(1{,}000\) are considered.
All four cases use the same three-dimensional spatial mesh containing
\(135{,}000\) cells and a uniform Cartesian velocity grid of
\(30\times24\times24\) points over
\(\bm{\xi}/\sqrt{RT_0}\in[-10,10]\times[-8,8]\times[-8,8]\).
GSIS-PBC and CIS use the same spatial mesh, velocity grid, and convergence criterion.
Each GSIS-PBC iteration consists of five kinetic iterations followed by ten PBC iterations.

\begin{figure}[t]
\centering
{\includegraphics[width=0.48\linewidth]{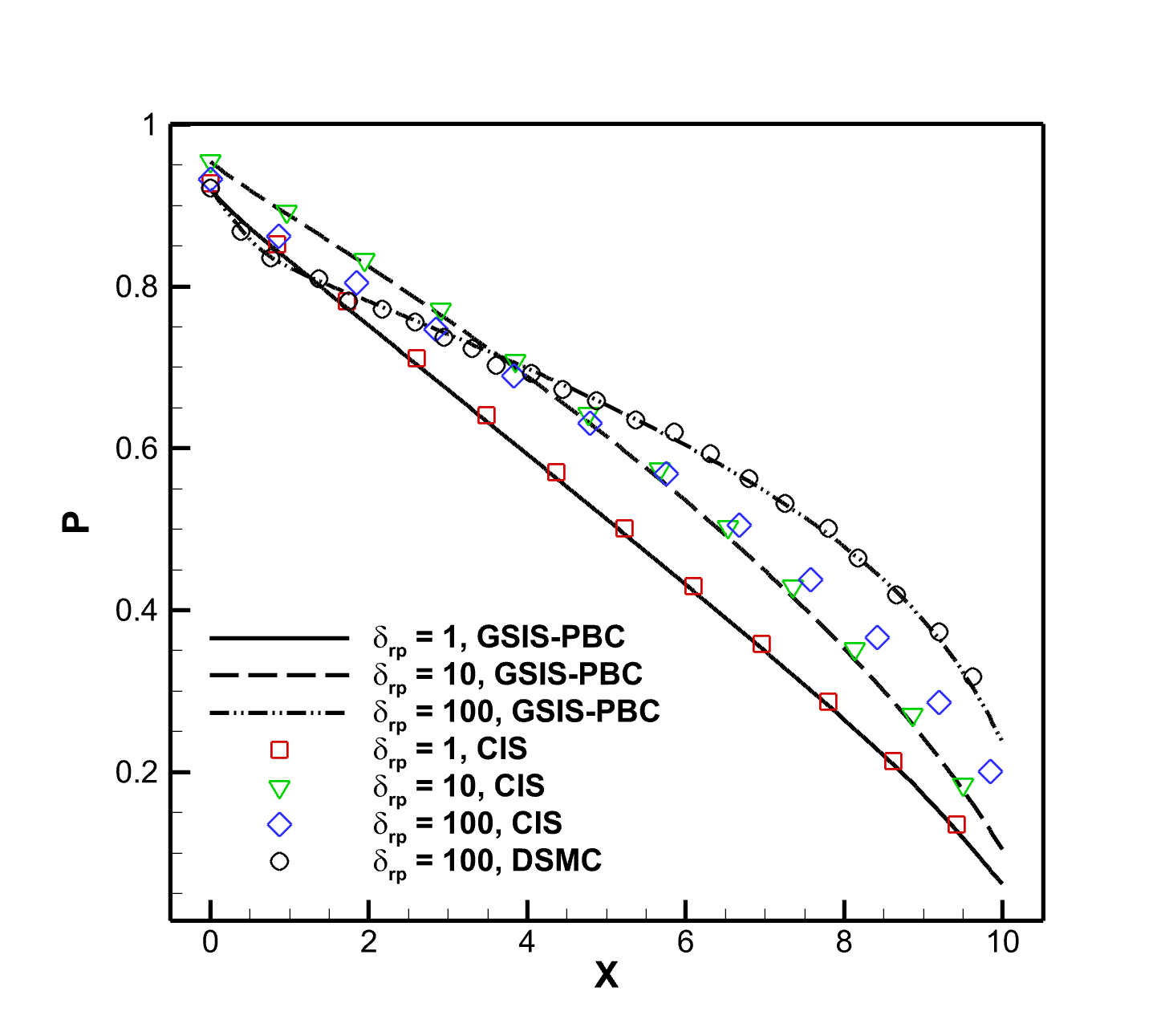}}
{\includegraphics[width=0.48\linewidth]{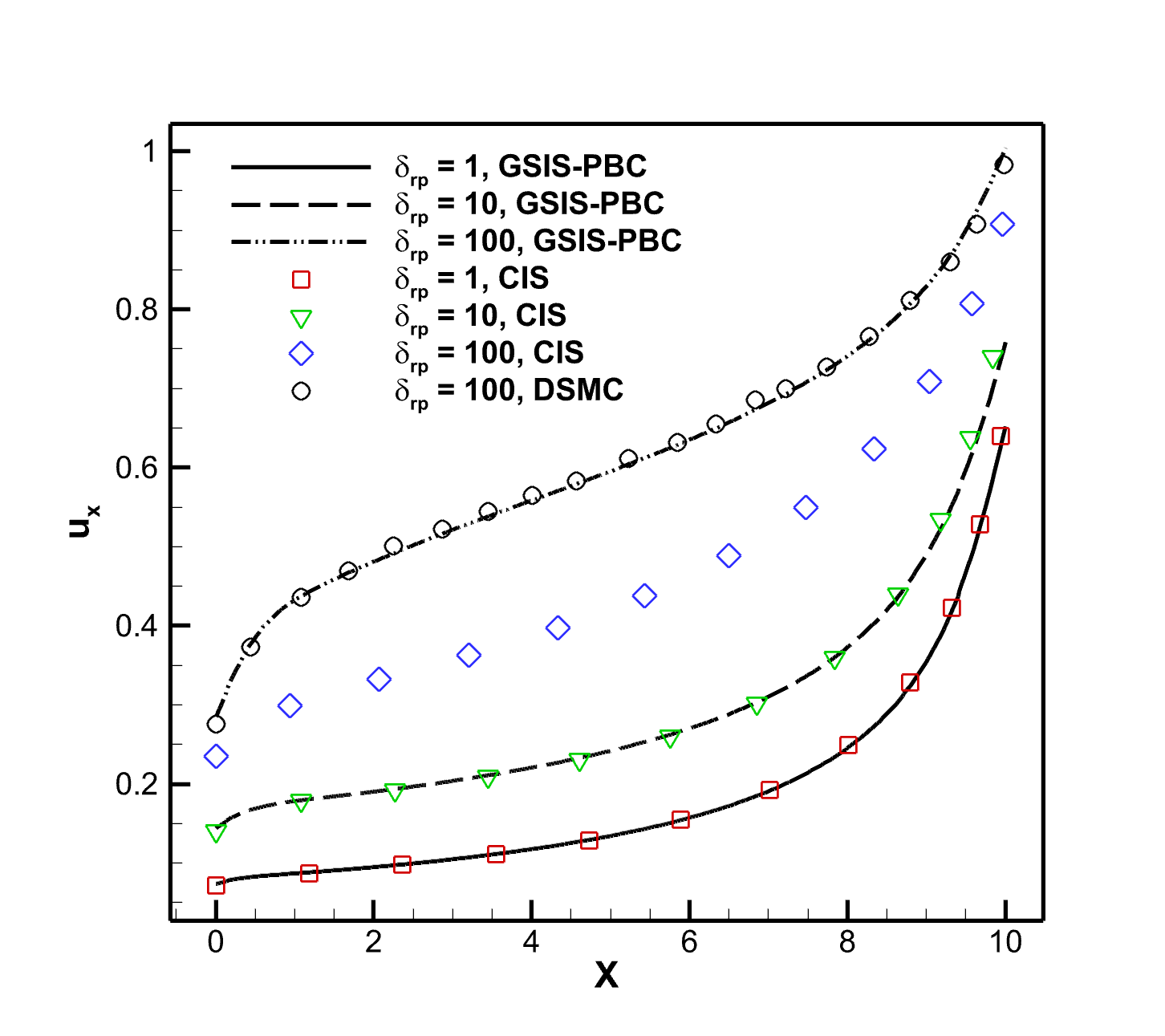}}
\caption{Profiles of pressure and axial velocity along the tube centreline from the inlet (\(X=0\)) to
the outlet (\(X=10\)) at
\(\delta_{\mathrm{rp}}=1\), \(10\), and \(100\). Lines denote GSIS-PBC,
black circles denote DSMC at \(\delta_{\mathrm{rp}}=100\), and the other
symbols denote CIS. Here \(X=x/L_{\mathrm{ref}}\) and \(P=p/p_{\mathrm{in}}\),
and the axial velocity is normalized by \(\sqrt{2RT_0}\).}
\label{fig:tube_centerline}
\end{figure}

\begin{table}[!htbp]
\centering
\caption{Mass-flow comparison and computational costs for tube flow
into vacuum. The reference values are DSMC results  for \(\delta_{\mathrm{rp}}=1\), \(10\), and \(100\), and a NSF
result computed by FLUENT for \(\delta_{\mathrm{rp}}=1{,}000\).
The two-dimensional axisymmetric DSMC simulations use approximately \(0.225\), \(0.899\), and \(4.2\) million spatial cells for \(\delta_{\mathrm{rp}}=1\), \(10\), and \(100\), respectively, with an average of 44 simulation particles per cell.
Using 640 cores, the corresponding wall-clock times are \(0.55\), \(2.19\), and \(10.2\,\mathrm{h}\).
Relative mass-flow errors are \(|\dot M-\dot M_{\mathrm{ref}}|/\dot M_{\mathrm{ref}}\), reported as percentages.
GSIS-PBC iteration counts refer to GSIS iterations, and
\(t\) denotes the total wall-clock time. The speedup of GSIS-PBC relative to CIS
is estimated by dividing the reported total wall-clock time of CIS by that of GSIS-PBC.
}
\label{tab:tube_results}
\small
\setlength{\tabcolsep}{2.7pt}
\begin{tabular}{@{}ccccccccccc@{}}
\toprule
\multirow{2}{*}{\(\delta_{\mathrm{rp}}\)} & \multirow{2}{*}{\(\dot M_{\mathrm{ref}}\)} & \multicolumn{4}{c}{GSIS-PBC} & \multicolumn{4}{c}{CIS} & \multirow{2}{*}{Speedup} \\
\cmidrule(lr){3-6}\cmidrule(lr){7-10}
 & & \(\dot M\) & Error (\%) & Iter. & \(t\) (h) & \(\dot M\) & Error (\%) & Iter. & \(t\) (h) & \\
\midrule
1 & 0.197 & 0.198 & 0.51 & 17 & 0.20 & 0.198 & 0.51 & 189 & 0.43 & \(2.2\times\) \\
10 & 0.330 & 0.326 & 1.21 & 9 & 0.11 & 0.348 & 5.45 & 484 & 1.05 & \(9.5\times\) \\
100 & 0.898 & 0.885 & 1.45 & 7 & 0.09 & 0.597 & 33.52 & \(2{,}796\) & 6.20 & \(68.9\times\) \\
% \(1{,}000\) & 1.117 & 1.263 & -- & 6 & 0.08 & -- & -- & \(5{,}669\) & 12.11 & \(151.4\times\) \\
\(1{,}000\) & 1.267 & 1.275 & 0.63 & 6 & 0.08 & 0.729 & 42.46 & \(7{,}847\) & 16.03 & \(200.4\times\) \\
\bottomrule
\end{tabular}
\end{table}

Figure~\ref{fig:tube_centerline} compares the centreline pressure and axial
velocity obtained with GSIS-PBC and CIS at \(\delta_{\mathrm{rp}}=1\),
\(10\), and \(100\), with DSMC results also shown for
\(\delta_{\mathrm{rp}}=100\). GSIS-PBC and CIS agree closely at
\(\delta_{\mathrm{rp}}=1\) and \(10\). At \(\delta_{\mathrm{rp}}=100\),
GSIS-PBC agrees well with DSMC, whereas CIS exhibits noticeable deviations due to the large numerical dissipation when coarse spatial grids are used,
particularly in the axial velocity.

The dimensionless mass flow rate \(\dot M\) is obtained by normalizing
the total mass flow rate by
\(\sqrt{\pi}L_{\mathrm{ref}}^2p_{\mathrm{in}}/\sqrt{2RT_0}\).
Table~\ref{tab:tube_results} compares the predicted mass flow rates
with the DSMC and NS reference results
and summarizes the iteration counts and total wall-clock times of
GSIS-PBC and CIS.
The relative mass-flow deviations of GSIS-PBC from DSMC are at most \(1.45\%\)
for \(\delta_{\mathrm{rp}}=1\), \(10\), and \(100\), while the deviation
from the NS result is \(0.63\%\) at \(\delta_{\mathrm{rp}}=1{,}000\).
In contrast, CIS underpredicts the reference mass flow rates by
\(33.52\%\) and \(42.46\%\) at \(\delta_{\mathrm{rp}}=100\) and
\(1{,}000\), respectively.
These cases span inlet Knudsen numbers from \(8.86\times10^{-4}\) to \(8.86\times10^{-1}\).
At small Knudsen numbers, traditional kinetic solvers lacking AP properties may introduce substantial numerical dissipation on coarse meshes, leading to underpredicted mass flow rates. 
% \leir{For example, in this case, the largest cell size is XXX times of the mean free path, so the CIS becomes inaccurate. The GSIS, however, can give accurate prediction due to its AP property.}
For example, at \(\delta_{\mathrm{rp}}=100\), the largest cell size is approximately 57 times the inlet mean free path, so the CIS becomes inaccurate. The GSIS, however, can give accurate predictions due to its AP property.

Using 96 cores, GSIS-PBC converges in 6--17 iterations, with
total wall-clock times ranging from \(0.08\) to \(0.20\,\mathrm{h}\).
For \(\delta_{\mathrm{rp}}=1\), \(10\), \(100\), and \(1{,}000\), the
wall-clock speedups over CIS are approximately \(2.2\), \(9.5\),
\(68.9\), and \(200.4\), respectively.
The increasing speedup as the inlet Knudsen number decreases indicates
a growing computational advantage of GSIS-PBC over CIS for the tested cases.

% \subsection{Micro-scale channel}
% Efficient DSBGK simulations of the low speed thermal transpiration gas flows through micro-channels. 
% Experiments on rarefied gas flows through tubes. 
% GSIS-PBC vs. Experiment

\subsection{Flows in dynamic gas lock}\label{subsec:dgl}

\begin{figure}[t]
\centering
\subfigure[Computational domain.]{\includegraphics[height=0.44\linewidth]{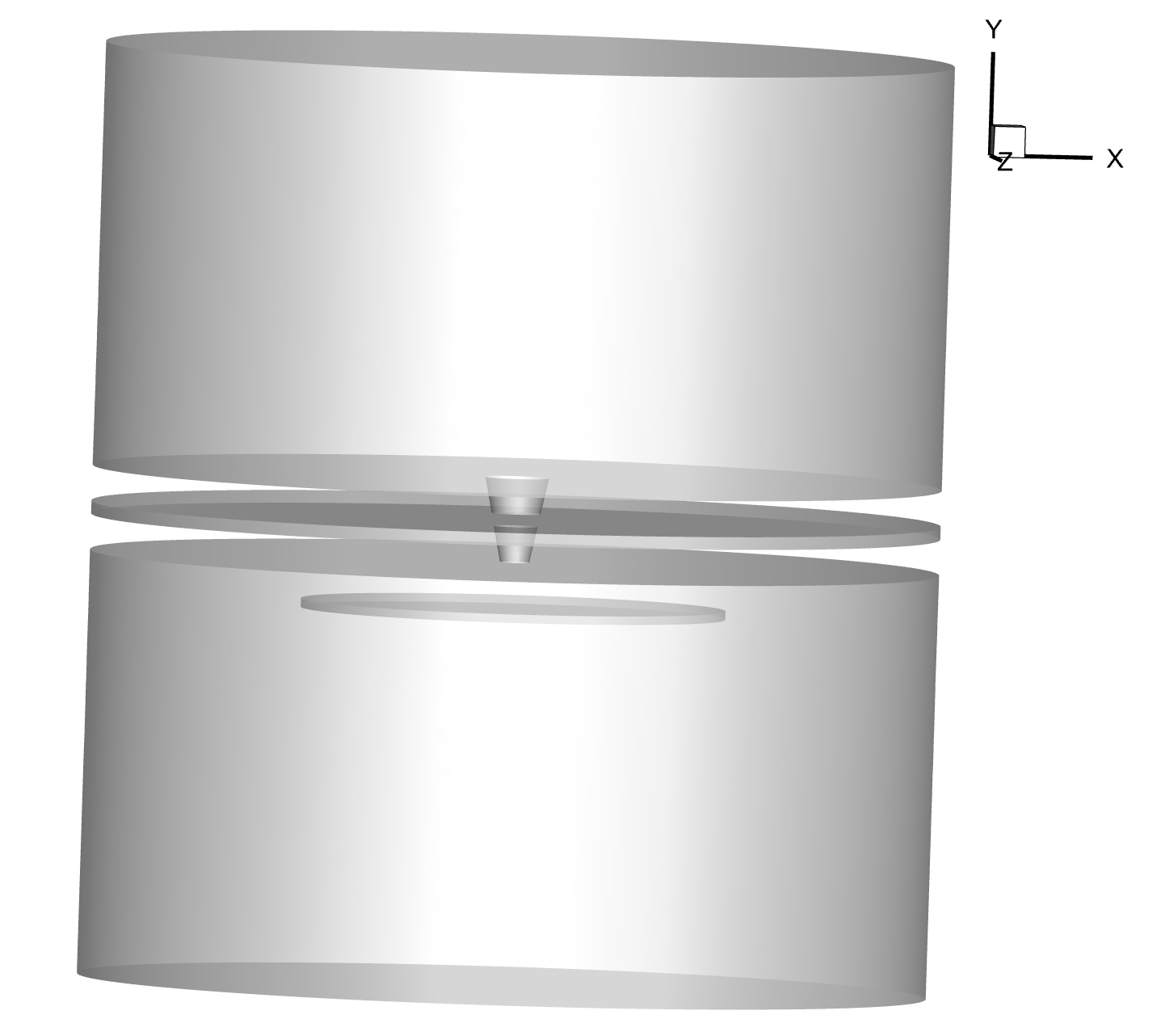}}
\subfigure[Cross-sectional geometry.]{\includegraphics[height=0.46\linewidth]{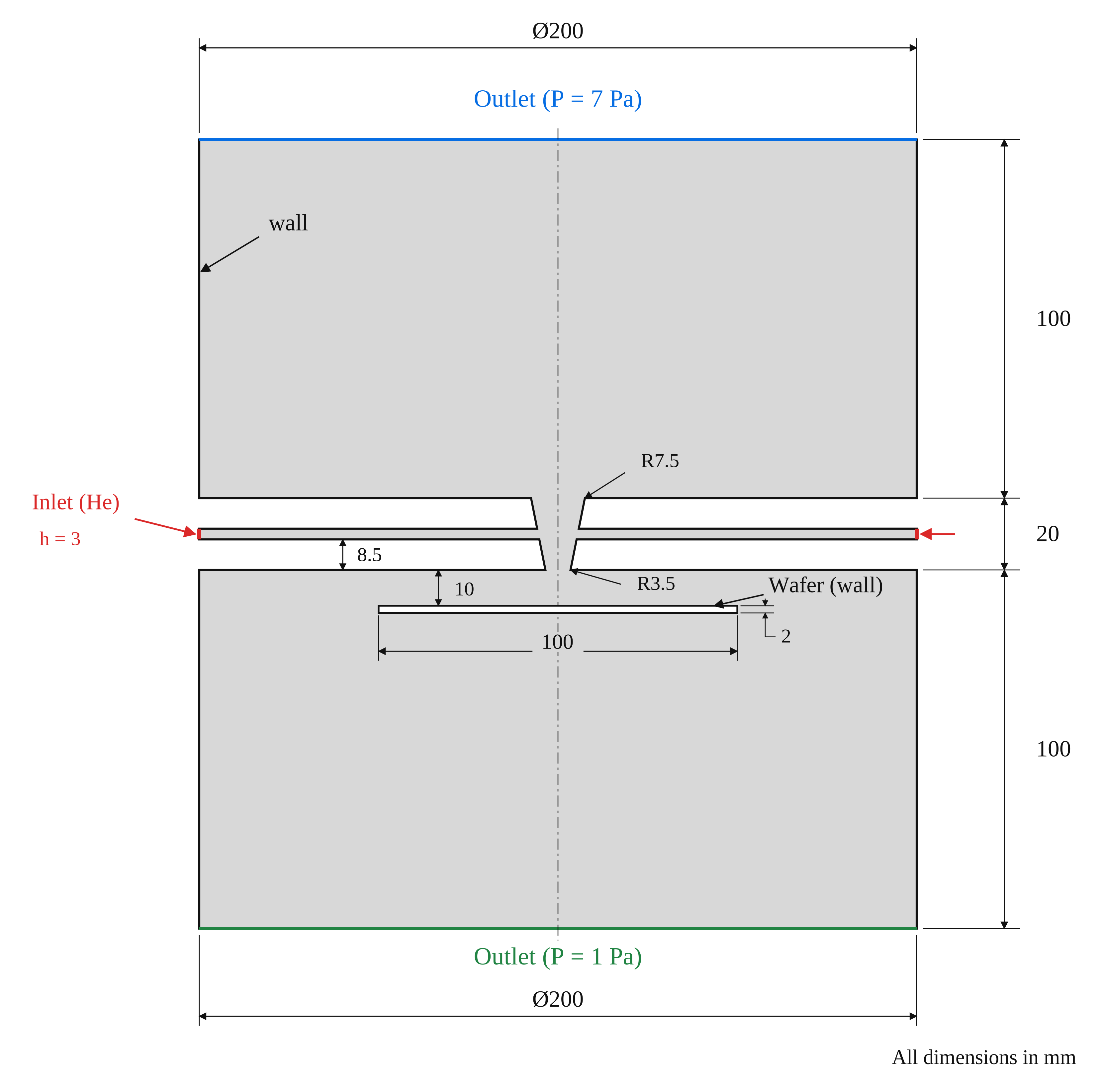}}
\caption{Geometry and boundary conditions of the dynamic gas lock. All dimensions are in millimetres.}
\label{fig:dgl_geometry}
\end{figure}

The computational cost of dynamic gas lock (DGL) simulations motivates the development of efficient kinetic solvers~\cite{he2026rarefied}.
This case examines how rapidly GSIS-PBC brings the boundary state and the interior flow field to convergence under prescribed mass-flow boundary conditions.
The geometry is adapted from Teng et al.~\cite{teng2023pollutant}.
The three-dimensional computational domain is constructed by revolving the meridional half-section through \(360^\circ\) about its symmetry axis, as shown in Fig.~\ref{fig:dgl_geometry}.
Two coaxial cylindrical chambers, each \(200\,\mathrm{mm}\) in diameter and \(100\,\mathrm{mm}\) in height, are connected by a tapered central passage.
Helium is supplied radially through a \(3\,\mathrm{mm}\)-high annular inlet between the chambers.
A circular wafer is located in the lower chamber.
The inlet gas and all walls, including the wafer, are maintained at \(T_0=300\,\mathrm{K}\).
Three inlet flow rates of \(10\), \(100\), and \(1{,}000\,\mathrm{sccm}\) are considered.
The upper and lower outlet pressures are fixed at \(7\) and \(1\,\mathrm{Pa}\).

The three-dimensional spatial mesh contains \(850{,}080\) cells, and the molecular velocity domain \(\bm{\xi}/\sqrt{RT_0}\in[-8,8]^3\) is discretized using a uniform Cartesian grid of \(24\times24\times24\) points.

\begin{figure}[!thbp]
\centering
\subfigure[Horizontal velocity.]{\includegraphics[width=0.48\linewidth]{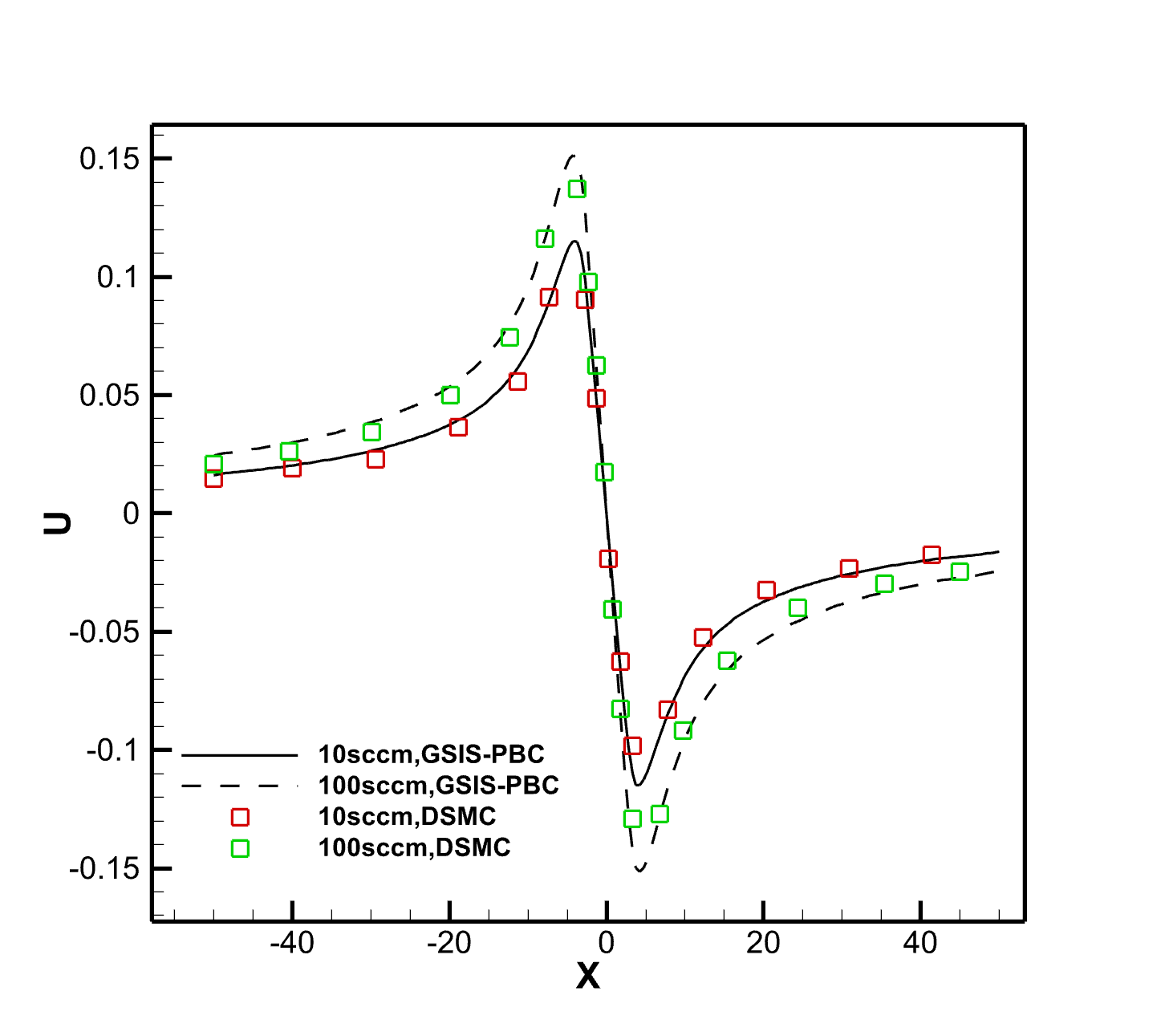}}
\subfigure[Axial velocity.]{\includegraphics[width=0.48\linewidth]{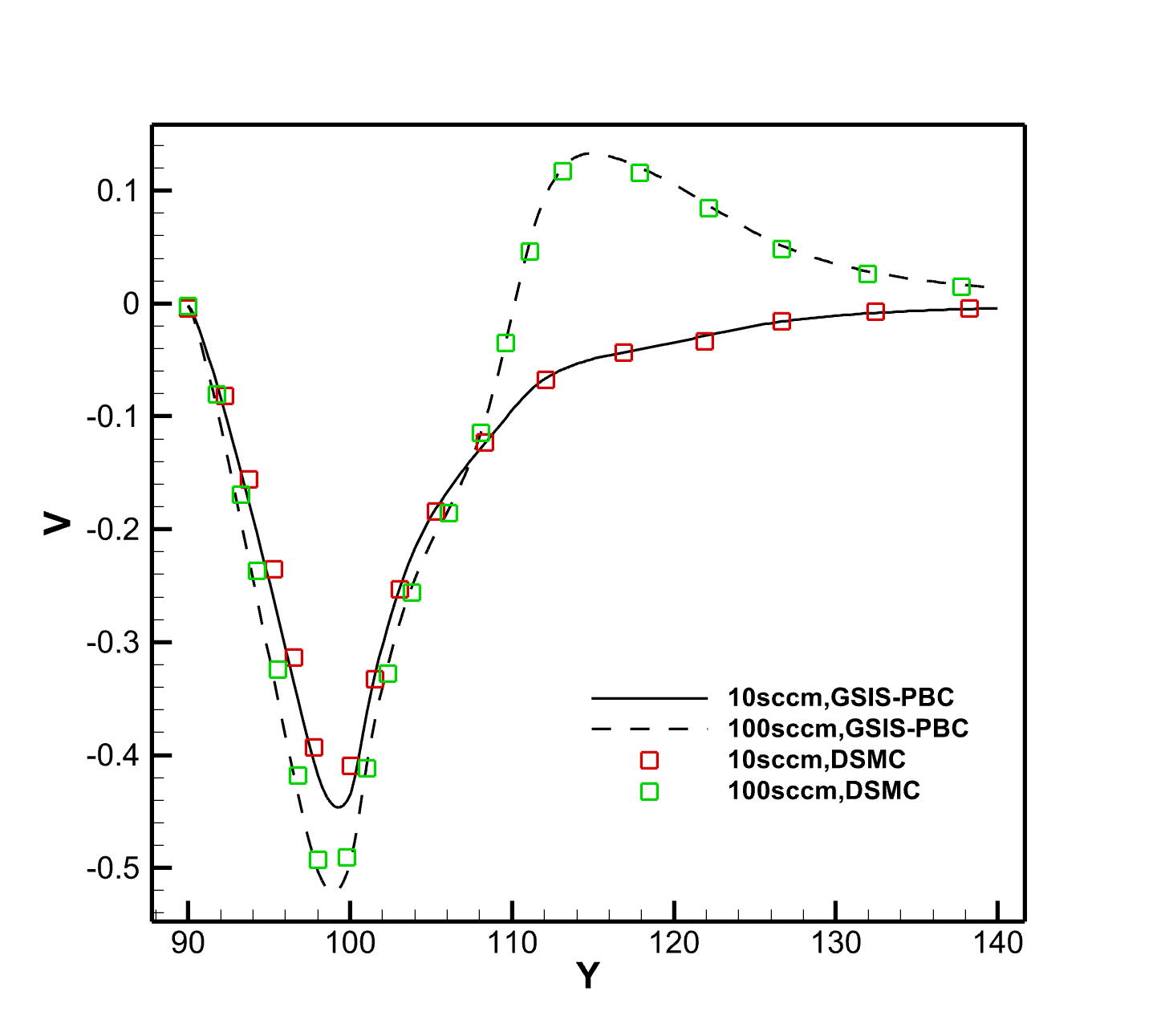}}
\caption{Velocity profiles in the \(Z=0\) plane of the dynamic gas lock at inlet flow rates of \(10\) and \(100\,\mathrm{sccm}\): (a) horizontal velocity along \(Y=95\); (b) axial velocity along \(X=0\). Coordinates are in millimetres, and \(U=u_x/\sqrt{2RT_0}\), \(V=u_y/\sqrt{2RT_0}\). Lines denote GSIS-PBC, and squares denote DSMC.}
\label{fig:dgl_profiles}
\end{figure}

\begin{figure}[!htbp]
\centering
\subfigure[\(10\,\mathrm{sccm}\).]{\includegraphics[trim=200bp 130bp 165bp 145bp,clip,width=0.47\linewidth]{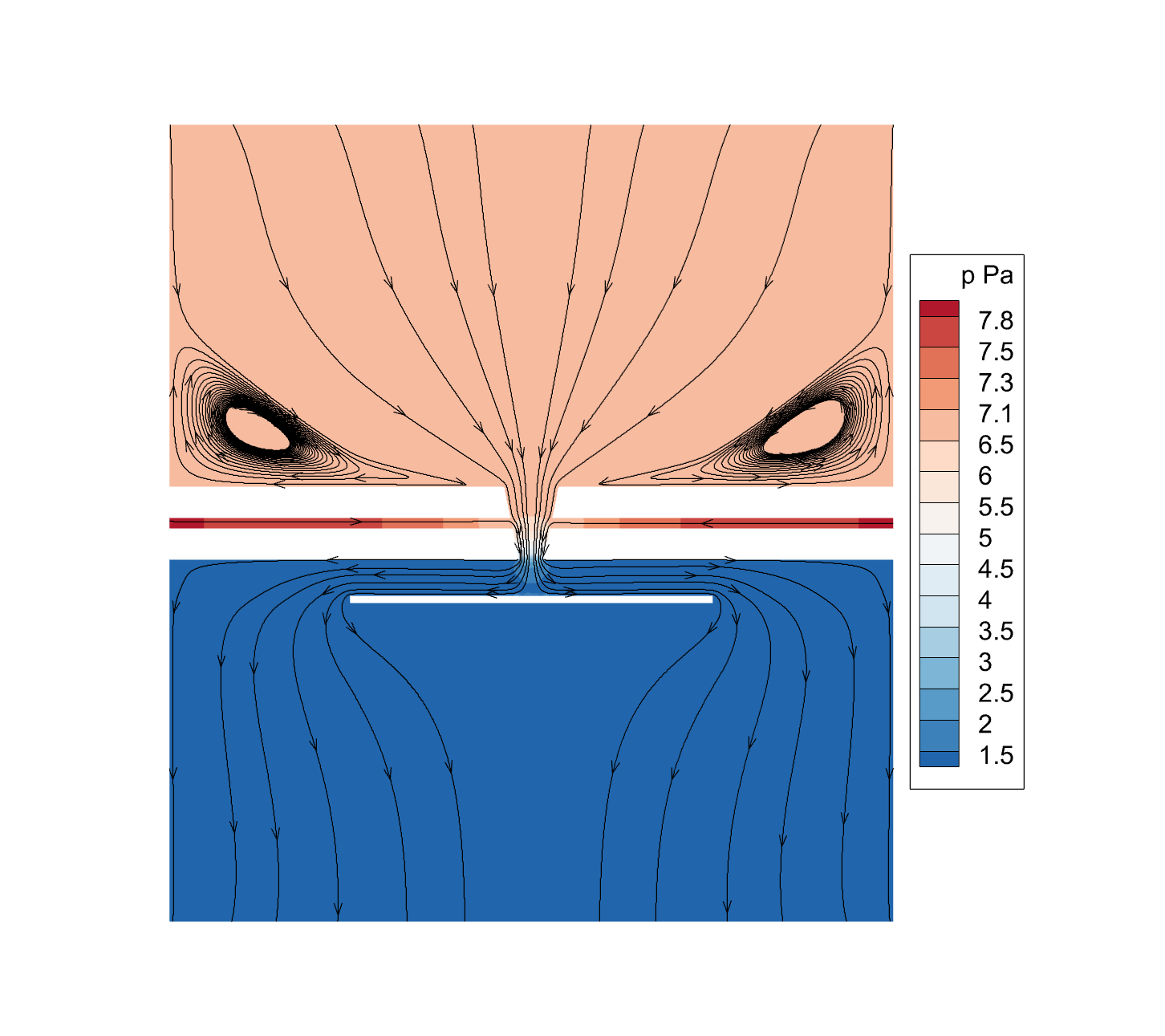}}
\hfill
\subfigure[\(100\,\mathrm{sccm}\).]{\includegraphics[trim=200bp 130bp 165bp 145bp,clip,width=0.47\linewidth]{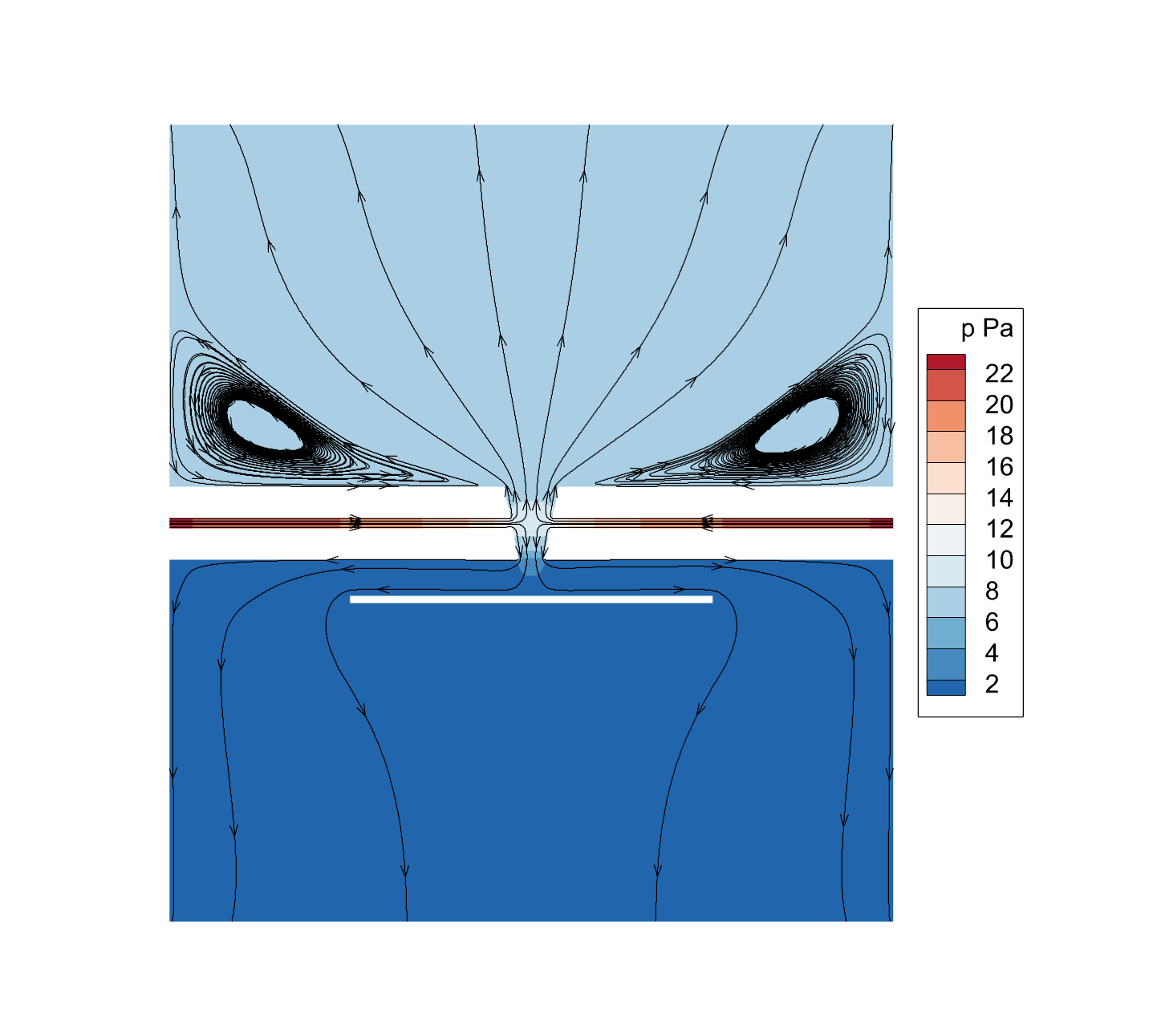}}
\caption{Pressure distributions and streamlines in the DGL.}
\label{fig:dgl_pressure_streamlines}
\end{figure}

% Figure~\ref{fig:dgl_fields} compares the velocity component \(u_x\) obtained with GSIS-PBC and DSMC at an inlet flow rate of \(100\,\mathrm{sccm}\).
Figure~\ref{fig:dgl_profiles} compares the horizontal and axial velocity profiles at inlet flow rates of \(10\) and \(100\,\mathrm{sccm}\).
The GSIS-PBC predictions agree well with the DSMC results for both cases.
The corresponding pressure distributions and streamlines are shown in Fig.~\ref{fig:dgl_pressure_streamlines}.
Table~\ref{tab:dgl_mass_flow} summarizes the net mass flow rates through the upper outlet, with positive values denoting flow out of the computational domain.
At \(10\,\mathrm{sccm}\), the streamlines indicate inflow through the upper opening, consistent with the negative net mass flow rate, whereas increasing the inlet flow rate to \(100\,\mathrm{sccm}\) produces net outflow.
The maximum relative deviation of GSIS-PBC from the reference mass flow rates is approximately \(1.25\%\).

% \begin{figure}[!htbp]
% \centering
% \includegraphics[width=0.8\linewidth]{./Figures/DGL_He_100sccm_ux_gsis_dsmc_cmp.png}
% \caption{Comparison of \(u_x\) in the dynamic gas lock at an inlet flow rate of \(100\,\mathrm{sccm}\). Filled contours: DSMC; black solid lines: GSIS-PBC.}
% \label{fig:dgl_fields}
% \end{figure}

\begin{table}[t]
\centering
\caption{Net mass flow rates through the upper outlet, expressed in sccm (standard cubic centimeter per minute). The reference values are obtained from DSMC for inlet flow rates of \(10\) and \(100\,\mathrm{sccm}\), and from Fluent for \(1{,}000\,\mathrm{sccm}\). Errors are relative to these reference values.}
\label{tab:dgl_mass_flow}
\small
\begin{tabular}{@{}cccc@{}}
\toprule
Inlet (sccm) & Reference (sccm) & GSIS-PBC (sccm) & Error (\%) \\
\midrule
10 & -18.73 & -18.66 & 0.37 \\
100 & 57.45 & 56.73 & 1.25 \\
\(1{,}000\) & 762.70 & 768.22 & 0.72 \\
\bottomrule
\end{tabular}
\end{table}

The DGL simulations are performed using 640 cores.
The GSIS with a density-based macroscopic solver~\cite{zhang2024efficient,zhang2026efficient}, denoted GSIS-DB, serves as the baseline for the convergence comparison.
Each GSIS-DB iteration consists of 2 kinetic iterations followed by 400 density-based macroscopic iterations, whereas each GSIS-PBC iteration consists of 10 kinetic iterations followed by 10 PBC iterations.
For inlet flow rates of \(10\) and \(100\,\mathrm{sccm}\), GSIS-PBC converges in 16 and 20 iterations, requiring \(0.35\) and \(0.44\,\mathrm{h}\), respectively.
In contrast, GSIS-DB requires at least \(2{,}818\) iterations and \(13.23\,\mathrm{h}\) for both cases.
The corresponding wall-clock speedups of GSIS-PBC are approximately \(60\) and \(30\), respectively.
At \(1{,}000\,\mathrm{sccm}\), GSIS-PBC converges in 9 iterations and \(0.21\,\mathrm{h}\).
For reference, the two-dimensional axisymmetric DSMC simulations use \(225{,}000\) spatial cells, with \(12\) and \(18\) million simulation particles for \(10\) and \(100\,\mathrm{sccm}\), and require \(4.12\) and \(12.55\,\mathrm{h}\), respectively.
These results demonstrate the rapid convergence of GSIS-PBC for the present DGL flows under prescribed mass-flow boundary conditions.

\subsection{Flows in a divertor in nuclear fusion}\label{subsec:divertor}

\begin{figure}[!t]
\centering
\subfigure[Computational domain.]{\includegraphics[width=0.49\linewidth]{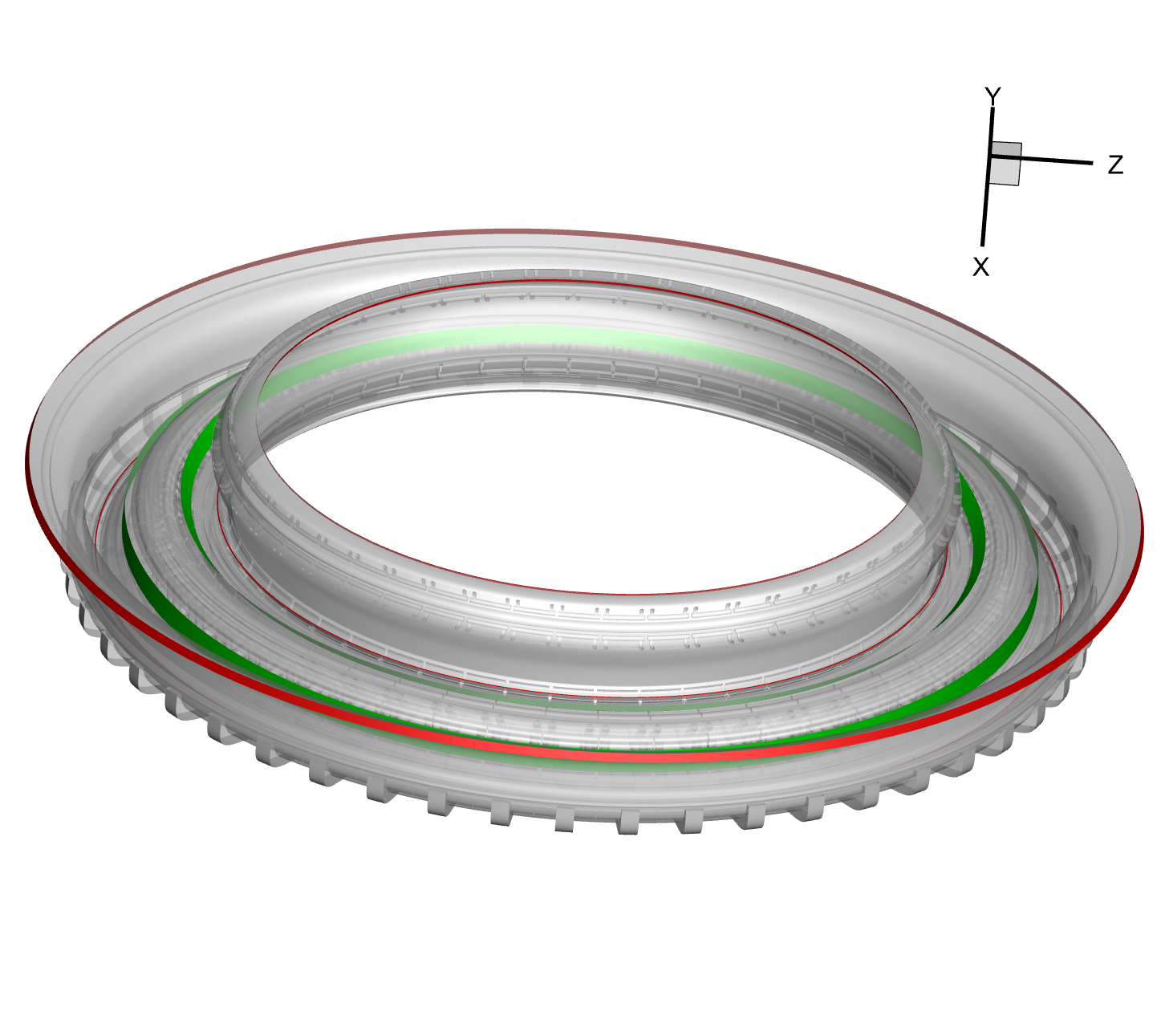}}
\hfill
\subfigure[Right-hand portion of the \(y=0\) cross-section.]{\includegraphics[width=0.49\linewidth]{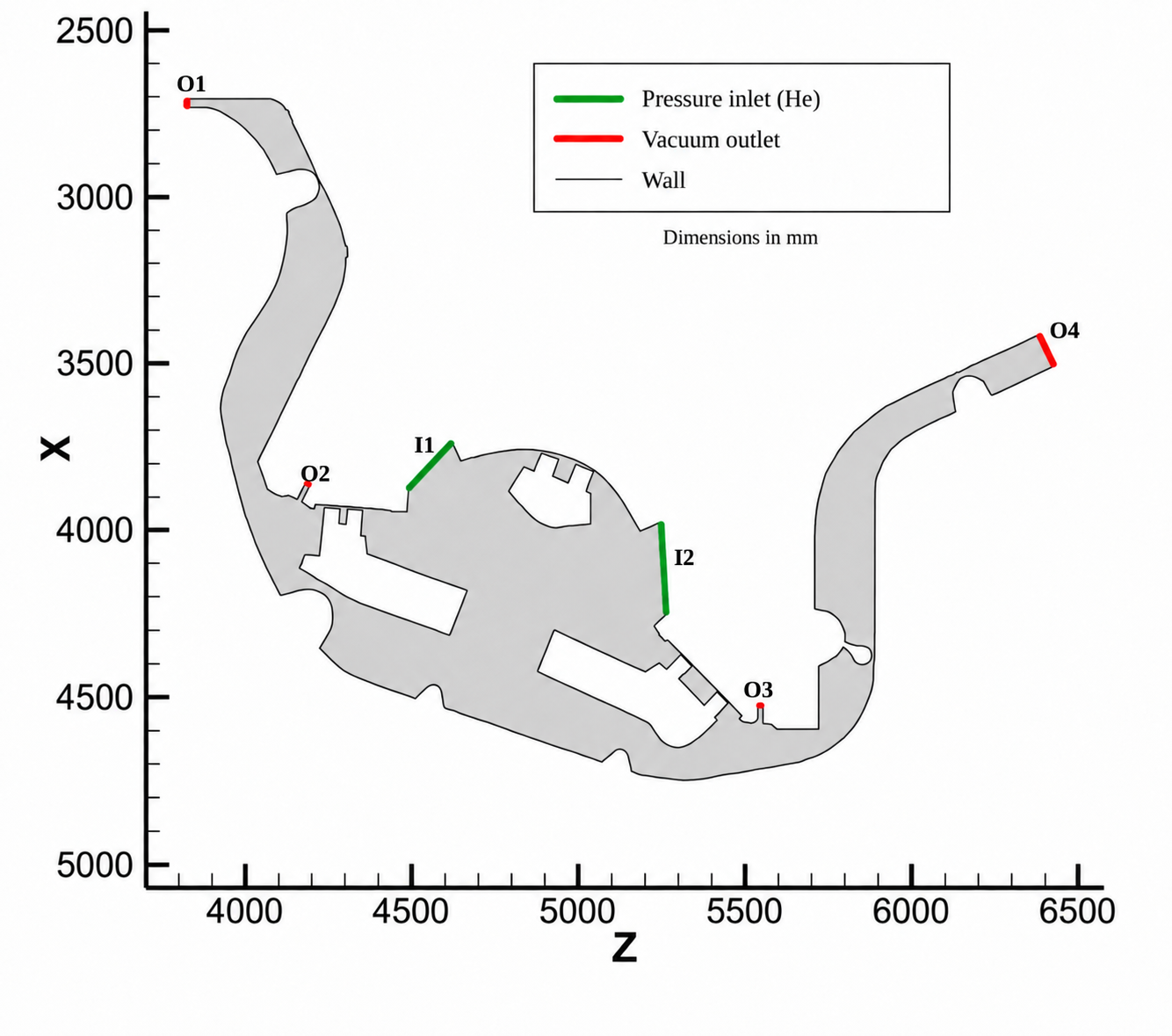}}
\par\bigskip
\subfigure[Polyhedral mesh.]{\includegraphics[width=0.49\linewidth]{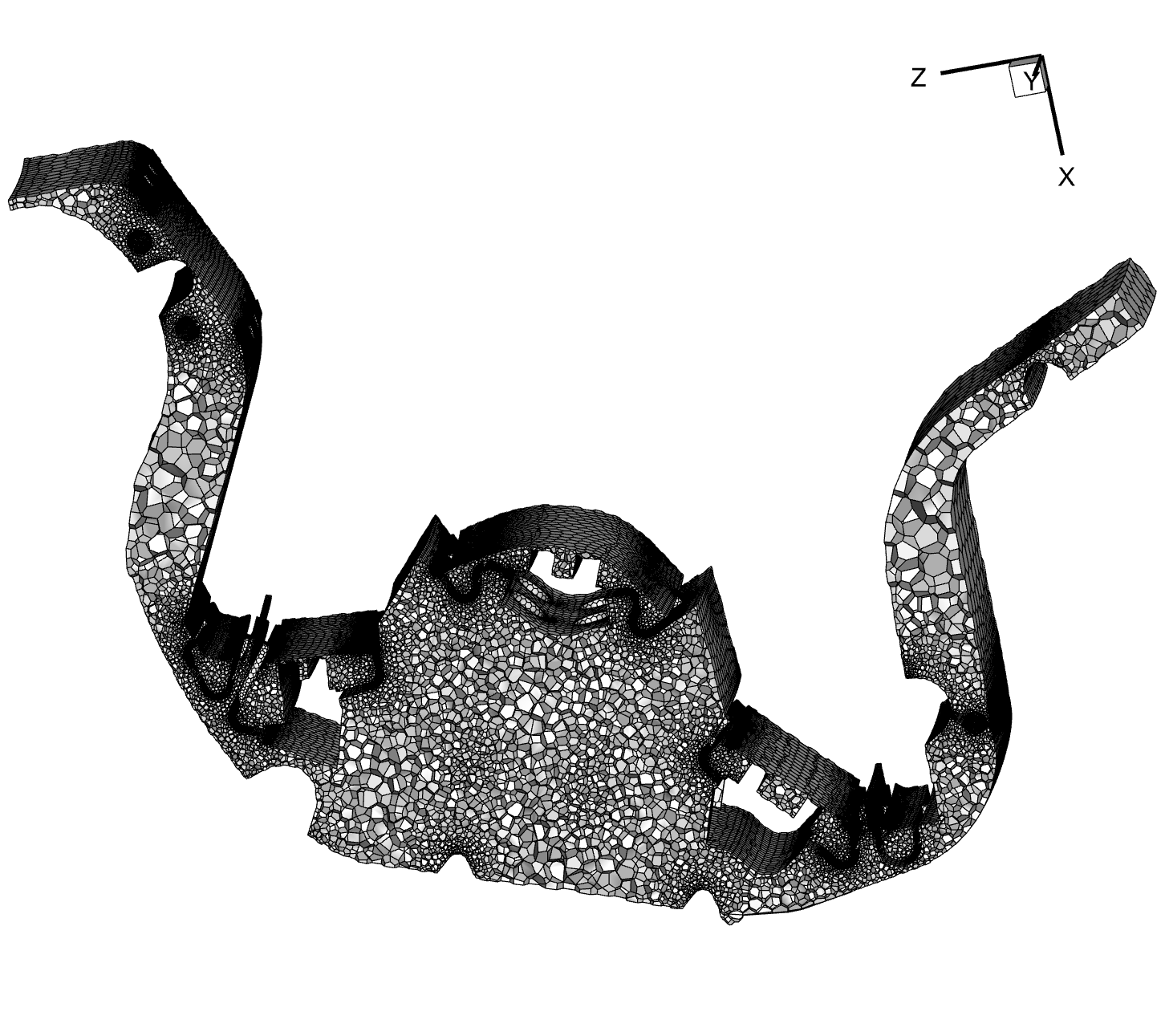}}
\hfill
\subfigure[Local pressure field and streamlines.]{\includegraphics[width=0.49\linewidth]{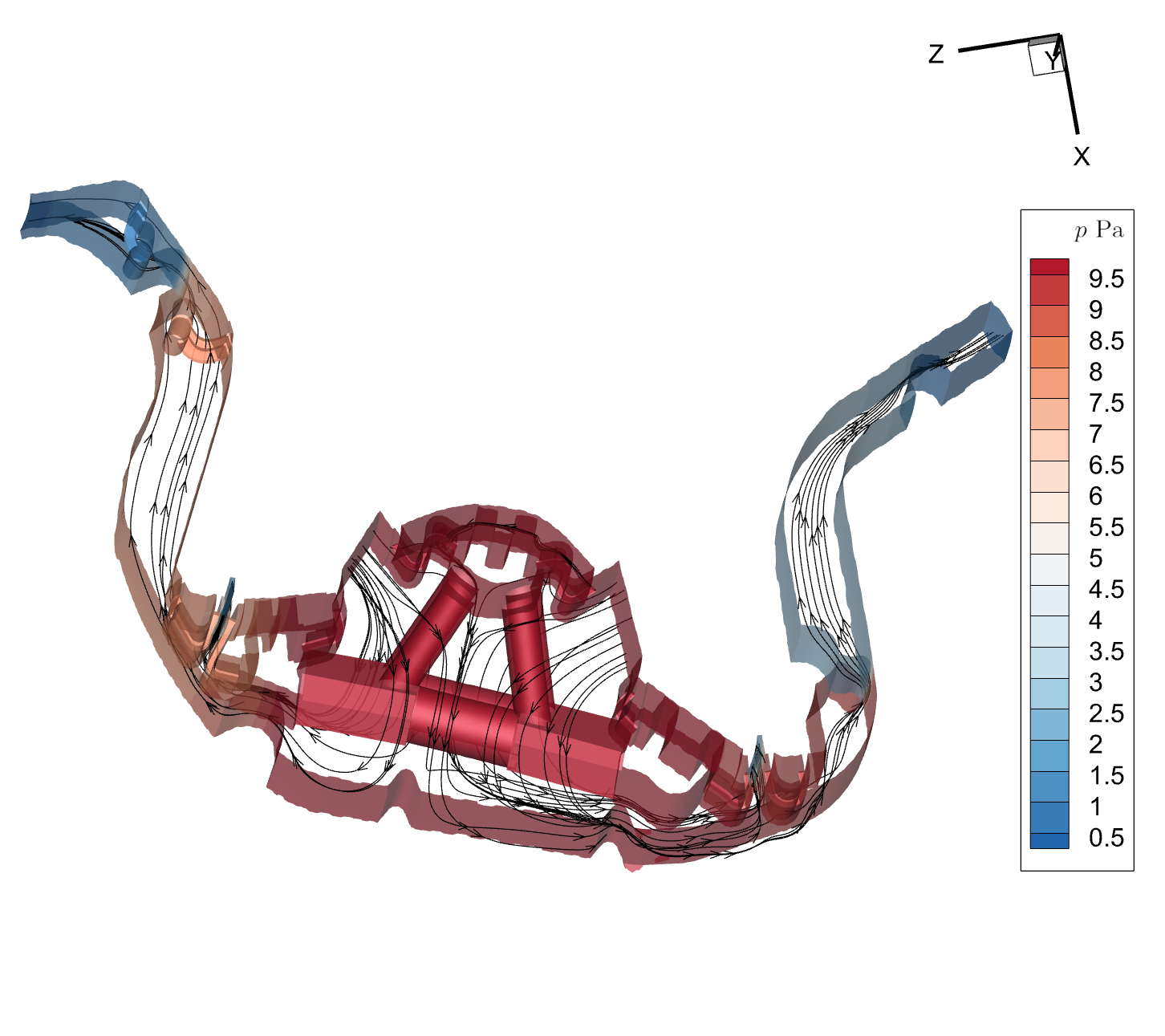}}
\caption{Computational domain, mesh, and local flow in the divertor. Green and red mark the pressure inlets and vacuum leakage outlets, respectively; the remaining boundaries are walls.}
\label{fig:divertor_overview}
\end{figure}

This case examines the convergence and computational efficiency of GSIS-PBC for large-scale three-dimensional gas flow in a complex divertor geometry. 
The computational geometry is derived from the divertor component of the publicly available simplified ITER tokamak model.\footnote{ITER Organization, \emph{Make your own tokamak with 3D printing!}, \url{https://www.iter.org/node/20687/make-your-own-tokamak-3d-printing}.} 
The full three-dimensional domain is shown in Fig.~\ref{fig:divertor_overview}(a), and the cross-sectional geometry and boundary labels are given in Fig.~\ref{fig:divertor_overview}(b).
Helium is supplied through two pressure inlets, I1 and I2, at \(10\,\mathrm{Pa}\) and \(420\,\mathrm{K}\). 
Taking \(L_{\mathrm{ref}}=1\,\mathrm{cm}\), the Knudsen number based on the inlet state is \(\Kn=0.298\).
All solid surfaces are stationary, fully diffuse reflecting walls maintained at \(420\,\mathrm{K}\). 
The pumping probability is set to zero, so the pumping surface is also treated as a wall. Vacuum boundary conditions are imposed at the leakage openings O1--O4.
The spatial mesh, shown in Fig.~\ref{fig:divertor_overview}(c), contains \(12{,}248{,}344\) polyhedral cells and \(80{,}495{,}318\) faces.
The molecular velocity space contains \(10{,}648\) discrete points.

\begin{figure}[!t]
\centering
\subfigure[Pressure.]{\includegraphics[width=0.49\linewidth]{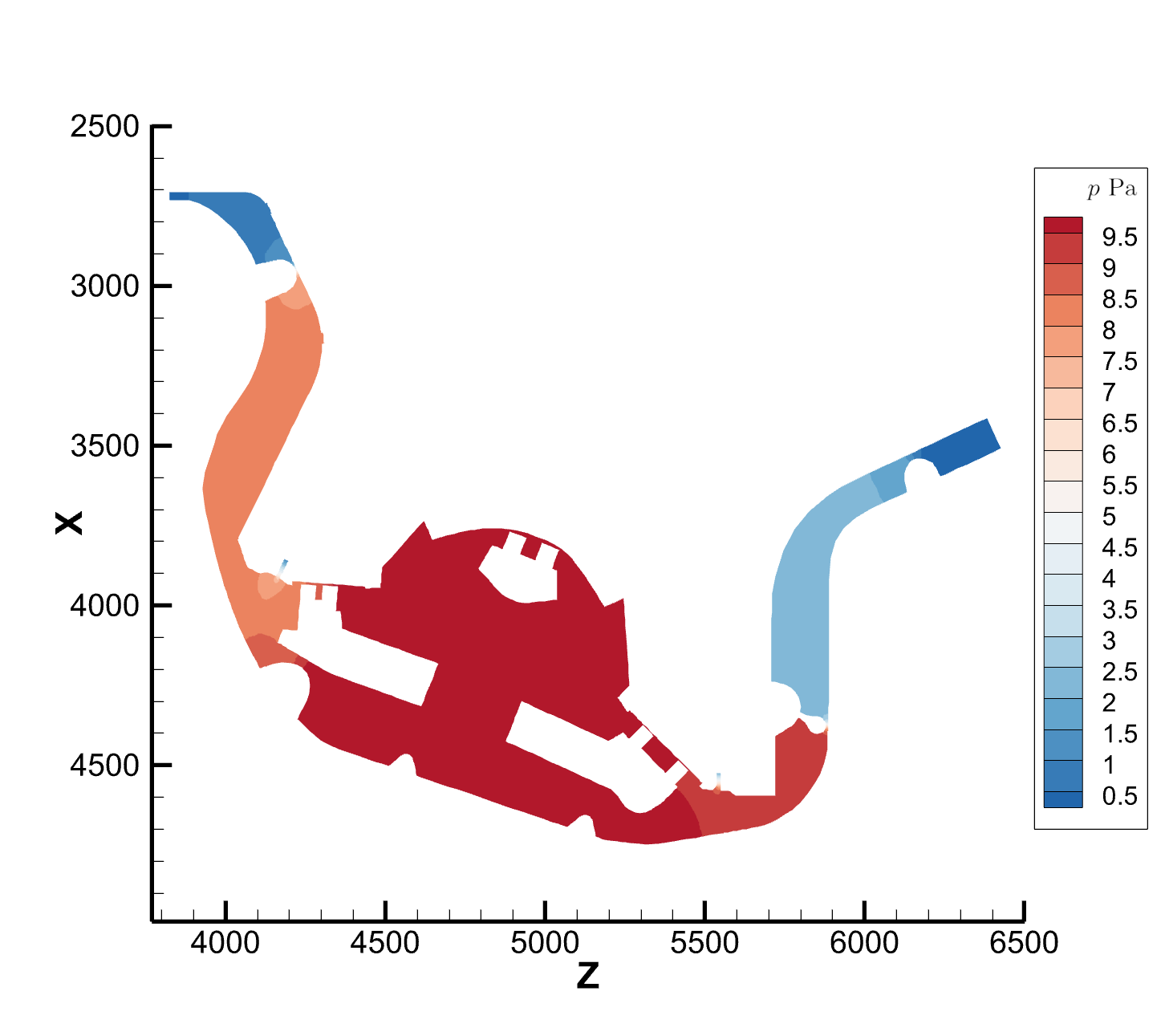}}
\hfill
\subfigure[\(z\)-velocity.]{\includegraphics[width=0.49\linewidth]{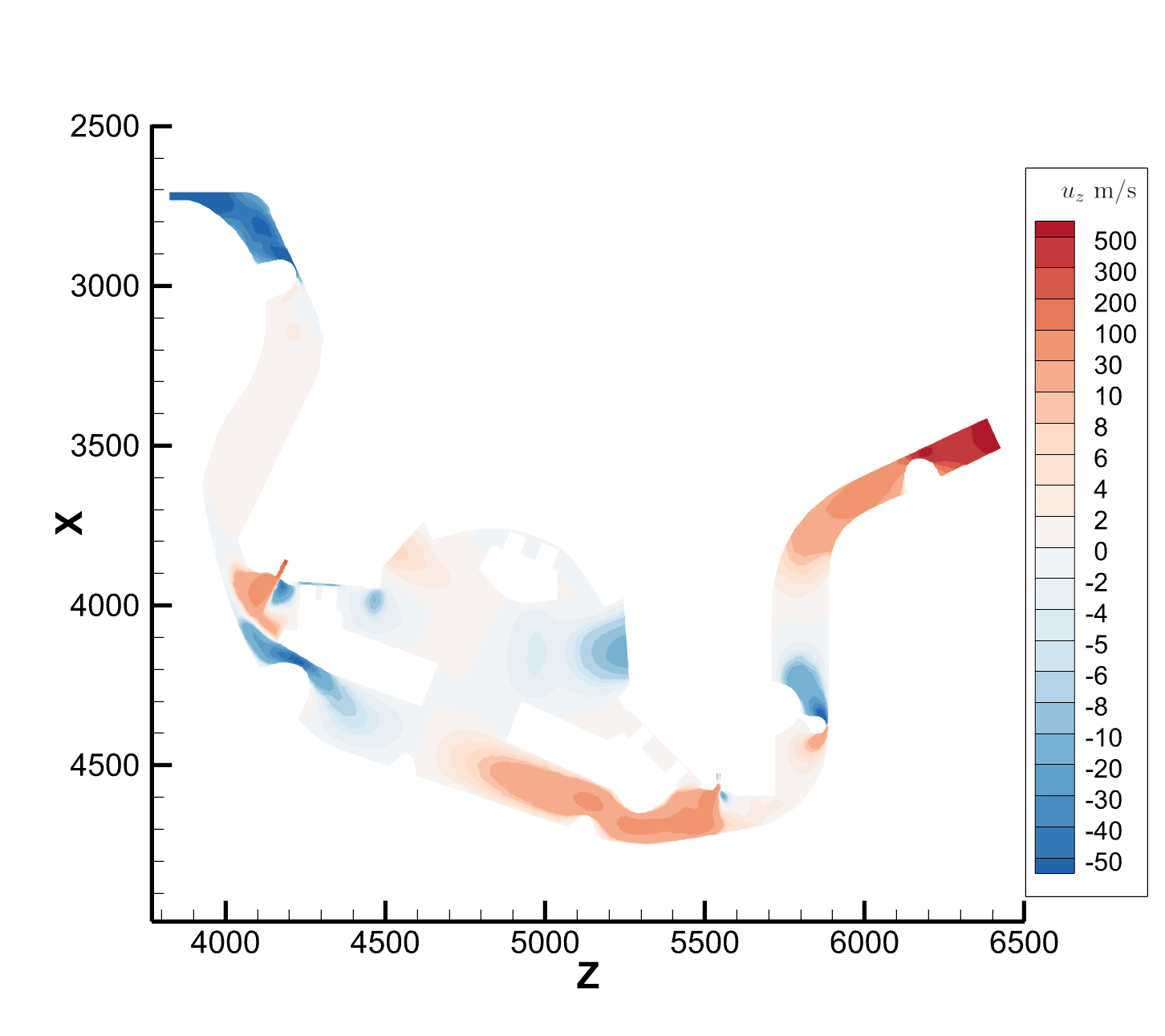}}
\caption{Cross-sectional pressure and \(z\)-velocity distributions in the divertor. Coordinates are in millimetres; pressure and velocity are in \(\mathrm{Pa}\) and \(\mathrm{m/s}\), respectively.}
\label{fig:divertor_fields}
\end{figure}

The streamlines in Fig.~\ref{fig:divertor_overview}(d) show how the inlet gas spreads through the central cavity into the connected leakage passages. The pressure remains relatively high in the cavity and decreases markedly towards the vacuum openings, as shown in Fig.~\ref{fig:divertor_fields}(a). The flow is predominantly at low Mach numbers, with \(\mathrm{Ma}\sim10^{-2}\) over most of the domain, while the velocity increases appreciably near the vacuum openings.

After initialization with 50 macroscopic and five kinetic iterations, each GSIS iteration comprises five kinetic and ten PBC iterations.
Using \(3{,}200\) CPU cores, GSIS-PBC converges in 11 iterations with a total wall-clock time of \(0.34\,\mathrm{h}\). The rapid convergence obtained for this predominantly low-Mach-number internal flow demonstrates the effectiveness of the pressure--velocity coupled macroscopic solver. 
% \leir{In Ref.~\cite{vasileiadis2016modeling}, DSMC simulations were employed, consuming XXX CPU hours to simulate one-eighteenth of the divertor. By applying symmetry boundary conditions, our GSIS-PBC would require only \(1/18\) of the computational time and memory overhead \cite{li2026multiscale}. In this sense, GSIS-PBC is nearly four orders of magnitude faster than DSMC.
% }
In Ref.~\cite{tantos2024neutral}, DSMC simulations consumed approximately \(6.9\times10^5\) CPU core-hours to simulate a \(10.8^\circ\) sector of the DTT divertor. By exploiting toroidal symmetry~\cite{li2026multiscale}, the GSIS-PBC cost for the same toroidal extent is estimated at approximately \(33\) CPU core-hours, giving a cross-case CPU-cost ratio of approximately \(2\times10^4\).
Moreover, under the same computational configuration as GSIS-PBC, GSIS-DB requires at least \(23.7\,\mathrm{h}\) to achieve inlet--outlet mass-flow balance, approximately 70 times the total wall-clock time required for GSIS-PBC to converge.

At convergence, the net inflow rates predicted by GSIS-PBC through I1 and I2 are \(6.338\times10^{-4}\) and \(1.237\times10^{-3}\,\mathrm{kg/s}\), respectively. The leakage outflow rates through O1--O4 are \(1.143\times10^{-4}\), \(3.945\times10^{-4}\), \(9.522\times10^{-4}\), and \(4.097\times10^{-4}\,\mathrm{kg/s}\), respectively.
Under the present boundary conditions, O3 accounts for approximately half of the total leakage flow and therefore warrants particular attention in studies of neutral-gas leakage control.

% \leir{maybe it is better to evaluate/estimate the advantage of GSIS-PBC over GSIS-DB }

\subsection{Flow around a sphere}\label{subsec:sphere}

\begin{figure}[h]
\centering
\subfigure[Computational domain.]{\includegraphics[width=0.48\linewidth]{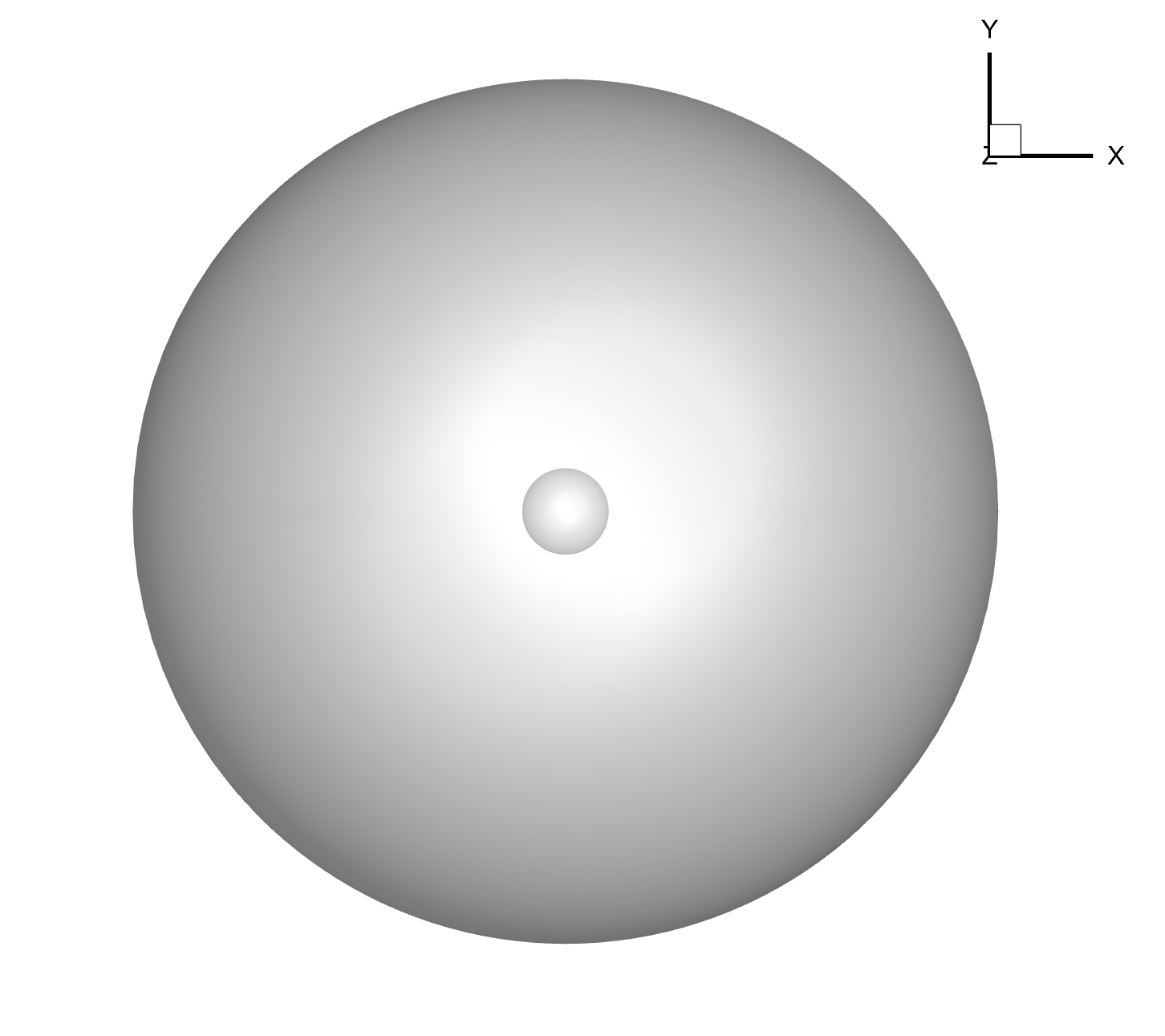}}
\subfigure[Cross-section of the spatial mesh.]{\includegraphics[width=0.48\linewidth]{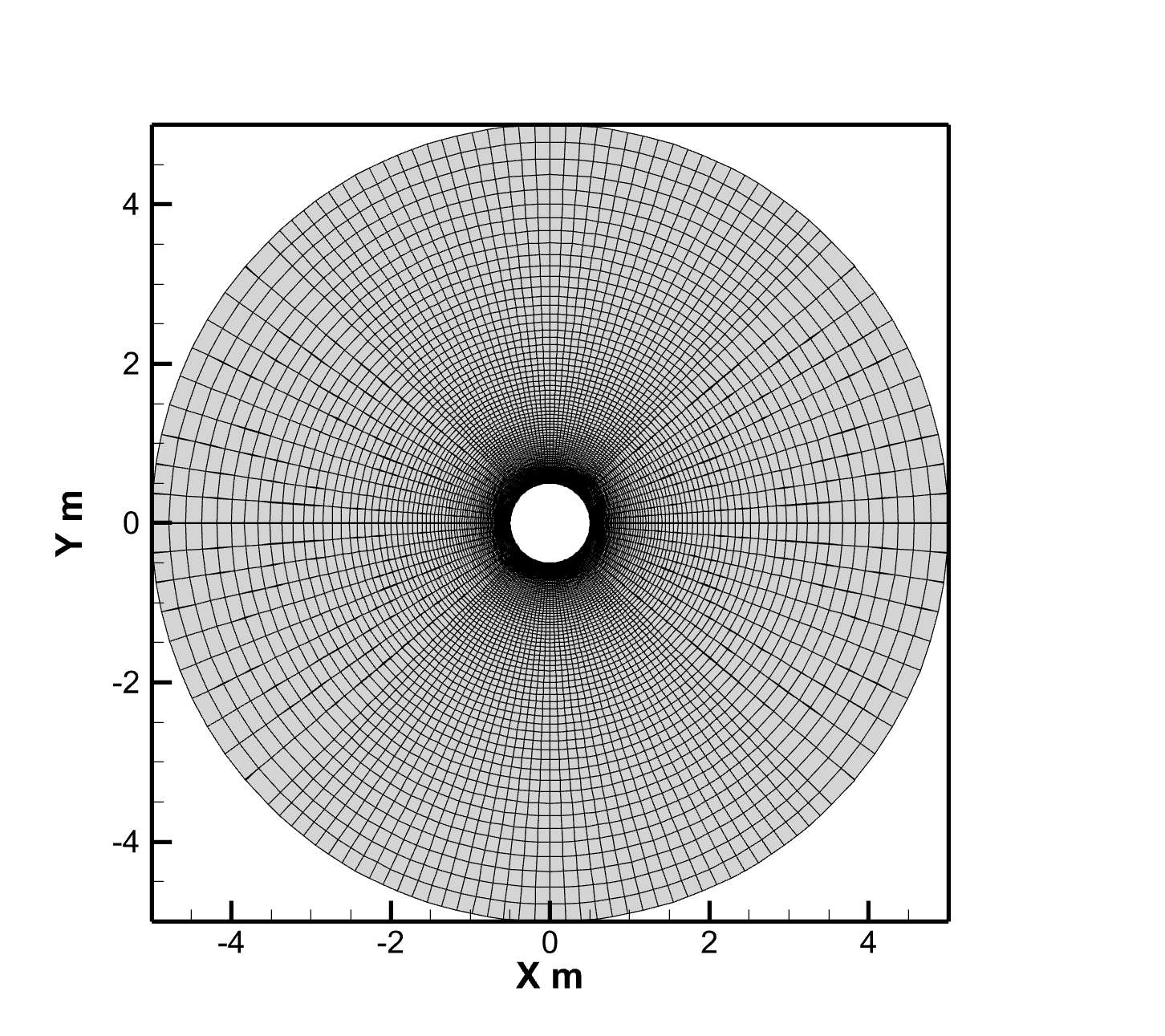}}
\caption{Geometry and spatial mesh for flow around a sphere. The sphere
diameter is \(1\,\mathrm{m}\), and the spherical outer boundary has a
radius of \(5\,\mathrm{m}\).}
\label{fig:sphere_geometry_mesh}
\end{figure}

Finally, to assess the applicability of GSIS-PBC at high flow speeds, we consider argon flow around a sphere.
The computational domain is bounded by a sphere of diameter
\(D=L_{\mathrm{ref}}=1\,\mathrm{m}\) and a concentric spherical
far-field boundary of radius \(5\,\mathrm{m}\), as shown in
Fig.~\ref{fig:sphere_geometry_mesh}.
The molecular mass of argon is \(m=6.63\times10^{-26}\,\mathrm{kg}\).
Its viscosity follows Eq.~\eqref{eq:viscosity_powerlaw}, with
\(\mu_{\mathrm{ref}}=2.117\times10^{-5}\,\mathrm{Pa\,s}\) at
\(T_{\mathrm{ref}}=273.15\,\mathrm{K}\) and \(\omega=0.81\).
The freestream temperature and pressure are
\(T_\infty=273.15\,\mathrm{K}\) and
\(p_\infty=0.1266\,\mathrm{Pa}\), respectively, corresponding to
\(\Kn=0.05\) based on the sphere diameter.
The freestream is directed along the positive \(x\)-axis, with
\(\mathrm{Ma}_\infty=2\), \(5\), and \(10\), corresponding to
velocities of \(615.8\), \(1{,}539.5\), and
\(3{,}079.0\,\mathrm{m/s}\), respectively.
The sphere surface is maintained at \(T_w=T_\infty\), with fully
diffuse reflection imposed.

The GSIS-PBC computations use a common spatial mesh containing
\(360,000\) cells, with a first-layer thickness of
\(0.005\,\mathrm{m}\) at the sphere surface. The molecular velocity
space is discretized by uniform Cartesian grids with
the resolutions listed in Table~\ref{tab:sphere_cost}.
Each calculation starts with 100 macroscopic iterations and five kinetic iterations to establish the initial flow field.
The subsequent GSIS iterations use ten PBC
iterations per macroscopic update, with six kinetic updates between
successive macroscopic updates for \(\mathrm{Ma}_\infty=2\) and \(5\),
and ten for \(\mathrm{Ma}_\infty=10\).

\begin{table}[t]
\centering
\caption{Computational settings and costs for sphere flow
at \(\Kn=0.05\). GSIS-PBC uses a three-dimensional domain, whereas DSMC
uses a two-dimensional axisymmetric formulation.
The fourth column gives the molecular velocity grid
resolution for GSIS-PBC and the total number of simulation particles for DSMC.
The iteration counts refer to GSIS iterations; times are total
wall-clock times.}
\label{tab:sphere_cost}
\small
\setlength{\tabcolsep}{4pt}
\begin{tabular}{@{}ccccccc@{}}
\toprule
\(\mathrm{Ma}_\infty\) & Method & Spatial cells & \begin{tabular}[c]{@{}c@{}}Velocity grid\\or total particles\end{tabular} & Cores & Iterations & Time (h) \\
\midrule
\multirow{2}{*}{2} & GSIS-PBC & \(360{,}000\) & \(30\times24\times24\) & 288 & 8 & 0.13 \\
 & DSMC & \(100{,}356\) & \(3{,}709{,}747\) & 200 & -- & 0.27 \\
\midrule
\multirow{2}{*}{5} & GSIS-PBC & \(360{,}000\) & \(30\times24\times24\) & 288 & 10 & 0.18 \\
 & DSMC & \(100{,}356\) & \(3{,}710{,}463\) & 200 & -- & 0.39 \\
\midrule
\multirow{2}{*}{10} & GSIS-PBC & \(360{,}000\) & \(40\times36\times36\) & 480 & 10 & 0.32 \\
 & DSMC & \(182{,}649\) & \(7{,}156{,}819\) & 200 & -- & 1.15 \\
\bottomrule
\end{tabular}
\end{table}

\begin{figure}[p]
\centering
\setlength{\subfigtopskip}{0pt}
\setlength{\subfigcapskip}{1pt}
\setlength{\subfigbottomskip}{0pt}
\subfigure[\(\mathrm{Ma}_\infty=2\), \(u_x\).]{\includegraphics[width=0.46\linewidth,trim=0 0 0 140bp,clip]{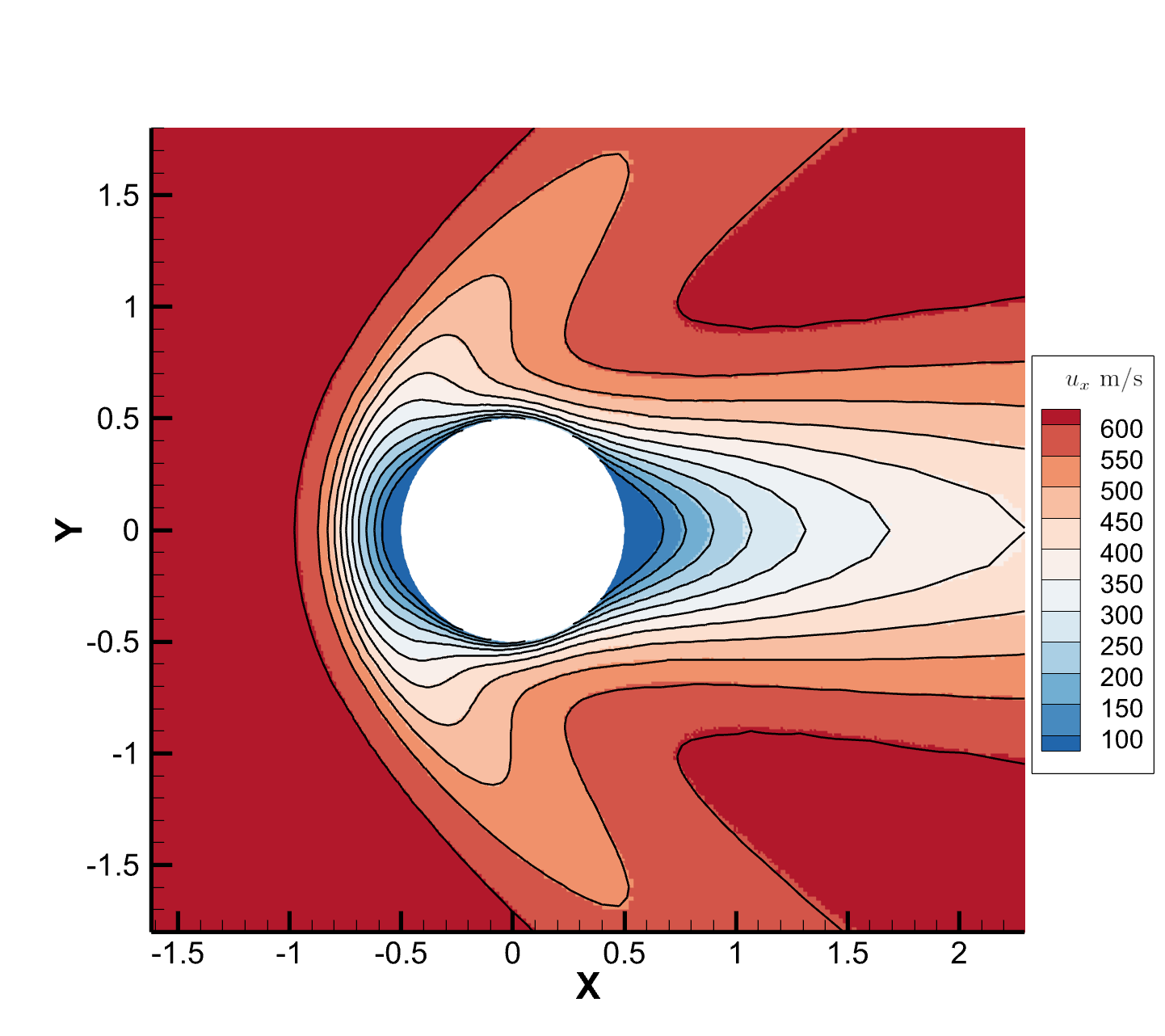}}
\hfill
\subfigure[\(\mathrm{Ma}_\infty=2\), \(T\).]{\includegraphics[width=0.46\linewidth,trim=0 0 0 140bp,clip]{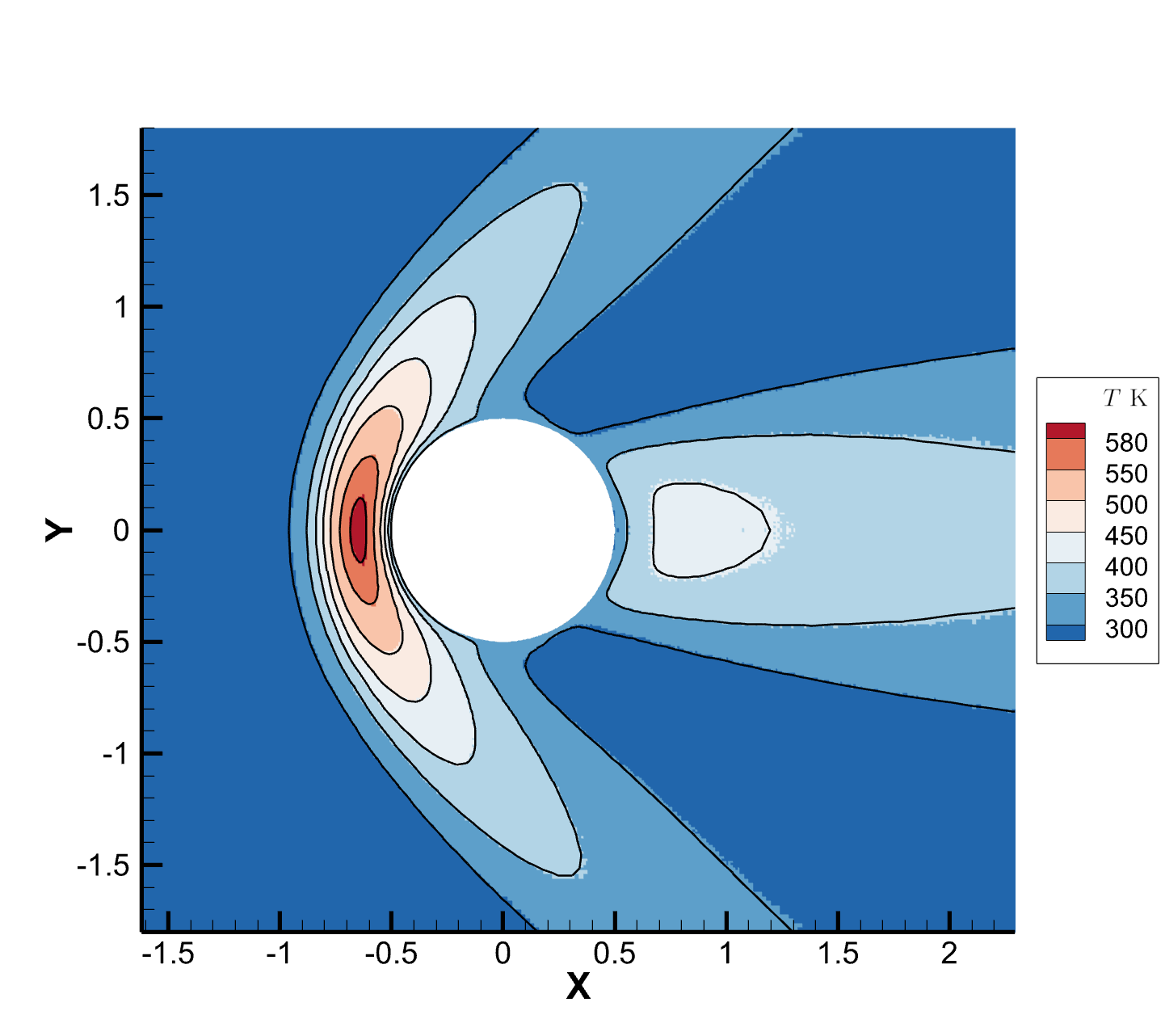}}
\par
\subfigure[\(\mathrm{Ma}_\infty=5\), \(u_x\).]{\includegraphics[width=0.46\linewidth,trim=0 0 0 140bp,clip]{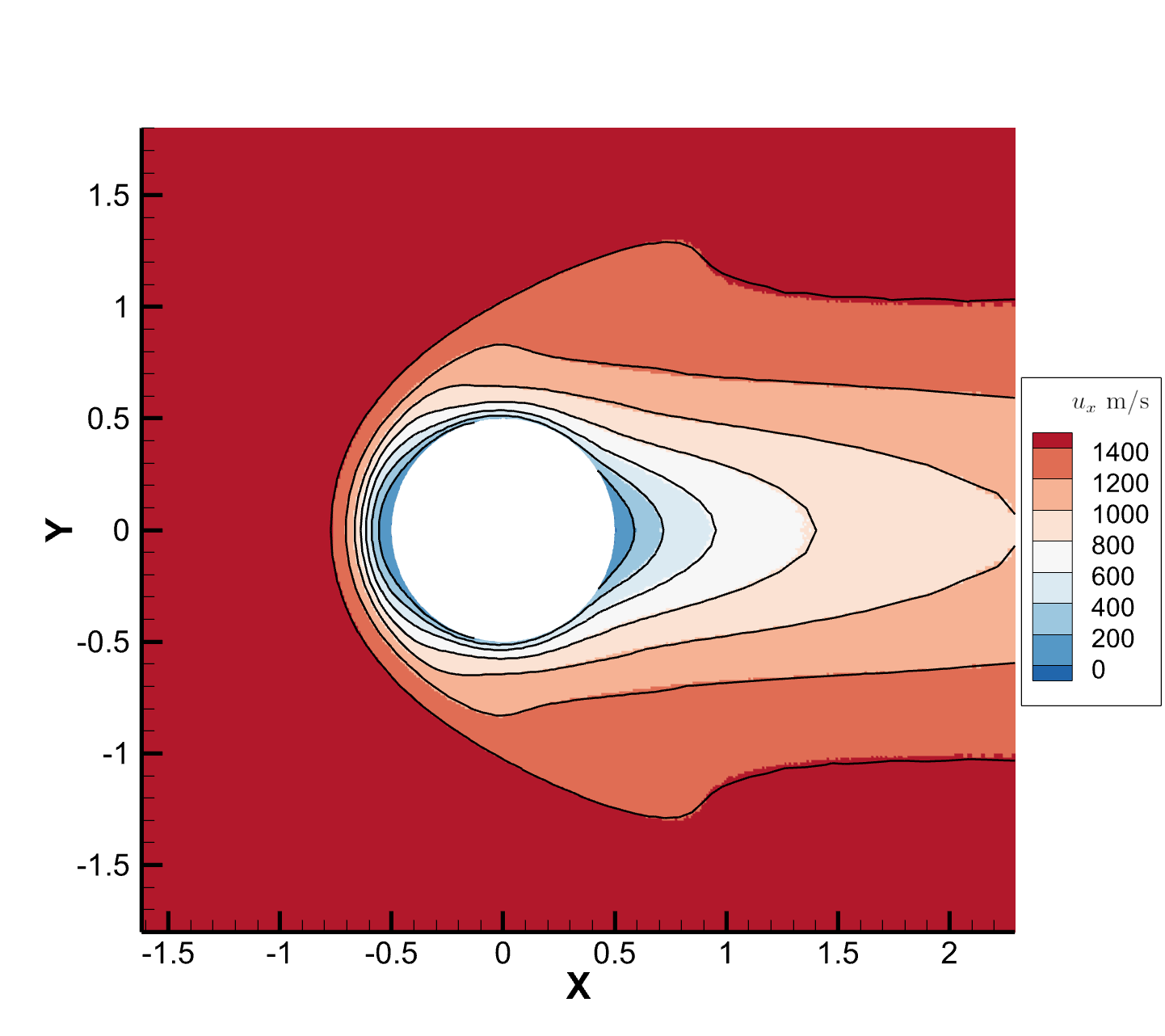}}
\hfill
\subfigure[\(\mathrm{Ma}_\infty=5\), \(T\).]{\includegraphics[width=0.46\linewidth,trim=0 0 0 140bp,clip]{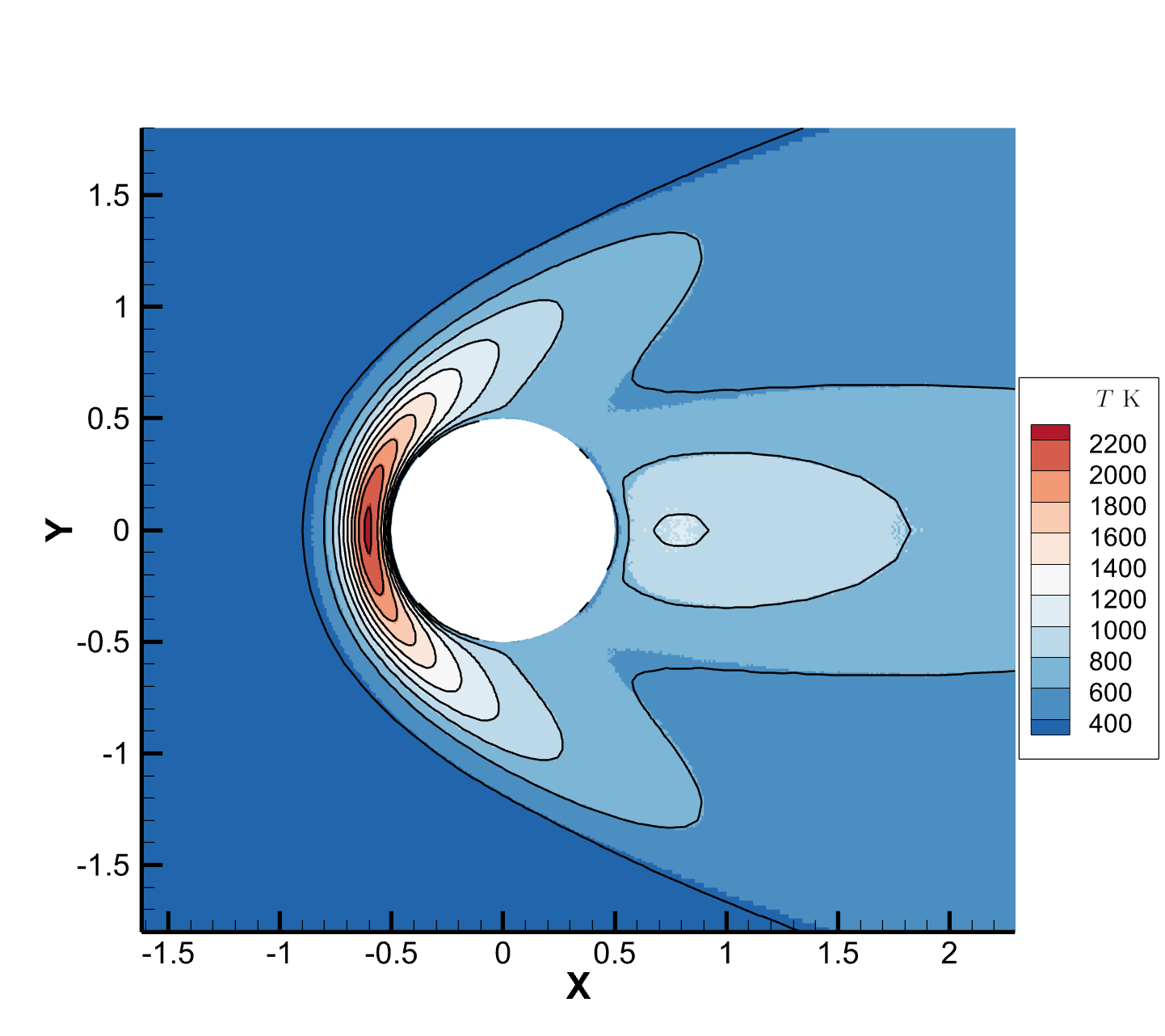}}
\par
\subfigure[\(\mathrm{Ma}_\infty=10\), \(u_x\).]{\includegraphics[width=0.46\linewidth,trim=0 0 0 140bp,clip]{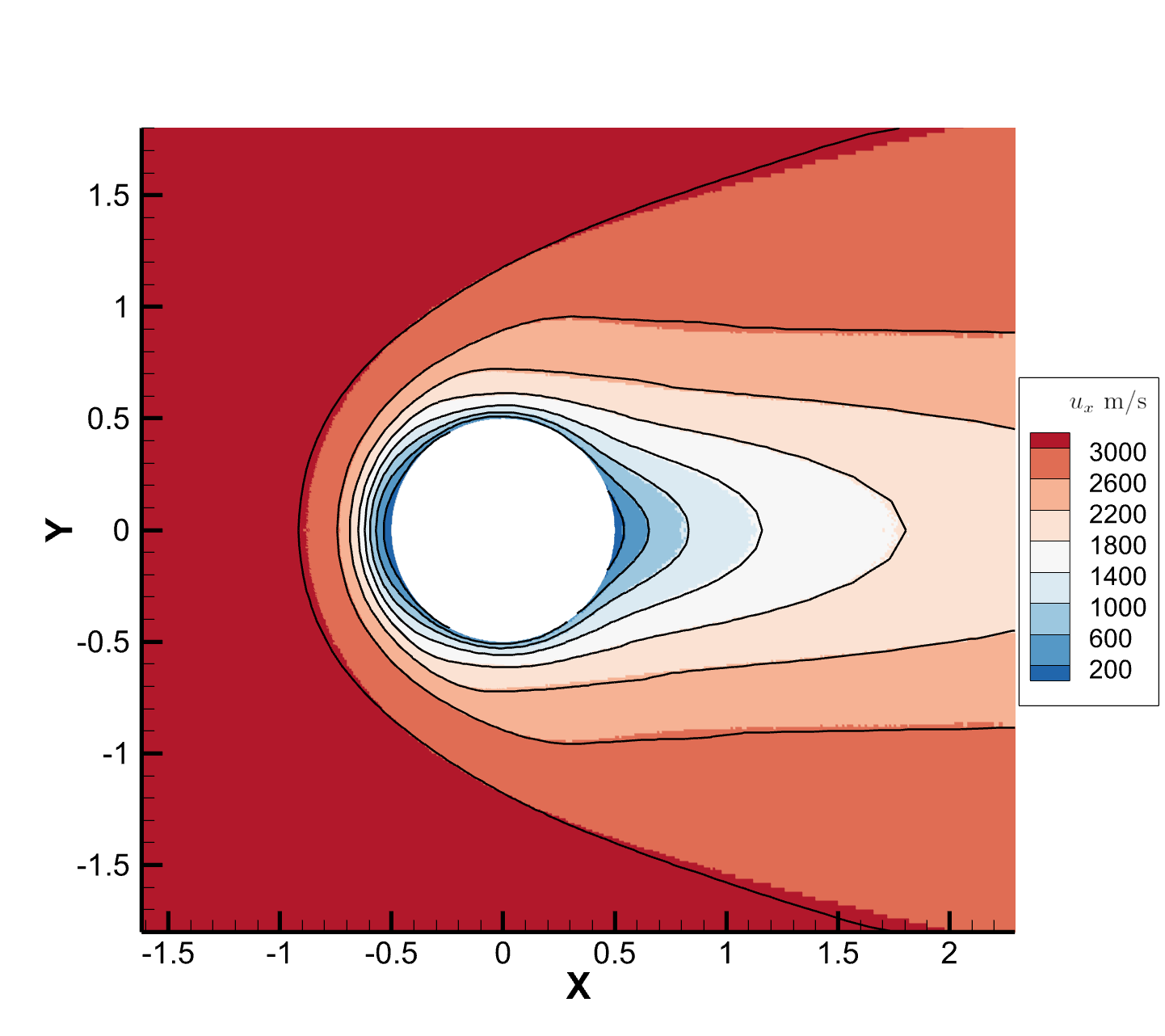}}
\hfill
\subfigure[\(\mathrm{Ma}_\infty=10\), \(T\).]{\includegraphics[width=0.46\linewidth,trim=0 0 0 140bp,clip]{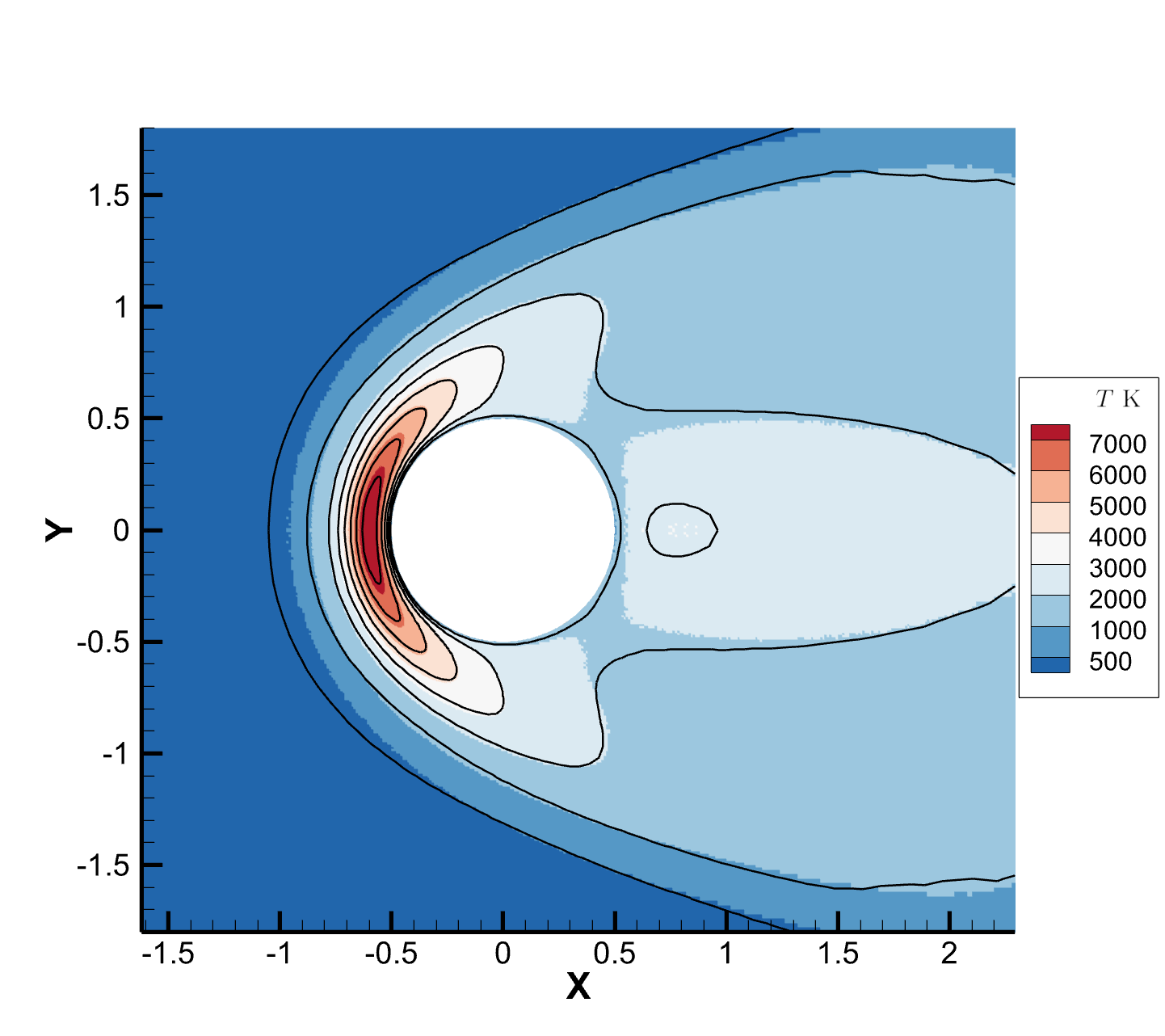}}
\caption{Comparisons of streamwise velocity (left column) and temperature
(right column) for argon flow around a sphere at \(\Kn=0.05\), at the $z=0$ plane.
Filled contours show the DSMC results, and black solid lines denote
GSIS-PBC. The spatial coordinates are in metres.}
\label{fig:sphere_fields}
\end{figure}

Figure~\ref{fig:sphere_fields} compares the streamwise velocity and
temperature fields predicted by GSIS-PBC with the DSMC results.
Good overall agreement is observed at all three Mach numbers.
GSIS-PBC predicts a slightly earlier temperature rise than DSMC on the
upstream side of the shock at \(\mathrm{Ma}_\infty=5\), and this difference
becomes more pronounced at \(\mathrm{Ma}_\infty=10\).
At \(\mathrm{Ma}_\infty=2\), the difference is less evident.
At higher Mach numbers, the Shakhov model's velocity-independent collision
frequency may allow energetic molecules to travel too far upstream,
exaggerating the pre-shock temperature rise~\cite{yuan2022capturing}.
The comparisons support the applicability of GSIS-PBC to the present
argon flows at freestream velocities up to Mach 10.

The computational settings and costs are summarized in
Table~\ref{tab:sphere_cost}. The GSIS-PBC calculations use 288 cores for
\(\mathrm{Ma}_\infty=2\) and \(5\), and 480 cores for
\(\mathrm{Ma}_\infty=10\); the DSMC calculations use 200 cores.
Steady macroscopic fields are obtained within ten GSIS iterations
for all three cases, demonstrating rapid convergence over the tested
Mach-number range. In terms of CPU core‑hours, the three‑dimensional GSIS‑PBC solver runs marginally faster than its two‑dimensional axisymmetric counterpart at Kn=0.05. If the axisymmetric GSIS formulation is adopted, the simulation runtime will be reduced by approximately one order of magnitude~\cite{wen2026symm}. As the Knudsen number decreases, the required simulation mesh size and computational time increase dramatically for DSMC, whereas the GSIS scheme retains the same order of computational cost.
% Mach-number range. \leir{In terms of CPU core‑hours, the three‑dimensional GSIS‑PBC solver runs marginally faster than its two‑dimensional axisymmetric counterpart at Kn=0.05. If the axisymmetric GSIS formulation is adopted, the simulation runtime will be reduced by approximately one order of magnitude~\cite{wen2026symm}. As the Knudsen number decreases, the required simulation mesh size and computational time increase dramatically for DSMC, whereas the GSIS scheme retains the same order of computational cost. }

\section{Conclusions}\label{sec:conclusion}

This work presents a pressure--based coupled general synthetic iterative scheme (GSIS--PBC) for efficient steady--state multiscale simulations of rarefied gas flows. By embedding a pressure--velocity coupled macroscopic solver within the GSIS framework and developing a consistent mesoscopic--macroscopic boundary treatment, the proposed method effectively mitigates the convergence deterioration exhibited by the original density--based GSIS for low--Mach--number internal flows, where large disparities between acoustic and convective time scales dominate. 

The accuracy and computational performance of GSIS--PBC are systematically validated against DSMC references across a suite of three--dimensional benchmark cases. Numerical results demonstrate that GSIS--PBC inherits the asymptotic--preserving property of GSIS and maintains satisfactory prediction accuracy for both internal low--speed flows and external high--Mach--number rarefied flows. For near--continuum internal flows, GSIS--PBC achieves up to two orders of magnitude wall--clock speed--up compared with the conventional iterative kinetic solver. For challenging internal problems such as dynamic gas--lock and fusion divertor flows, GSIS--PBC substantially outperforms the original density--based GSIS and yields a speed--up of approximately four orders of magnitude relative to DSMC.

Future research will integrate the GSIS--PBC solver with adjoint shape optimization techniques \cite{zhang2026fast}  to facilitate multiscale design--oriented simulations for practical engineering systems, such as EUV lithography chambers, dynamic gas locks, and neutral--gas exhaust components in fusion facilities.

\section*{Declaration of competing interest}  The authors declare that they have no known competing financial interests or personal relationships that could have appeared to influence the work reported in this paper.

\section*{Acknowledgments} This work is supported by the National Natural Science Foundation of China (125B2052). The authors acknowledge the computing resources from the Center for Computational Science and Engineering at the Southern University of Science and Technology.

\section*{Data availability} Data will be made available on request.

\appendix
\section{Kinetic boundary implementation}
\label{app:kinetic_boundary}

This appendix specifies the incoming distribution
\(f_{b,k}^{\mathrm{bc}}\) in Eq.~\eqref{eq:cis_kinetic_boundary_distribution}
for the boundary conditions used in this work. The boundary normal and the
half-spaces \(\Xi_b^\pm\) follow Eq.~\eqref{eq:cis_boundary_half_spaces}. The
outgoing distribution on \(\Xi_b^+\) is reconstructed from the adjacent cell,
whereas the formulas below prescribe the incoming distribution on
\(\Xi_b^-\).

\subsection{Prescribed mass-flow and fully diffuse wall boundaries}
\label{app:kinetic_bc_massflow}

Let \(\Gamma\) be a boundary on which the nonnegative inward mass-flow rate
\(\dot M_\Gamma\) and the incoming Maxwellian temperature \(T_\Gamma\) are
prescribed. The total area and uniform inward mass-flux density are
\begin{equation}
 A_\Gamma=\sum_{b\in\Gamma}A_b,
 \qquad
 j_\Gamma=\frac{\dot M_\Gamma}{A_\Gamma}\geq0.
 \label{eq:app_massflow_density}
\end{equation}
The unit-density equilibrium at \(T_\Gamma\) is represented by
\(\bm W_\Gamma^{(1)}=(RT_\Gamma,\bm0,T_\Gamma)^{\mathsf T}\). Define the
discrete outgoing mass-flux density and the incoming mass-flux density of this
unit-density equilibrium by
\begin{equation}
 j_b^{+}
 =\sum_{\bm\xi_k\in\Xi_b^+}\varpi_k
 (\bm\xi_k\cdot\bm n_b)
 f_{i,k}^{\mathrm{rec}},
 \qquad
 j_{\Gamma,b}^{-,(1)}
 =\sum_{\bm\xi_k\in\Xi_b^-}\varpi_k
 (\bm\xi_k\cdot\bm n_b)
 f^{\mathrm{eq}}(\bm W_\Gamma^{(1)};\bm\xi_k).
 \label{eq:app_massflow_discrete_half_fluxes}
\end{equation}
The incoming Maxwellian amplitude and distribution are
\begin{equation}
 \rho_{\Gamma,b}
 =\frac{-j_\Gamma-j_b^{+}}
 {j_{\Gamma,b}^{-,(1)}},
 \qquad
 f_{b,k}^{\mathrm{bc}}
 =\rho_{\Gamma,b}
 f^{\mathrm{eq}}(\bm W_\Gamma^{(1)};\bm\xi_k),
 \quad \bm\xi_k\in\Xi_b^-.
 \label{eq:app_massflow_kinetic_distribution}
\end{equation}
Equation~\eqref{eq:app_massflow_kinetic_distribution} enforces the prescribed
facewise mass-flux density in the discrete velocity space. All walls considered in
this work are stationary; fully diffuse reflection at temperature \(T_w\) is
obtained by setting \(j_\Gamma=0\) and \(T_\Gamma=T_w\).

\subsection{Pressure inlet and outlet}
\label{app:kinetic_bc_pressure}

The pressure boundaries use an incoming reservoir Maxwellian constructed from
the reconstructed interior state
\(\bm W_{i,b}=(p_{i,b},\bm u_{i,b},T_{i,b})^{\mathsf T}\). The subscript
\((i,b)\) denotes a quantity reconstructed from the adjacent cell \(i\) to the
interior side of boundary face \(b\); \(\rho_{i,b}\) and \(c_{s,i,b}\) denote
the density and local speed of sound evaluated from this state, respectively.
For an outlet with prescribed pressure \(p_{\mathrm{out}}\), the corrected
normal velocity and reservoir state are
\begin{equation}
 \Delta u_{\mathrm{out}}
 =\frac{p_{i,b}-p_{\mathrm{out}}}{\rho_{i,b}c_{s,i,b}},
 \qquad
 \bm W_b^{\mathrm{out}}
 =\left(
 p_{\mathrm{out}},
 \bm u_{i,b}+\Delta u_{\mathrm{out}}\bm n_b,
 T_{i,b}\right)^{\mathsf T}.
 \label{eq:app_pressure_outlet_state}
\end{equation}
For an inlet with prescribed pressure \(p_{\mathrm{in}}\) and temperature
\(T_{\mathrm{in}}\), the corresponding quantities are
\begin{equation}
 \Delta u_{\mathrm{in}}
 =\frac{p_{i,b}-p_{\mathrm{in}}}{\rho_{i,b}c_{s,i,b}},
 \qquad
 \bm W_b^{\mathrm{in}}
 =\left(
 p_{\mathrm{in}},
 \bm u_{i,b}+\Delta u_{\mathrm{in}}\bm n_b,
 T_{\mathrm{in}}\right)^{\mathsf T}.
 \label{eq:app_pressure_inlet_state}
\end{equation}
With \(\bm W_b\) selected from
Eqs.~\eqref{eq:app_pressure_outlet_state} and
\eqref{eq:app_pressure_inlet_state}, the incoming distribution is
\begin{equation}
 f_{b,k}^{\mathrm{bc}}
 =f^{\mathrm{eq}}(\bm W_b;\bm\xi_k),
 \qquad \bm\xi_k\in\Xi_b^-.
 \label{eq:app_pressure_kinetic_distribution}
\end{equation}

\subsection{Far-field boundary}
\label{app:kinetic_bc_farfield}

For the prescribed state
\(\bm W_\infty=(p_\infty,\bm u_\infty,T_\infty)^{\mathsf T}\), the incoming
distribution is
\begin{equation}
 f_{b,k}^{\mathrm{bc}}
 =f^{\mathrm{eq}}(\bm W_\infty;\bm\xi_k),
 \qquad \bm\xi_k\in\Xi_b^-.
 \label{eq:app_farfield_kinetic_distribution}
\end{equation}

\section{Kinetic-consistent macroscopic boundary implementation}
\label{app:macroscopic_boundary}

This appendix gives the closed-form evaluation of the Maxwellian half-space
moments in Eq.~\eqref{eq:pbc_boundary_half_moment} and the boundary-specific
expressions for \(\widetilde{\boldsymbol F}_b^-\) in
Eq.~\eqref{eq:pbc_boundary_incoming_update}.

\subsection{Maxwellian half-space moments}
\label{app:half_space_moments}

Complete the outward unit normal \(\boldsymbol n_b\) with two orthogonal unit tangents
\(\boldsymbol t_{1,b}\) and \(\boldsymbol t_{2,b}\) to form a local orthonormal frame. For a
molecular velocity \(\boldsymbol\xi\), let
\begin{equation}
 \xi_n=\boldsymbol\xi\cdot\boldsymbol n_b,
 \qquad
 \xi_{t1}=\boldsymbol\xi\cdot\boldsymbol t_{1,b},
 \qquad
 \xi_{t2}=\boldsymbol\xi\cdot\boldsymbol t_{2,b}.
 \label{eq:app_local_molecular_velocity}
\end{equation}
In this frame, the local representation of the equilibrium half-space moment
in Eq.~\eqref{eq:pbc_boundary_half_moment} is
\begin{equation}
 \boldsymbol F_{\mathrm{eq,loc}}^{\pm}
 (\boldsymbol W;\boldsymbol n_b)
 =\int_{\Xi_b^\pm}
 \xi_n
 \begin{bmatrix}
 1, &\xi_n, &\xi_{t1}, &\xi_{t2}, &\tfrac12|\boldsymbol\xi|^2
 \end{bmatrix}^{\mathsf T}
 f^{\mathrm{eq}}(\boldsymbol W;\boldsymbol\xi)\,
 \mathrm d\boldsymbol\xi.
 \label{eq:app_half_space_definition}
\end{equation}
This integral has the closed form
\begin{equation}
 \boldsymbol F_{\mathrm{eq,loc}}^{\pm}
 =\rho
 \begin{bmatrix}
 I_1^{\pm}\\
 I_2^{\pm}\\
 u_{t1}I_1^{\pm}\\
 u_{t2}I_1^{\pm}\\
 \dfrac12\left[I_3^{\pm}
 +(u_{t1}^2+u_{t2}^2+2a^2)I_1^{\pm}\right]
 \end{bmatrix}.
 \label{eq:app_closed_half_space_moments}
\end{equation}
Here $a=\sqrt{RT}$, $u_n=\boldsymbol u\cdot\boldsymbol n_b$, and
$u_{tr}=\boldsymbol u\cdot\boldsymbol t_{r,b}$ for $r=1,2$. Define
$\eta=u_n/(\sqrt{2}a)$ and
\begin{equation}
 \mathcal C_{\pm}(\eta)
 =\frac12\left[1\pm\operatorname{erf}(\eta)\right],
 \qquad
 \mathcal G(\eta)=\frac{\exp(-\eta^2)}{\sqrt{2\pi}}.
 \label{eq:app_half_space_functions}
\end{equation}
The required normal moments are
\begin{equation}
 \begin{aligned}
  I_1^{\pm}
  &=u_n\mathcal C_{\pm}
    \pm a\mathcal G,\\
  I_2^{\pm}
  &=(u_n^2+a^2)\mathcal C_{\pm}
    \pm u_na\mathcal G,\\
  I_3^{\pm}
  &=(u_n^3+3u_na^2)\mathcal C_{\pm}
    \pm a
    (u_n^2+2a^2)\mathcal G.
 \end{aligned}
 \label{eq:app_normal_half_space_moments}
\end{equation}
Let the five-moment rotation from the local frame to the global Cartesian
frame be
\begin{equation}
 \boldsymbol P=
 \begin{bmatrix}
  1&0&0&0&0\\
  0&n_x&t_{1x}&t_{2x}&0\\
  0&n_y&t_{1y}&t_{2y}&0\\
  0&n_z&t_{1z}&t_{2z}&0\\
  0&0&0&0&1
 \end{bmatrix}.
 \label{eq:app_five_moment_rotation}
\end{equation}
Thus, the global half-space moment defined in
Eq.~\eqref{eq:pbc_boundary_half_moment} is recovered as
\begin{equation}
 \boldsymbol F_{\mathrm{eq}}^{\pm}
 (\boldsymbol W;\boldsymbol n_b)
 =\boldsymbol P\boldsymbol F_{\mathrm{eq,loc}}^{\pm}
 (\boldsymbol W;\boldsymbol n_b).
 \label{eq:app_global_half_space_moment}
\end{equation}

\subsection{Prescribed mass-flow and fully diffuse wall boundaries}
\label{app:bc_massflow}

For the prescribed mass-flow boundary in \ref{app:kinetic_bc_massflow}, the
target outward mass flux through face \(b\) is \(-j_\Gamma A_b\). Given the
outgoing boundary flux \(\bm F_b^+\), the incoming Maxwellian density and flux
are
\begin{equation}
 \widetilde{\rho}_{\Gamma,b}
 =\frac{j_\Gamma A_b+(\bm F_b^+)_\rho}
 {A_b\sqrt{RT_\Gamma/(2\pi)}},
 \qquad
 \widetilde{\bm F}_b^-\!\left(\bm W_{i,b},\bm F_b^+\right)
 =A_b\widetilde{\rho}_{\Gamma,b}
 \bm F_{\mathrm{eq}}^{-}(\bm W_\Gamma^{(1)};\bm n_b).
 \label{eq:app_massflow_incoming_flux}
\end{equation}
Here the subscript \(\rho\) denotes the mass-flux component.
By construction, the total boundary flux defined in
Eq.~\eqref{eq:pbc_consistent_boundary_flux} satisfies the prescribed facewise
mass-flow condition, \((\bm F_b^m)_\rho=-j_\Gamma A_b\), at every PBC
iteration.

For a stationary fully diffuse wall, setting \(j_\Gamma=0\) and
\(T_\Gamma=T_w\) in Eq.~\eqref{eq:app_massflow_incoming_flux} gives the
incoming diffuse-reflection flux and enforces zero normal mass flow.

\subsection{Pressure inlet and outlet}
\label{app:bc_pressure}

The pressure-outlet and pressure-inlet reservoir states
\(\bm W_b^{\mathrm{out}}\) and \(\bm W_b^{\mathrm{in}}\) are defined in
Eqs.~\eqref{eq:app_pressure_outlet_state} and
\eqref{eq:app_pressure_inlet_state}, respectively. The continuous incoming flux is
\begin{equation}
 \widetilde{\bm F}_b^-\!\left(\bm W_{i,b},\bm F_b^+\right)
 =A_b
 \begin{cases}
  \bm F_{\mathrm{eq}}^{-}(\bm W_b^{\mathrm{out}};\bm n_b),
  & \text{pressure outlet},\\[1mm]
  \bm F_{\mathrm{eq}}^{-}(\bm W_b^{\mathrm{in}};\bm n_b),
  & \text{pressure inlet}.
 \end{cases}
 \label{eq:app_pressure_incoming_flux}
\end{equation}

\subsection{Far-field boundary}
\label{app:bc_farfield}

With the prescribed state \(\bm W_\infty\) defined in
\ref{app:kinetic_bc_farfield}, the continuous incoming flux is
\begin{equation}
 \widetilde{\bm F}_b^-\!\left(\bm W_{i,b},\bm F_b^+\right)
 =A_b\bm F_{\mathrm{eq}}^{-}(\bm W_\infty;\bm n_b).
 \label{eq:app_farfield_incoming_flux}
\end{equation}
It is independent of the interior state and remains fixed during the PBC
iterations.

Equations~\eqref{eq:app_massflow_incoming_flux}, 
\eqref{eq:app_pressure_incoming_flux}, and 
\eqref{eq:app_farfield_incoming_flux} give the boundary-specific incoming 
fluxes in 
Eq.~\eqref{eq:pbc_boundary_incoming_update}. Together 
with Eq.~\eqref{eq:app_global_half_space_moment}, they complete the boundary 
update in Eqs.~\eqref{eq:pbc_boundary_outgoing_update}--
\eqref{eq:pbc_consistent_boundary_flux}.

\section{Analytical Jacobians of the boundary fluxes}
\label{app:boundary_jacobians}

Only the current continuous half-space fluxes in
Eqs.~\eqref{eq:pbc_boundary_outgoing_update} and
\eqref{eq:pbc_boundary_incoming_update} are differentiated; the bracketed
discrete consistency corrections remain fixed during the PBC
iterations. Consistent with the cell-centred boundary linearization in
Section~\ref{subsubsec:pbc_pressure_velocity},
\(\boldsymbol W_{i,b}=\boldsymbol W_i\) is used in the Jacobian. The
full-state form of Eq.~\eqref{eq:pbc_boundary_pu_jacobian} is
\begin{equation}
 \begin{aligned}
  \boldsymbol J_{b,i}
  % ={}&\frac{\partial\boldsymbol F_b}{\partial\boldsymbol W_i}
  ={}&\left(
   \boldsymbol I_5+
   \frac{\partial\widetilde{\boldsymbol F}_b^-
   (\boldsymbol W_i,\boldsymbol F_b^+)}
        {\partial\boldsymbol F_b^+}
  \right)
  \frac{\partial\widetilde{\boldsymbol F}_b^+(\boldsymbol W_i)}
       {\partial\boldsymbol W_i}
  +\frac{\partial\widetilde{\boldsymbol F}_b^-
   (\boldsymbol W_i,\boldsymbol F_b^+)}
        {\partial\boldsymbol W_i}
  \in\mathbb R^{5\times5}.
 \end{aligned}
 \label{eq:app_full_boundary_flux_jacobian}
\end{equation}
Its upper-left \(4\times4\) block gives the pressure--velocity Jacobian in
Eq.~\eqref{eq:pbc_boundary_pu_jacobian}, and its final column is used in
Eq.~\eqref{eq:pbc_boundary_temperature_jacobian}.

\subsection{Maxwellian half-space Jacobian}
\label{app:half_space_jacobian}

Each boundary Jacobian uses the Maxwellian half-space derivative. Let
\(\boldsymbol h^\pm\) denote the column vector in
Eq.~\eqref{eq:app_closed_half_space_moments}, so that
\(\boldsymbol F_{\mathrm{eq,loc}}^\pm=\rho\boldsymbol h^\pm\). Differentiating
Eqs.~\eqref{eq:app_closed_half_space_moments} and
\eqref{eq:app_normal_half_space_moments} gives
\begin{equation}
 \begin{aligned}
  \boldsymbol h_{u_n}^\pm
  &=
  \begin{bmatrix}
   \mathcal C_\pm\\
   2I_1^\pm\\
   u_{t1}\mathcal C_\pm\\
   u_{t2}\mathcal C_\pm\\
   \dfrac12\!\left[
    3I_2^\pm+(u_{t1}^2+u_{t2}^2+2a^2)\mathcal C_\pm
   \right]
  \end{bmatrix},
  &
  \boldsymbol h_{u_{t1}}^\pm
  &=
  \begin{bmatrix}
   0\\0\\I_1^\pm\\0\\u_{t1}I_1^\pm
  \end{bmatrix},\\[2mm]
  \boldsymbol h_{u_{t2}}^\pm
  &=
  \begin{bmatrix}
   0\\0\\0\\I_1^\pm\\u_{t2}I_1^\pm
  \end{bmatrix},
  &
  \boldsymbol h_T^\pm
  &=
  \begin{bmatrix}
   I_{1,T}^\pm\\
   R\mathcal C_\pm\\
   u_{t1}I_{1,T}^\pm\\
   u_{t2}I_{1,T}^\pm\\
   \dfrac12\!\left[
    5RI_1^\pm+(u_{t1}^2+u_{t2}^2+2a^2)I_{1,T}^\pm
   \right]
  \end{bmatrix},
 \end{aligned}
 \label{eq:app_local_half_range_derivatives}
\end{equation}
where \(I_{1,T}^\pm=\pm a\mathcal G/(2T)\). Combining these derivatives with
\(\rho=p/(RT)\) and the rotation \(\boldsymbol P\) in
Eq.~\eqref{eq:app_five_moment_rotation} gives the half-space Jacobian in
global Cartesian coordinates:
\begin{equation}
 \boldsymbol J_{\mathrm{eq}}^\pm
 (\boldsymbol W;\boldsymbol n_b)
 =\boldsymbol P\,\rho
 \begin{bmatrix}
  \dfrac{\boldsymbol h^\pm}{p}&
  \boldsymbol h_{u_n}^\pm&
  \boldsymbol h_{u_{t1}}^\pm&
  \boldsymbol h_{u_{t2}}^\pm&
  \boldsymbol h_T^\pm-\dfrac{\boldsymbol h^\pm}{T}
 \end{bmatrix}
 \boldsymbol P^{\mathsf T}.
 \label{eq:app_global_half_range_jacobian}
\end{equation}
It follows from Eq.~\eqref{eq:pbc_boundary_continuous_outgoing_flux} that
\(\partial\widetilde{\boldsymbol F}_b^+/\partial\boldsymbol W_i
=A_b\boldsymbol J_{\mathrm{eq}}^+(\boldsymbol W_i;\boldsymbol n_b)\).

\subsection{Prescribed mass-flow and fully diffuse wall boundaries}
\label{app:massflow_boundary_jacobian}

Let
\(\boldsymbol e_\rho^{\mathsf T}
=\begin{bmatrix}1&0&0&0&0\end{bmatrix}\) select the mass-flux component.
Differentiating Eq.~\eqref{eq:app_massflow_incoming_flux} gives the incoming
flux derivatives and the resulting full boundary Jacobian as
\begin{equation}
 \begin{aligned}
  \frac{\partial\widetilde{\boldsymbol F}_b^-}
       {\partial\boldsymbol F_b^+}
  &=\sqrt{\frac{2\pi}{RT_\Gamma}}
    \boldsymbol F_{\mathrm{eq}}^-
    (\boldsymbol W_\Gamma^{(1)};\boldsymbol n_b)
    \boldsymbol e_\rho^{\mathsf T},
  \qquad
  \frac{\partial\widetilde{\boldsymbol F}_b^-}
       {\partial\boldsymbol W_i}
  =\boldsymbol 0,\\[1mm]
  \boldsymbol J_{b,i}
  &=\left[
    \boldsymbol I_5
    +\sqrt{\frac{2\pi}{RT_\Gamma}}
     \boldsymbol F_{\mathrm{eq}}^-
     (\boldsymbol W_\Gamma^{(1)};\boldsymbol n_b)
     \boldsymbol e_\rho^{\mathsf T}
   \right]
   A_b\boldsymbol J_{\mathrm{eq}}^+
   (\boldsymbol W_i;\boldsymbol n_b).
 \end{aligned}
 \label{eq:app_massflow_boundary_jacobian}
\end{equation}
The stationary fully diffuse wall has the same Jacobian with
\(T_\Gamma=T_w\).

\subsection{Pressure inlet and outlet}
\label{app:pressure_boundary_jacobian}

For either pressure boundary, let \(\Delta u\) denote the corresponding
normal-velocity correction in
Eqs.~\eqref{eq:app_pressure_outlet_state} and
\eqref{eq:app_pressure_inlet_state}. The required derivatives are
\begin{equation}
 \frac{\partial\Delta u}{\partial p_i}
 =\frac{1}{\rho_i c_{s,i}}-\frac{\Delta u}{p_i},
 \qquad
 \frac{\partial\Delta u}{\partial T_i}
 =\frac{\Delta u}{2T_i},
 \qquad
 \frac{\partial\Delta u}{\partial\boldsymbol u_i}
 =\boldsymbol 0.
 \label{eq:app_pressure_velocity_correction_derivatives}
\end{equation}
Accordingly, the two pressure-reservoir Jacobians are
\begin{equation}
 \begin{aligned}
  \boldsymbol J_{b,\mathrm{out}}^{\mathrm{res}}
  &=\frac{\partial\boldsymbol W_b^{\mathrm{out}}}
          {\partial\boldsymbol W_i}
  =
  \begin{bmatrix}
   0&\boldsymbol0_{1\times3}&0\\
   \boldsymbol n_b\dfrac{\partial\Delta u_{\mathrm{out}}}{\partial p_i}
   &\boldsymbol I_3
   &\boldsymbol n_b\dfrac{\partial\Delta u_{\mathrm{out}}}{\partial T_i}\\
   0&\boldsymbol0_{1\times3}&1
  \end{bmatrix},\\[2mm]
  \boldsymbol J_{b,\mathrm{in}}^{\mathrm{res}}
  &=\frac{\partial\boldsymbol W_b^{\mathrm{in}}}
          {\partial\boldsymbol W_i}
  =
  \begin{bmatrix}
   0&\boldsymbol0_{1\times3}&0\\
   \boldsymbol n_b\dfrac{\partial\Delta u_{\mathrm{in}}}{\partial p_i}
   &\boldsymbol I_3
   &\boldsymbol n_b\dfrac{\partial\Delta u_{\mathrm{in}}}{\partial T_i}\\
   0&\boldsymbol0_{1\times3}&0
  \end{bmatrix}.
 \end{aligned}
 \label{eq:app_pressure_reservoir_jacobians}
\end{equation}
The incoming pressure-boundary flux is independent of
\(\boldsymbol F_b^+\), i.e.,
\(\partial\widetilde{\boldsymbol F}_b^-/\partial\boldsymbol F_b^+
=\boldsymbol 0\). Substitution of the corresponding reservoir
Jacobian gives
\begin{equation}
 \boldsymbol J_{b,i}
 =A_b
 \begin{cases}
  \boldsymbol J_{\mathrm{eq}}^+(\boldsymbol W_i;\boldsymbol n_b)
  +\boldsymbol J_{\mathrm{eq}}^-(\boldsymbol W_b^{\mathrm{out}};\boldsymbol n_b)
   \boldsymbol J_{b,\mathrm{out}}^{\mathrm{res}},
  &\text{pressure outlet},\\[1mm]
  \boldsymbol J_{\mathrm{eq}}^+(\boldsymbol W_i;\boldsymbol n_b)
  +\boldsymbol J_{\mathrm{eq}}^-(\boldsymbol W_b^{\mathrm{in}};\boldsymbol n_b)
   \boldsymbol J_{b,\mathrm{in}}^{\mathrm{res}},
  &\text{pressure inlet}.
 \end{cases}
 \label{eq:app_pressure_boundary_jacobian}
\end{equation}

\subsection{Far-field boundary}
\label{app:farfield_boundary_jacobian}

The far-field incoming flux is independent of both \(\boldsymbol W_i\) and
\(\boldsymbol F_b^+\). Therefore,
\begin{equation}
 \frac{\partial\widetilde{\boldsymbol F}_b^-}
      {\partial\boldsymbol W_i}
 =\boldsymbol0,
 \qquad
 \frac{\partial\widetilde{\boldsymbol F}_b^-}
      {\partial\boldsymbol F_b^+}
 =\boldsymbol0,
 \qquad
 \boldsymbol J_{b,i}
 =A_b\boldsymbol J_{\mathrm{eq}}^+
 (\boldsymbol W_i;\boldsymbol n_b).
 \label{eq:app_farfield_boundary_jacobian}
\end{equation}

\bibliographystyle{elsarticle-num}

\bibliography{ref}

\end{document}